\documentclass[a4paper,fleqn,usenatbib]{mnras}

\usepackage{newtxtext,newtxmath}
\usepackage[T1]{fontenc}
\usepackage{ae,aecompl}

\usepackage{graphicx}	
\usepackage{amsmath}	
\usepackage{float}
\usepackage{color}
\usepackage{xcolor}
\usepackage{colortbl}
\usepackage{nicefrac}
\usepackage{hyperref}
\usepackage{array}
\newcolumntype{$}{>{\global\let\currentrowstyle\relax}}
\newcolumntype{^}{>{\currentrowstyle}}
\newcommand{\rowstyle}[1]{\gdef\currentrowstyle{#1}%
  #1\ignorespaces
}

\newcommand{\teh}{\textsc{T-800}}
\newcommand{\johnc}{\textsc{John Connor}}
\newcommand{\sarahc}{\textsc{Sarah Connor}}
\newcommand{\darwin}{\textsc{darwin}}
\newcommand{\rext}{\ensuremath{r^{\text{ext}}}}
\newcommand{\rextf}{\ensuremath{r^{\text{ext}}_{\text{final}}}}
\newcommand{\rint}{\ensuremath{r^{\text{int}}}}
\newcommand{\radd}{\ensuremath{r_{\text{d}}}}
\newcommand{\FD}{\ensuremath{\mathfrak{F}_{\text{D}}}}
\newcommand{\mFD}{\ensuremath{\langle\mathfrak{F}_{\text{D}}\rangle}}
\newcommand{\cs}{\ensuremath{c_{\text{s}}}}
\newcommand{\uss}{\ensuremath{u_{\text{ss}}}}
\newcommand{\ussi}{\ensuremath{u_{\text{ss},\infty}}}
\newcommand{\vD}{\ensuremath{v_{\text{D}}}}
\newcommand{\vzeta}{\ensuremath{v_{\zeta}}}
\newcommand{\ddg}{\ensuremath{\delta_{\text{dg}}}}
\newcommand{\mddg}{\ensuremath{\langle\ddg\rangle}}
\newcommand{\vDeq}{\ensuremath{\mathring{v}_{\text{D}}}}
\newcommand{\vDinf}{\ensuremath{v_{\text{D},\infty}}}
\newcommand{\dissc}{\ensuremath{\mathcal{K}}}
\newcommand{\agr}{\ensuremath{a_{\text{gr}}}}
\newcommand{\agrt}{\ensuremath{\widetilde{a}_{\text{gr}}}}
\newcommand{\taui}{\ensuremath{\tau^{-1}}}
\newcommand{\tauG}{\ensuremath{\tau_{\text{gr}}}}
\newcommand{\tauGi}{\ensuremath{\tau^{-1}_{\text{gr}}}}
\newcommand{\tausp}{\ensuremath{\tau^{-1}_{\text{ns}}}}
\newcommand{\taudc}{\ensuremath{\tau^{-1}_{\text{dc}}}}

\newcommand{\rhog}{\ensuremath{\rho_{\text{g}}}}

\newcommand{\rhod}[1][]{%
  \ifthenelse{\equal{#1}{}}{\ensuremath{\rho_{\text{d}}}}{\ensuremath{\rho_{\text{d},#1}}}
}

\newcommand{\sign}{\ensuremath{\text{sign}}}
\newcommand{\rhoint}{\ensuremath{\rho_{\text{m}}}}
\newcommand{\mH}{\ensuremath{m_{\text{H}}}}
\newcommand{\massp}{\ensuremath{m_{\text{p}}}}
\newcommand{\NA}{\ensuremath{N_{\text{A}}}}
\newcommand{\ted}{\ensuremath{T_{\text{d}}}}
\newcommand{\tedg}{\ensuremath{T_{\text{d,grey}}}}
\newcommand{\teg}{\ensuremath{T_{\text{g}}}}
\newcommand{\ter}{\ensuremath{T_{\text{r}}}}
\newcommand{\fdrag}{\ensuremath{f_{\text{drag}}}}
\newcommand{\fradd}{\ensuremath{f_{\text{rad,d}}}}
\newcommand{\fgravd}{\ensuremath{f_{\text{grav,d}}}}
\newcommand{\fin}{\ensuremath{f_{\text{in}}}}
\newcommand{\kappanu}{\ensuremath{\kappa_{\nu}}}
\newcommand{\kappanugd}{\ensuremath{\kappa_{\nu,\text{g}+\text{d}}}}
\newcommand{\kappaJHS}{\ensuremath{\kappa_{\text{J,H,S}}}}
\newcommand{\kappaJ}{\ensuremath{\kappa_{\text{J}}}}
\newcommand{\kappaH}{\ensuremath{\kappa_{\text{H}}}}
\newcommand{\kappaS}{\ensuremath{\kappa_{\text{S}}}}

\newcommand{\kappaR}{\ensuremath{\kappa_{\text{R}}}}
\newcommand{\kappadnu}{\ensuremath{\kappa_{\text{d},\nu}}}

\newcommand{\kappadS}{\ensuremath{\kappa_{\text{d,S}}}}

\newcommand{\chinu}{\ensuremath{\chi_{\nu}}}
\newcommand{\chiJHS}{\ensuremath{\chi_{\text{J,H,S}}}}
\newcommand{\chiJ}{\ensuremath{\chi_{\text{J}}}}
\newcommand{\chiH}{\ensuremath{\chi_{\text{H}}}}
\newcommand{\chiS}{\ensuremath{\chi_{\text{S}}}}

\newcommand{\chiR}{\ensuremath{\chi_{\text{R}}}}
\newcommand{\chigrey}{\ensuremath{\chi}}
\newcommand{\Pg}{\ensuremath{P_{\text{g}}}}
\newcommand{\Qnu}{\ensuremath{Q_{X,\nu}}}
\newcommand{\Qpnu}{\ensuremath{Q_{X,\nu}^{\prime}}}

\newcommand{\Qpnuext}{\ensuremath{Q_{\text{ext},\nu}^{\prime}}}

\newcommand{\Qnuabs}{\ensuremath{Q_{\text{abs},\nu}}}
\newcommand{\Qnuabsp}{\ensuremath{Q_{\text{abs},\nu}^{\prime}}}
\newcommand{\Qnuabspr}{\ensuremath{Q_{\text{abs},\nu}(\text{pr})}}

\newcommand{\QnuabsH}{\ensuremath{Q_{\text{abs},\nu}(\chiR)}}
\newcommand{\Qnuext}{\ensuremath{Q_{\text{ext},\nu}}}
\newcommand{\Qnusca}{\ensuremath{Q_{\text{sca},\nu}}}

\newcommand{\cost}{\ensuremath{\langle\cos\theta\rangle}}
\newcommand{\kB}{\ensuremath{k_{\text{B}}}}
\newcommand{\Sourcef}{\ensuremath{S_{\text{g}}}}
\newcommand{\B}{\ensuremath{B}}
\newcommand{\Bnu}{\ensuremath{B_{\nu}}}
\newcommand{\Bdnu}{\ensuremath{B_{\text{d},\nu}}}
\newcommand{\fedd}{\ensuremath{f_{\text{Edd}}}}
\newcommand{\kms}{\ensuremath{\,\mbox{km}\,\mbox{s}^{-1}}}

\newcommand{\drhog}{\ensuremath{\ddg\mathfrak{F}_{\text{D}}}}
\newcommand{\drhogf}{\ensuremath{\displaystyle\ddg\mathfrak{F}_{\text{D}}}}
\newcommand{\lav}{\ensuremath{l_{\text{av}}}}
\newcommand{\uinf}{\ensuremath{u_{\infty}}}
\newcommand{\vinf}{\ensuremath{v_{\infty}}}
\newcommand{\nC}{\ensuremath{n_{\text{C}}}}
\newcommand{\fcond}{\ensuremath{f_{\text{cond}}}}
\newcommand{\fcondPC}{\ensuremath{f_{\text{cond},PC}}}

\newcommand{\vdrinf}{\ensuremath{v_{\text{D},\infty}}}
\newcommand{\ncore}{\ensuremath{N_{\text{C}}}}
\newcommand{\ngrid}{\ensuremath{N_{\text{D}}}}
\newcommand{\nnu}{\ensuremath{N_{\nu}}}
\newcommand{\teff}{\ensuremath{T_{\text{eff}}}}
\newcommand{\deltauvelp}{\ensuremath{\Delta\,u_{\text{p}}}}
\newcommand{\mdotu}{\ensuremath{M_{\sun}\,\text{yr}^{-1}}}
\newcommand{\sistd}{\ensuremath{\sigma_{\text{s}}}}
\newcommand{\StefanB}{\ensuremath{\sigma_{\text{SB}}}}
\newcommand{\mdot}{\ensuremath{\dot{M}}}
\newcommand{\mmdot}{\ensuremath{\langle\mdot\rangle}}
\newcommand{\dmdot}{\ensuremath{\dot{M}_{\text{d}}}}
\newcommand{\mdmdot}{\ensuremath{\langle\dmdot\rangle}}
\newcommand{\mmdmdot}{\ensuremath{\langle\dmdot/\mdot\rangle}}
\newcommand{\muinf}{\ensuremath{\langle u_{\infty}\rangle}}
\newcommand{\mfcond}{\ensuremath{\langle f_{\text{cond}}\rangle}}
\newcommand{\mdrhog}{\ensuremath{\langle\drhog\rangle}}
\newcommand{\mdrhogf}{\ensuremath{\left\langle\drhogf\right\rangle}}

\newcommand{\mvdrinf}{\ensuremath{\langle\vdrinf\rangle}}
\newcommand{\mradd}{\ensuremath{\langle r_{\text{d}}\rangle}}
\newcommand{\mvD}{\ensuremath{\langle\vD\rangle}}
\newcommand{\mvDinf}{\ensuremath{\langle\vDinf\rangle}}

\newcommand{\Rs}{\ensuremath{R_{\star}}}
\newcommand{\Ms}{\ensuremath{M_\star}}
\newcommand{\Ls}{\ensuremath{L_\star}}
\newcommand{\Lsun}{\ensuremath{L_{\sun}}}
\newcommand{\Msun}{\ensuremath{M_{\sun}}}
\newcommand{\CtO}{\ensuremath{\log\left(C-O\right)+12}}
\newcommand{\rfluc}{\ensuremath{\hat{r}}}
\newcommand{\pP}{\ensuremath{P}}
\newcommand{\dens}{\ensuremath{\text{g}\,\text{cm}^{-3}}}
\newcommand{\surft}{\ensuremath{\sigma_{\text{grain}}}}
\newcommand{\ergcm}{\ensuremath{\text{erg}\,\text{cm}^{-2}}}
\newcommand{\aC}{\ensuremath{\xi_{\text{C}}}}
\newcommand{\aCt}{\ensuremath{\xi_{\text{C}_2}}}
\newcommand{\aCtH}{\ensuremath{\xi_{\text{C}_2\text{H}}}}
\newcommand{\aCtHt}{\ensuremath{\xi_{\text{C}_2\text{H}_2}}}

\newcommand{\EIKone}{\ensuremath{^{\text{EI01}}K_1}}
\newcommand{\EIKtwo}{\ensuremath{^{\text{EI01}}K_2}}
\newcommand{\EIzeta}{\ensuremath{^{\text{EI01}}\zeta}}
\newcommand{\EIeta}{\ensuremath{^{\text{EI01}}\eta}}
\newcommand{\tauV}{\ensuremath{\tau_{\text{V}}}}
\newcommand{\rcond}{\ensuremath{r_{\text{c}}}}

\newcommand{\tsc}{\ensuremath{_{\text{c}}}}
\newcommand{\tsm}{\ensuremath{_{\text{m}}}}
\newcommand{\tsw}{\ensuremath{_{\text{w}}}}

\newcommand{\greq}{\mathrel{\overset{\makebox[0pt]{\mbox{\normalfont\tiny\sffamily grey}}}{=}}}

\def\lsim{\!\!\!\phantom{\le}\smash{\buildrel{}\over
  {\lower2.5dd\hbox{$\buildrel{\lower2dd\hbox{$\displaystyle<$}}\over
                               \sim$}}}\,\,}
\def\gsim{\!\!\!\phantom{\ge}\smash{\buildrel{}\over
  {\lower2.5dd\hbox{$\buildrel{\lower2dd\hbox{$\displaystyle>$}}\over
                              \sim$}}}\,\,}

\newcommand{\rSaHoa}{Paper~I}
\newcommand{\rSaHob}{Paper~II}
\newcommand{\rSaHoc}{Paper~III}
\newcommand{\rSa}{Paper~IV}
\newcommand{\rMWH}{M10}
\newcommand{\rMH}{MH11}
\newcommand{\rHM}{MH15}
\newcommand{\rENHAW}{E14}
\newcommand{\rGS}{GS14}
\newcommand{\rHGAJ}{H03}

\newcommand{\rEI}{EI01}
\newcommand{\rIE}{IE10}
\newcommand{\rBEM}{B19}

\title[3-component modelling of C-rich AGB star winds -- V]{Three-component modelling of C-rich AGB star winds -- V. Effects of frequency-dependent radiative transfer including drift\thanks{Dedicated to Agda Sandin.}}

\author[C. Sandin and L. Mattsson]{
Christer Sandin,\thanks{E-mail: ChristerSandin@yahoo.se}
Lars Mattsson
\\
Nordita, KTH Royal Institute of Technology and Stockholm University, Roslagstullsbacken 23, SE-106 91 Stockholm, Sweden
}

\date{Accepted 2020 September 01. Received 2020 September 01; in original form 2020 June 19.}

\pubyear{2020}

\begin{document}
\label{firstpage}
\pagerange{\pageref{firstpage}--\pageref{lastpage}}
\maketitle

\begin{abstract}
Stellar winds of cool carbon stars enrich the interstellar medium with significant amounts of carbon and dust. We present a study of the influence of two-fluid flow on winds where we add descriptions of frequency-dependent radiative transfer. Our radiation hydrodynamic models in addition include stellar pulsations, grain growth and ablation, gas-to-dust drift using one mean grain size, dust extinction based on both the small particle limit and Mie scattering, and an accurate numerical scheme. We calculate models at high spatial resolution using 1024 gridpoints and solar metallicities at 319 frequencies, and we discern effects of drift by comparing drift models to non-drift models. Our results show differences of up to 1000 per cent in comparison to extant results. Mass-loss rates and wind velocities of drift models are typically, but not always, lower than in non-drift models. Differences are larger when Mie scattering is used instead of the small particle limit. Amongst other properties, the mass-loss rates of the gas and dust, dust-to-gas density ratio, and wind velocity show an exponential dependence on the dust-to-gas speed ratio. Yields of dust in the least massive winds increase by a factor four when drift is used. We find drift velocities in the range $10$--$67\,\kms$, which is drastically higher than in our earlier works that use grey radiative transfer. It is necessary to include an estimate of drift velocities to reproduce high yields of dust and low wind velocities.
\end{abstract}


\section{Introduction}\label{sec:introduction}
Winds of AGB stars are believed to be driven by radiation pressure on dust grains, which create an outflow when they in turn drag the gas in the atmosphere along. These winds are relatively slow ($\sim10\,\kms$), but mass-loss rates can be high ($\sim10^{-5}\,\mdotu$) owing to high densities. The type of dust forming in AGB-star atmospheres depends on the chemical composition of the gas: oxygen-rich stars (C/O~$<1$) form mostly silicate-type grains \citep[but also iron dust can form in significant quantities, see][]{MaDeDi.:19}, whilst carbon-rich stars (C/O~$>1$) form mainly amorphous carbon (amC) grains and smaller amounts of grains of SiC and polycyclic aromatic hydrocarbons (PAHs). The latter type of stars is usually referred to as ``carbon stars'' and represents evolved stars with initial masses in the range $1.5$--$4\,\Msun$ that undergo so-called thermal pulses. That is,  after the helium shell runs out of fuel, the star derives its energy from hydrogen burning in a thin shell; eventually, accumulated helium from the hydrogen burning ignites, causing a helium shell flash. During the thermal pulses, which last a few hundred years, material from the inner regions is mixed into the outer layers. This process is referred to as dredge-up and changes the surface composition of the star; in particular, this is how an oxygen-rich AGB star evolves into a carbon star. The amount of carbon expelled by carbon stars is significant and they may thus play role for the evolution of carbon (and carbonaceous dust) in the universe, although it cannot be ruled out that massive stars may be equally important \citep[e.g.,][]{GuKaOl.:99,Ma:10}. The carbon production of carbon stars is important and understanding the wind-formation mechanisms is essential to the full picture.

Radiatively accelerated dust grains exert a drag force on the gas; this drag force depends on how well grains couple to the gas, which in turn depends on the radiation flux, the density and temperature of the gas, as well as the extinction and cross section of dust grains. In case gas and dust are perfectly coupled, often referred to as complete momentum coupling, the momentum gained by dust grains from the radiation pressure is immediately transferred (or ``shared'') with the gas. In this case, gas and dust move at the equilibrium drift velocity. Not only is drift ignored when dust and gas are assumed to move with the same velocity [position coupling (PC)], but also the mass of dust particles.

The full system of radiation hydrodynamics including a description of stellar pulsations and drift using one mean dust velocity is, for the first time, modelled by \citet[hereafter \rSaHoa]{SaHo:03}, \citet[hereafter \rSaHob]{SaHo:03b}, and \citet[hereafter \rSaHoc]{SaHo:04} -- these works are summarised with \citet{CSa:03} -- who use grey RT and find that models including drift form more dust in the form of larger grains. The numerical approach of these models is improved with the work presented in \citet[hereafter \rSa]{Sa:08}. See, for example, {\rSaHoa} (and references therein) for a list of earlier studies of cool star stellar winds that consider drift. Otherwise, \citet{LiLaBe:01} present the only existing wind model that includes both drift and frequency-dependent radiative transfer of the dust component (in a stationary formulation).

\citet[hereafter \rEI]{ElIv:01} and \citet[hereafter \rIE]{IvEl:10} deserve an honourable mention, as the authors prove the importance of drift analytically. Although they make several simplifying assumptions, their conclusions are robust: drift and reddening play crucial roles in shaping the velocity structure of dusty winds and the mass-loss rate must be strongly correlated with drift velocities.

With the simulation code we present here, {\teh}, we extend our gas-dust drift models with frequency-dependent gas and dust opacities, where we can choose between the small particle limit (SPL) and Mie scattering \citep[following our approach in][hereafter \rMH]{MaHo:11}, and we also calculate RT using a Feautrier-based solver \citep[using the readily available description of][hereafter \rHM]{HuMi:15} that allows the calculation of models with as many gridpoints as are unprecedented in the respect that there are no similar stellar-wind models that include as much physics and can operate with the exceptional numerical accuracy needed to deal with drift. Our results vitiate the current consensus that drift plays a minor role \citep[footnote~6]{HoOl:18}.

Here, we re-evaluate the wind formation for a set of the model parameters in \citet[hereafter {\rMWH}]{MaWaHo:10} and \citet[hereafter {\rENHAW}]{ErNoHoArWa:14}. Our aim is to scrutinise basic differences between drift and position coupled (PC) models. We first describe the physics and numerical features of our models in Section~\ref{sec:features}. The modelling procedure and results are then presented in Section~\ref{sec:procedure}. We discuss the influence of drift on our results in Section~\ref{sec:discussion} and summarise the paper with our conclusions in Section~\ref{sec:conclusions}.

\section{Model features and improvements}\label{sec:features}
As in the four previous papers in this series, we consider three interacting physical components in the outer atmosphere and wind of the star: gas, dust, and radiation field. The three components are described by a system of equations that conserve and describe the interchange of mass, energy, and momentum. We call our simulation code {\teh} and our simulation code for calculating initial structures {\johnc} (our analysis tool is similarly named {\sarahc}). Whilst the physical system is described in part in {\rSaHoa}--{\rSaHob} and references therein, a number of adjustments merit a more complete description of the current capabilities. In comparison to other AGB star wind models, {\teh} share most features with the {\darwin} PEDDRO-type\footnote{Pulsation-enhanced dust-driven outflow (PEDDRO), see for example \citet{HoOl:18}} models that originate with \citet{HoGaArJo:03}.

We describe the hydrodynamic equations and the physical terms of {\teh} in the following three subsections: the gas component, Section~\ref{sec:gasterms}; the dust component, Section~\ref{sec:dustterms}; the radiation field, Section~\ref{sec:radfield}; and the numerical method, Section~\ref{sec:numerics}. All used abbreviations and symbols are collected in Tables~\ref{app:tababbreviations}--\ref{app:tabcglossary}.

\subsection{The gas component}\label{sec:gasterms}
Matter is present in either gas or dust phase. Five equations describe the gas phase: the equation of integrated mass, the equation of continuity, the equation of motion, the equation of inner energy, and the equation of number density of the condensible material:
\begin{eqnarray}
m_r&=&\int_0^r4\pi (r')^2\rhog\,\text{d}r'\label{eq:integratedmass}\\
\frac{\partial}{\partial t}\rhog+\nabla\cdot(\rhog u)&=&-m_1\mathcal{S}\label{eq:gcont}\\
\frac{\partial}{\partial t}(\rhog u)+\nabla\cdot(\rhog u\,u)&=&-\nabla\Pg-\frac{Gm_r}{r^2}\rhog+\frac{4\pi}{c}\kappa_H\rhog H-\nonumber\\
&&-m_1\mathcal{S}v+\fdrag\label{eq:gmot}\\
        \frac{\partial}{\partial t}(\rhog e)+\nabla\cdot(\rhog e\,u)&=&-\Pg\nabla\cdot u+4\pi\rhog(\kappa_JJ-\kappa_S\Sourcef)+\nonumber\\
        &&+\frac{1+\varepsilon}{2}\vD\fdrag\label{eq:gene}\\
\frac{\partial}{\partial t}\nC+\nabla\cdot(\nC u) &=& -\mathcal{S},
\end{eqnarray}
where $m_r$ is the integrated mass at radius $r$, $\rhog$ the gas density, $t$ the time, $u$ the gas velocity, $m_1$ the dust monomer mass, $\mathcal{S}$ the net condensation rate (see below), {\Pg} the pressure, $G$ the gravitational constant, $c$ the light speed, {\kappaH} (\kappaJ, \kappaS) the gas opacity weighted with the first radiative moment $H$ (the zeroth radiative moment $J$, the source function \Sourcef), $v$ the dust velocity, {\fdrag} the drag force, $e$ the specific internal energy of the gas, $\varepsilon$ the fraction of specular collisions between gas and dust particles, $\vD=v-u$ the drift velocity, and {\nC} the number density of the condensible material. As before, an ideal gas law is used for the equation of state
\begin{equation}
\teg=\left(\gamma-1\right)\frac{\mu\NA\massp}{R}e,\quad\text{and}\quad \Pg=\left(\gamma-1\right)\rhog e,
\end{equation}
where {\teg} is the gas temperature, $\gamma=5/3$ is the ratio of specific heats, $\mu=1.26$ is the mean molecular weight, {\NA} is Avogadros constant, and {\massp} is the proton mass. Assuming local thermal equilibrium, the source function equals the Planck function, $\Sourcef=\B$. Abundances of carbon and oxygen are initially specified as (log) number fractions relative to hydrogen, i.e. $\epsilon_{\text{X}}=\log_{10}\left(n_{\text{X}}/n_{\text{H}}\right)+12$, and the carbon-to-oxygen-ratio is $C/O=n_{\text{C}}/n_{\text{O}}=10^{\epsilon_{\text{C}}}/10^{\epsilon_{\text{O}}}$. However, the ratio changes differently across the radial domain as dust forms.

The number densities of the molecules in the gas phase that are part of the grain formation are calculated in an equilibrium chemistry of H, H$_2$, C, C$_2$, C$_2$H, and C$_2$H$_2$. Partial pressures of single H and C atoms are calculated according to the description in \citep[chapter~10.3, hereafter \rGS]{GaSe:14}, and we use dissociation constants (\dissc) of \citet{ShHu:90}.

\subsection{The dust component}\label{sec:dustterms}
Five equations describe the dust phase: four moment equations $K_{0}$--$K_{3}$ of the grain-size distribution function (\rGS), and the equation of motion of the dust.
\begin{eqnarray}
\frac{\partial}{\partial t}K_j+\nabla\cdot(K_jv) &=& 
\frac{j}{3}\frac{1}{\taui}K_{j-1}+N_{\text{l}}^{j/3}J_{\star}\label{eq:dmom}\\
\frac{\partial}{\partial t}(\rhod v)+\nabla\cdot(\rhod v\,v)&=&
-\frac{Gm_r}{r^2}\rhod+\frac{4\pi}{c}\chi_\text{H}H+\nonumber\\
&&+m_1\mathcal{S}v-\fdrag\label{eq:dmot}
\end{eqnarray}
where $0\le j\le3$, $\rhod=m_1K_{3}$ the dust density, {\taui} the net grain growth rate, $N_{\text{l}}$ is the lower size-limit of macroscopic grains (we use $N_{\text{l}}=1000$ carbon atoms), $J_{\star}$ is the nucleation rate, and {\chiH} the extinction coefficient weighted with the first radiative moment $H$. The net grain growth is $\tau^{-1}=\tauG^{-1}-\tau_{\text{dc}}^{-1}-\tau_{\text{ns}}^{-1}$,
where {\tauGi} is (homogeneous and heterogeneous) grain growth, {\taudc} grain decay (by evaporation and chemical sputtering), and {\tausp} ablation (by non-thermal sputtering), see {\rSaHoc} for details. Dust particles are assumed to be spherical grains of amC. Effects of non-zero drift velocities are included in the grain growth and decay terms according to the description in {\rSaHoc}.

The moments of the size distribution allow calculation of average properties of the dust, including the total number density of dust grains $n_{\text{d}}=K_{0}$, the mean grain radius $\langle r_{\text{d}}\rangle=r_0K_{1}/K_{0}$ ($r_0$ is the monomer radius), the mean grain cross section $\langle\sigma\rangle=\pi r_0^2K_{2}/K_{0}$,\footnote{In earlier papers, we have used $\langle\sigma\rangle=\pi r_0^2K_{1}^2/K_{0}^2$, which results in slightly more problematic models in regions where dust vanishes ($K_{0}\rightarrow 0$).} and the mean grain size $\langle N\rangle=K_{3}/K_{0}$. The monomer radius is defined as \citep[equation 2.2]{GaKeSe:84}
\begin{eqnarray}
r_0=\left(\frac{3Am_{\text{p}}}{4\pi\rhoint}\right)^{\frac{1}{3}},
\end{eqnarray}
where $A$ is the atomic weight of the dust-forming species, and $\rhoint$ the intrinsic density of dust grains.
The net condensation rate $\mathcal{S}$ equals the right-hand side of equation~(\ref{eq:dmom}) when $j=3$, $\mathcal{S}=\tauGi K_{2}+N_{\text{l}}J_{\star}$. The term $m_1\mathcal{S}v$ accounts for momentum that is moved to the dust from the gas phase when dust forms; this term appears with a different sign in both equations of motion. The (average dust velocity) drag force is the same as in {\rSaHoa}, and we only consider specular collisions ($\varepsilon=1$):
\begin{eqnarray}
\fdrag=\langle\sigma\rangle\rhog n_{\text{d}}\frac{\vD^2}{2}C_{\text{D}}^{\text{LA}}=\pi r_0^2\frac{K_{2}}{K_{0}}\rhog K_{0}\frac{\vD^2}{2}\times\nonumber\\
        \frac{2}{\vD}\Bigg[\left(\vzeta^2+\vD^2\right)^{\frac{1}{2}}
          +\frac{1}{3}\pi^{\frac{1}{2}}(1-\varepsilon)\sqrt{\frac{2\kB}{\mu \mH}}T_{\text{d}}^{\frac{1}{2}}\Bigg],\label{eq:drag}
\end{eqnarray}
where $C_{\text{D}}^{\text{LA}}$ is the limits approximation of the drag coefficient (\rSaHoa, equation~23), {\ted} the dust temperature (equation~\ref{eq:tdust}), and {\vzeta} the thermal velocity $\vzeta=\sqrt{\zeta\teg}$, and $\zeta=128\kB/\left(9\pi\mu\mH\right)$, where {\mH} is the mass of a hydrogen atom.

In models that use position coupling (PC) instead of gas-dust drift, all terms in the dust equation of motion (equation~\ref{eq:dmot}) are added to the gas equation of motion (equation~\ref{eq:gmot}), and the equation is replaced with the relation $v=u$.

\subsection{The radiation field component}\label{sec:radfield}
The radiation field is described by the zeroth and first moment equations of the radiative transfer equation
\begin{eqnarray}
\frac{1}{c}\frac{\partial}{\partial t}J+\frac{1}{c}\nabla\cdot(Ju)&=&
        -\nabla\cdot H-\frac{1}{c}K\nabla\cdot u+\frac{u}{c}\frac{3K-J}{r}-\nonumber\\
      &&-\rhog(\kappaJ J-\kappaS\Sourcef)\label{eq:radenergy}\\
        \frac{1}{c}\frac{\partial}{\partial t}H+\frac{1}{c}\nabla\cdot(Hu)&=&-\frac{1}{q}\nabla\left(qK\right)-\nonumber\\
        &&-\frac{1}{c}H\nabla u-\left(\kappaH\rhog+\chiH\right)H,\label{eq:radflux}
\end{eqnarray}
where $J$ represents the radiative energy density, $H$ the radiative energy flux, $K$ is the second moment of the specific intensity, and $q$ the sphericality (equation~\ref{eq:q}). The two moment equations depend on three moments of the radiation field, which is why it is necessary to solve the RT equation to calculate the Eddington factor {\fedd}, which gives $K=\fedd J$.

We describe our approach to solve the RT equation next in Section~\ref{sec:rt}. The interactions between the radiation field and the gas and dust are described separately in Sections \ref{sec:rtgas} and \ref{sec:rtdust}. Regarding the frequency-dependent problem of RT, we follow the guidelines of \citet[\S 82]{MiWe:84}.

\subsubsection{Radiative transfer}\label{sec:rt}
We solve the equation of RT in spherical geometry, without any frequency redistribution. Up to now, all our models have used the time-independent spherical-geometry method of \citet{Yo:80}, \citet{Ba:88}, and \citet[chapter~9]{BoLaRoYo:07}; the method considers rays that are individually first integrated inwards and then outwards. Whilst this approach has worked well in models using grey RT and reasonably well in models using frequency-dependent RT, we found that it sometimes is unstable.

Following the literature \citep[\rHM]{MiWe:84,An:02}, we chose to write a new solver based on a differential-equation technique that uses a Feautrier-type solution along individual impact parameters. We used the description of (\rHM, chapter~19.1); because of numerical inaccuracies in optical shells that are more thin, we found it is necessary to either use quadruple precision when solving the resulting tridiagonal system of equations or rewrite the equations according to the mention in \citet[section~3.9.1]{No:82} and description in \citet[appendix~A]{RyHu:91}.

The inputs to the solver are the total absorption coefficient $\kappanugd=\kappanu+\chinu/\rhog$, where {\kappanu} and {\chinu} are the frequency-dependent gas opacity and the dust extinction efficiency, respectively; the source function 
\begin{equation}
\Sourcef{}_{,\nu}=\frac{\rhog\kappanu\Bnu(\teg)+\chinu\Bnu(\ted)}{\rhog\kappanu+\chinu},
\end{equation}
where {\Bnu} is the Planck function; the radiative temperature at the outer boundary, $T^{\text{ext}}_{\text{r}}=0\,$K; and a frequency-dependent expression for the radiative flux at the inner boundary $H^{\text{int}}$, which is placed where the diffusion limit applies (\rHM, equation~11.176),
\begin{equation}
H_{\nu}^\text{int}=\left(\frac{\kappaR}{\kappanu}\right)\left(\frac{\partial\Bnu/\partial\teg}{\partial B/\partial\teg}\right)H^{\text{int}},\label{eq:Hint}
\end{equation}
where {\kappaR} is the Rosseland mean opacity (equation~\ref{eq:rosseland}). The RT calculations yield mean-intensity-like and flux-like variables that are integrated over impact parameters according to the description in \citet{Yo:80} to yield the three radiative moments $J_\nu$, $H_\nu$, and $K_\nu$. And these moments are in turn used in the equations in the next two sections to calculate frequency-integrated properties.

The Eddington factor {\fedd} and the sphericality factor $q$ are calculated as
\begin{equation}
\fedd=\frac{\int_{0}^{\infty}K_{\nu}\text{d}\nu}{\int_{0}^{\infty}J_{\nu}\text{d}\nu}=\frac{\overline{K}}{\overline{J}},\label{eq:edd}
\end{equation}
and
\begin{equation}
\ln q=\int_{r_{\text{C}}}^{r}\frac{3\overline{K}-\overline{J}}{r^{\prime}\overline{K}}\text{d}r^{\prime}=\int_{r_{\text{C}}}^{r}\frac{3\fedd-1}{r^{\prime}\fedd}\text{d}r^{\prime},\label{eq:q}
\end{equation}
where $r_{\text{C}}$ is the radius at the inner boundary. The frequency-integrated zeroth and second radiative moments are over-lined here to indicate that these properties are calculated from the output of the RT calculations and not from the hydrodynamic equations.

The computing time using $N_{\nu}$ frequencies, {\ngrid} gridpoins, and $N_{\text{C}}$ core rays scales as $t\propto N_{\nu}\left(\ngrid N_{\text{C}}+\sum_iN_{\text{D},i}\right)\simeq N_{\nu}\ngrid^2$ (for $\ngrid\gg N_{\text{C}}$). The RT calculations can with advantage be executed in parallel for individual frequencies since no frequency redistribution is used.

\subsubsection{The interaction between radiation field and gas}\label{sec:rtgas}
For each wavelength, the RT equation is solved for the previous time step using the radial structure of the gas density, opacity, and temperature. The first three radiative moments $J_{\nu}$, $H_{\nu}$, and $K_{\nu}$ that result from the calculations are integrated over frequency to find the following moment-weighted gas opacities
\begin{eqnarray}
\kappaJ&=&\frac{\int_{0}^{\infty}\kappanu J_{\nu}\text{d}\nu}{\int_{0}^{\infty}J_{\nu}\text{d}\nu},\quad
\kappaH=\frac{\int_{0}^{\infty}\kappanu H_{\nu}\text{d}\nu}{\int_{0}^{\infty}H_{\nu}\text{d}\nu},\nonumber\\
\kappaS&=&\frac{\int_{0}^{\infty}\kappanu\Bnu\text{d}\nu}{\int_{0}^{\infty}\Bnu\text{d}\nu},\quad
\kappaR=\frac{\displaystyle\int_{0}^{\infty}\frac{\partial\Bnu}{\partial\teg}\text{d}\nu}{\displaystyle\int_{0}^{\infty}\frac{1}{\kappanu}\frac{\partial\Bnu}{\partial\teg}\text{d}\nu},\label{eq:rosseland}
\end{eqnarray}
where {\kappaS} is the Planck mean opacity. Gas opacities are provided as tabulated values \kappanu(\rhog, \teg, $\nu$). The tables are created with the \textsc{coma} code \citep{BAr:00,ArGiNo.:09} and include updates regarding abundances (solar composition), frequency interpolation and resolution (B.~Aringer, priv.comm.). The frequency integral ($0\le\nu<\infty$) is simply taken as the range of frequencies that is available in the tabulated data.

\subsubsection{The interaction betweeen radiation field and dust}\label{sec:rtdust}
The extinction coefficients of the dust that correspond to the weighted gas opacities are calculated as follows using the radial structure of the dust extinction coefficient, the dust temperature, and the mean grain radius (\mradd):
\begin{eqnarray}
\chiJ&=&\frac{\int_{0}^{\infty}\chinu J_{\nu}\text{d}\nu}{\int_{0}^{\infty}J_{\nu}\text{d}\nu},\quad
\chiH=\frac{\int_{0}^{\infty}\chinu H_{\nu}\text{d}\nu}{\int_{0}^{\infty}H_{\nu}\text{d}\nu},\nonumber\\
\kappadS&=&\frac{\int_{0}^{\infty}\kappadnu\Bdnu\text{d}\nu}{\int_{0}^{\infty}\Bdnu\text{d}\nu},\quad
\chiR=\frac{\displaystyle\int_{0}^{\infty}\frac{\partial\Bdnu}{\partial\ted}\text{d}\nu}{\displaystyle\int_{0}^{\infty}\frac{1}{\chinu}\frac{\partial\Bdnu}{\partial\ted}\text{d}\nu},\label{eq:chi}
\end{eqnarray}
where {\Bdnu} is the Planck function at the dust temperature {\ted}, $\kappadnu=\chinu/\rhog$, and we also have that (\rMH)
\begin{equation}
  \chinu=\pi\langle\Qnuabsp(\agr)\rangle\int_0^{\infty}\agr^3n(\agr)\text{d}\agr=\pi r_0^3K_{3}\Qnuabsp\left(\agrt\right),
\end{equation}
where {\agr} is the grain radius, $\agrt\equiv\mradd$ the mean grain radius, and $\Qnuabsp=\Qnuabs/\agr$ the absorption efficiency. The absorption efficiency {\Qnuabs} used to calculate {\chiJ} and {\kappaS} is
\begin{equation}
  \Qnuabs=\Qnuext-\Qnusca,
\end{equation}
and the absorption efficiency {\Qnuabspr} used to calculate {\chiH} accounting for radiation pressure is
\begin{equation}
  \Qnuabspr=\Qnuext-\cost_{\nu}\Qnusca,
\end{equation}
where {\Qnuext} is the extinction efficiency, {\Qnusca} the scattering efficiency, and $\cost_\nu$ the average scattering angle. Finally, the Rosseland dust extinction {\chiR} is calculated assuming $\Qnusca=0$, which is why $\QnuabsH=\Qnuext$. The frequency integral is also here taken as the range of frequencies that is available in the table.

Using Mie scattering, we followed the approach of {\rMH}. Whilst calculating $\Qnuext(\ted, \mradd)$, $\Qnusca(\ted, \mradd)$, and {$\cost_{\nu}(\ted, \mradd)$}, we did not as in {\rMH} use \textsc{bhmie} of \citet{BoHu:83}\footnote{The code \textsc{bhmie} modified by B. Draine and others is available at \href{https://www.astro.princeton.edu/~draine/scattering.html}{https://www.astro.princeton.edu/$\sim$draine/scattering.html}.}, instead we implemented the theory as described by {\rGS} (chapter~7.3). The calculation of these extinction coefficients requires tabulated values of the refractive indices $n_{\nu}$ (phase velocity) and $k_{\nu}$ (extinction coefficient). Assuming SPL, we use \citep[$m_\nu=n_\nu+ik_\nu$; e.g.][equation~2.30]{Wi:72i}
\begin{eqnarray}
\Qpnuext\simeq-\frac{8\pi\nu}{c}\Im\left(\frac{1-m_{\nu}^2}{2+m_{\nu}^2}\right)\;[\text{cm}^{-1}],
\end{eqnarray}
and assume that the average scattering angle vanishes, $\cost=0$.

We use the approach of {\rGS} (chapter~8.3; also see \rHGAJ) to calculate a dust temperature as
\begin{equation}
\ted^4=\frac{1}{\rhog}\frac{\chiJ}{\kappadS}\ter^4=\frac{1}{\rhog}\frac{\displaystyle\int_{0}^{\infty}\chinu J_{\nu}\text{d}\nu}{\displaystyle\int_{0}^{\infty}J_{\nu}\text{d}\nu}\frac{\displaystyle\int_{0}^{\infty}\Bdnu\text{d}\nu}{\displaystyle\int_{0}^{\infty}\kappadnu\Bdnu\text{d}\nu}\frac{\pi J}{\StefanB},\label{eq:tdust}
\end{equation}
where {\StefanB} is the Stefan-Boltzmann constant and the radiative temperature $\ter=(\pi J/\StefanB)^\frac{1}{4}$. We also compare our new results with grey calculations where we have assumed that $\tedg=\ter$ and that (see equation~4 in \rSaHoa)
\begin{eqnarray}
\chigrey=\pi r_0^3K_{3}\times4.4\ter.\label{eq:chigrey}
\end{eqnarray}

\subsection{Numerical method}\label{sec:numerics}
The eleven partial differential equations -- equations~(\ref{eq:gcont})--(\ref{eq:dmot}) and (\ref{eq:radenergy})--(\ref{eq:radflux}) -- are discretised in the volume-integrated conservation form on a staggered mesh, which together with the integrated-mass equation (equation~\ref{eq:integratedmass}) and the adaptive grid equation \citep[see, e.g.][section~3.2]{DoHo:91} form a system of thirteen equations. The non-linear system of equations is solved implicitly using a Newton-Raphson algorithm where the Jacobian of the system is inverted by the Henyey method. The flow of mass, energy, and momentum between the gridpoints is described with a second-order volume weighted van Leer advection scheme \citep{DoPiSt.:06}; details of the implemented advection scheme are described in {\rSa}. We give further details of the basics of the numerical method used with {\teh} in, for example, {\rSaHoa} and references therein.

\subsubsection{Using the adaptive grid equation or a mostly fixed grid}\label{sec:grid}
The fundamental problem with using a single adaptive grid equation for two fluids (gas and dust) or three components (gas, dust, and radiation field) is that it is very difficult to trace features in both the gas and the dust at the same time using a limited number of gridpoints \ngrid.

In comparison to our earlier work, we do not use the adaptive-grid equation to resolve shocks or other features. Instead, we fix the outer parts of the grid and set $\ngrid=1024$. In this approach, we do not let the outer boundary of the grid expand from the value used in the initial-model calculations to the pre-determined outer boundary of the wind calculations (as we do in \rSaHoa--{\rSa} and is done in all other implementations of stellar winds using the adaptive grid equation); the radial range of the initial model is instead set to the full range of the wind right away where the outer boundary typically is set to $40\Rs$, but with slower winds we use $20\Rs$. We calculate a grid using a logarithmic distribution of gridpoints. In some models, we find that it is necessary to increase the spatial resolution in the centre parts to achieve convergence, which is why we doubled the number of gridpoints inside of $r_{950}$ of the initial gridpoint 950 and re-distribute the remaining ($1024-2\times(1024-950)=$) 876 gridpoints in the region outside of $r_{950}$.

Owing to a grid that is fixed where there is dust, we use the same amount of artificial viscosity in both PC and drift models; according to the description in {\rSaHoa} (equations 13 and 14), we set the length scale to
\begin{eqnarray}
  \lav=3.5\times10^{-3}r.\label{eq:avisc}
\end{eqnarray}
We also do not use any artificial mass diffusion. (Notably, in {\rSaHoc} and {\rSa}, we use both artificial mass diffusion and a length scale twice as high in the presented models, where gridpoints move about to some extent throughout the model domain.) With one model setup, we instead tested using $\lav=r$, with negligible differences in resulting physical structures.

We compare some of the new models with models that use the adaptive grid equation, where the gas density and energy are resolved; these models use $\ngrid=100$, $\lav=r$, and extend out to $25\Rs$ (see Section~\ref{sec:resolution}).

\subsubsection{Modelling stellar pulsations at the inner boundary}
We model effects of stellar pulsations on the atmosphere and wind region using a \textit{piston} boundary condition, which is a sinusoidal and radially varying inner boundary that is placed above the region where the, so-called, $\kappa$-mechanism supposedly originates. The piston boundary is described with the period $P$ and the amplitude {\deltauvelp}. The adaptive grid equation is used to allow gridpoints where $r<2\Rs$ to stretch with the piston.

We do not model any inflow of mass through the inner boundary, as, for example, \citet[section~4.2]{SiIcDo:01} and \citet[section~2.5.1]{Wo:06} do. Abundances at the inner boundary remain unaffected by depletion of carbon owing to the efficient formation of dust using drift as the models do not include any mixing of the gas between lower layers. The amount of mass in the modelled envelope is $0.14$--$2.3$\% of the total mass including the core (see Table~\ref{tab:resmodpar}) before the dynamical modelling of the stellar wind begins. As the model domain is rather quickly depleted of material owing to the stellar wind, long-term modelling, covering thousands of years, is problematic in the current approach.

\subsubsection{Additional considerations in the new models}
Rosseland and Planck mean opacities and extinction coefficients are calculated when the tabularised frequency-dependent gas opacity and dust extinction data are loaded; the mean extinction coefficients are calculated for a set of pre-defined temperatures and grain radii. The equation of RT is solved for the previous time step in the first iteration, before the non-linear system of equations is solved for the current iteration. {\teh} executes the RT calculations in parallel for individual frequencies using a hybrid approach that makes use of both \textsc{OpenMP} and MPI.\footnote{Speedup tests show that it is more efficient to only use MPI with clusters where hyperthreads are used to have two threads share a core.} The frequency-integrated weighted gas opacities and dust extinction coefficients are divided with the Rosseland and Planck mean opacities and extinction coefficients at the first iteration to calculate opacity ratios $k_X^{\text{it}=1}$ as follows
\begin{equation}
  k_{\text{g,J}}^{\text{it}=1}=\frac{\kappaJ}{\kappaS},\quad
  k_{\text{g,H}}^{\text{it}=1}=\frac{\kappaH}{\kappaR},\quad
  k_{\text{d,J}}^{\text{it}=1}=\frac{\chiJ}{\chiS},\quad
  k_{\text{d,H}}^{\text{it}=1}=\frac{\chiH}{\chiR}.
\end{equation}
Only the Rosseland and Planck mean opacities and extinction coefficients are calculated in subsequent iterations, and those values are then multiplied with the opacity ratios $k_X^{\text{it}=1}$ of the first iteration to get the respective value in the current iteration.

\section{Modelling procedure and results}\label{sec:procedure}
We first describe our modelling procedure in Section~\ref{sec:modeling}, and the physics setup and choice of model parameter sets in Section~\ref{sec:setup}. We present our results in Section~\ref{sec:results}.

\subsection{Modelling procedure}\label{sec:modeling}
Our modelling procedure consists of four separate stages. We begin by calculating a hydrostatic dust-free initial model using {\johnc} (Appendix~\ref{app:initial}). The initial model spans the radial range $\left[\rint,\rext\right]\approx\left[0.9\Rs,1.8\Rs\right]$. The inner radius {\rint} is set as small as possible, but the model will not converge if it is lower than a model-dependent threshold radius. Simultaneously, the external radius {\rext} cannot be too large in the steep hydrostatic structure as that also prevents the model from converging. We have found empirically that the external radius is well selected such that $\rhog(\rext)=10^{-6}\rhog(\rint)$.

The model domain of the converged initial model is extended to instead use the outer radius $\rextf=40\Rs$ before it is saved to a file. Models with an expected low terminal velocity ($\uinf\lsim10\,\kms$) instead use $\rextf=20\Rs$ to save calculation time. Physical properties are not used in the radial range $\rext<r<\rextf$, which is used as a ``gridpoint reservoir'' in the expansion stage.

In the relaxation stage, the initial model of {\johnc} is relaxed using the system of equations of {\teh}, still without simulating stellar pulsations or allowing dust to form. A model has relaxed to be hydrostatic (using the dynamic code) when the time step becomes higher than $10^{14}\,$s. In the expansion stage, the piston is switched on from zero to full amplitude in 2 pulsation periods. Gridpoints $i$ where $r_i<2\Rs$ thereby move along with the inner boundary, whilst remaining gridpoints are held fixed. Simultaneously with the piston activation, all dust equations and terms are switched on. Dust begins to form in the cooler outer layers of the initial model, i.e. at $r\approx2\Rs$; dust grains absorb radiative momentum and are accelerated outwards whereby they drag the gas along, thus initiating a wind. The expansion of the physical region proceeds until the outer region has moved from {\rext} and reached {\rextf} of the fixed grid.

In the final wind stage, the wind model is evolved for a time interval of about $12$--$200\,P$. The modelled time interval depends on the time that is needed for transients in the expansion phase to leave the model domain through the outer boundary. Temporally averaged properties are measured only after this time. The interval is longer with models that do not develop periodic or close to periodic structures.

\begin{figure}
\includegraphics{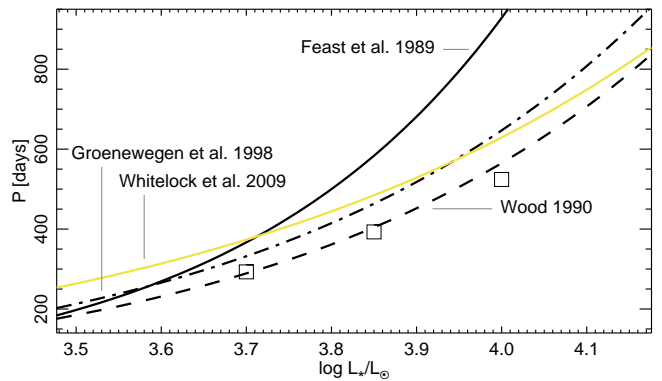}
\caption{Period-luminosity relations (lines) and the periods we use ($\square$).\label{fig:pl}}
\end{figure}

\subsection{Physics setup and selection of model parameters}\label{sec:setup}
Our approach is to use a physical setup that mostly is identical to that of {\rMWH}. We used 319 frequencies in the RT calculations, whilst {\rMWH} use 64 frequencies. The abundances are set to solar values, with the exception of the carbon-to-oxygen ratio that is an input parameter. We assume $\deltauvelp=4\,\kms$ and use $\Ms=1.0\,\Msun$. We use the same pulsation periods as before, but note that these are somewhat different from the period relation of \citet[][see equation~2.44 in \citealt{LaCa:99}, where the form below was achieved using equation~\ref{eq:rstellar} and assuming $\teff/T_{\text{eff},\sun}=0.572$ and $\Ms=1.0\Msun$]{Wo:90} and the period-luminosity ($P$-\Ls) relations of \citet{FeGlWhCa:89} and \citet{GrWhSmKe:98}
\begin{eqnarray}
m_{\text{bol}}(\text{Feast\,1989})&=&-1.86\log P+18.76,\\
m_{\text{bol}}(\text{Wood\,1990})^*&=&-2.58\log P+20.33,\,\,\text{and}\\
m_{\text{bol}}(\text{Groenewegen\,1998})&=&-2.59\log P+20.52.
\end{eqnarray}
Here, we used the same distance modulus to the Large Magellanic Cloud as \citet{GrWhSmKe:98}, $\mu_{\text{D}}=18.50$, and the bolometric magnitude of the sun $M_{\text{bol},\sun}=4.74$. The bolometric magnitude $m_{\text{bol}}$ is converted to a luminosity using $2.5\log(\Ls/\Lsun)=M_{\text{bol},\sun}+\mu_{\text{D}}-m_{\text{bol}}$. The resulting $P$-\Ls-relations are shown in Fig.~\ref{fig:pl} along with the values we use. (For reference, the figure also shows the $P$-\Ls-relation for O-rich Miras of \citealt{WhMeFe.:09}.) The values we use match the relation of \citet{Wo:90} the best.

\begin{figure}
\includegraphics{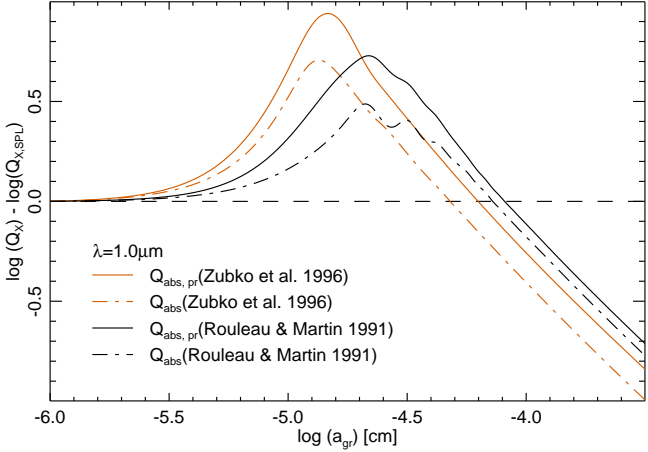}
\caption{Absorption efficiency factors $\Qnuabs(\text{pr})$ and {\Qnuabs} relative to the corresponding SPL values plotted versus the grain radius {\agr} (cf.\ fig.~3 in {\rMH}). The properties are shown for $\lambda=1\mu m$ using amC data of \citet{RoMa:91} (black lines) and \citet{ZuMeCoBu:96} (red lines).
\label{fig:mie}}
\end{figure}

The dust consists of amC, where the intrinsic dust density $\rhoint=1.85\,\dens$, the surface tension $\surft=1400\,\ergcm$, and the tabulated values ($n_{\nu}$ and $k_{\nu}$) used to calculate the dust extinction are taken from \citet{RoMa:91}. We assume SPL, but calculate three models using Mie scattering as comparison. The importance of using Mie scattering in place of SPL is evident when comparing the absorption efficiencies of the two approaches, see Fig.~\ref{fig:mie} (cf. fig.~3 in \rMH). The ratio $\Qnuabs(\text{pr})/\Qnuabs(\text{SPL})\ga1.1$ in the grain radius interval 0.039--0.75\,$\mu$m, where it is at most 5.4 times higher when $\agr=0.22\,\mu$m. Differences are even larger when instead using the amC data of \citet{ZuMeCoBu:96}; the peak is shifted to somewhat lower interval 0.027--0.58\,$\mu$m, where the peak ratio is 8.7 times higher when $\agr=0.15\,\mu$m.

Furthermore, amongst other parameters, \citet{AnHoGa:03} study the role of the sticking coefficients to properties of stellar winds that use frequency-dependent RT and find that mass-loss rates, terminal velocities, and the degree of condensation change. More recent models \citep[beginning with][]{MaHoHe:07} use sticking coefficients that are set to unity as these form a stellar wind easier owing to larger grains than when the original lower values are used. We use the original values that are used in all models in this series of articles $\Xi_{0.34}$ ($\aC=0.37$, $\aCt=0.34$, $\aCtH=0.34$, $\aCtHt=0.34$). A comparison of results of three models that use both $\Xi_{0.34}$ and $\Xi_{1.00}$ ($\aC=1.00$, $\aCt=1.00$, $\aCtH=1.00$, $\aCtHt=1.00$) shows non-trivial differences (Appendix~\ref{app:stick}).

\begin{table}
  \caption{Model parameters, see Section~\ref{sec:setup} for further details. The model name is given in Column~1. The following five columns specify: stellar luminosity {\Ls}, effective temperature {\teff}, chemistry (initial $C/O$ ratio), and pulsation period $P$. The last column shows the fraction of the stellar mass contained in the radial domain of the initial model. The stellar mass $M_\star$ is nearly $1.0\,\Msun$ in all models, and the pulsation amplitude $\deltauvelp=4\,\kms$.}
\label{tab:resmodpar}
\begin{tabular}{lccccc}\hline\hline\\[-1.8ex]
model &$\log\left(\Ls\right)$       &{\teff}     & {\CtO} & \pP & $\displaystyle\frac{M_\text{e}}{M_\star}$\\
      &$[L_{\sun}]$&$[\mbox{K}]$& & $[\mbox{d}]$& [\%]\\[1.0ex]\hline\\[-1.8ex]
L3.70T24E88 & 3.70 & 2400 & 8.80 & 295 & 0.60\\
L3.70T24E91 & 3.70 & 2400 & 9.10 & 295 & 0.50\\
L3.85T24E85 & 3.85 & 2400 & 8.50 & 393 & 1.32\\
L3.85T25E88 & 3.85 & 2400 & 8.80 & 393 & 1.19\\
L3.85T24E91 & 3.85 & 2400 & 9.10 & 393 & 0.91\\
L4.00T24E82 & 4.00 & 2400 & 8.20 & 524 & 2.27\\
L4.00T24E85 & 4.00 & 2400 & 8.50 & 524 & 1.47\\
L4.00T24E88 & 4.00 & 2400 & 8.80 & 524 & 0.75\\
L4.00T24E91 & 4.00 & 2400 & 9.10 & 524 & 0.42\\[0.5ex]
L3.70T26E85 & 3.70 & 2600 & 8.50 & 295 & 0.37\\
L3.70T26E88 & 3.70 & 2600 & 8.80 & 295 & 0.52\\
L3.85T26E85 & 3.85 & 2600 & 8.50 & 393 & 0.63\\
L3.85T26E88 & 3.85 & 2600 & 8.80 & 393 & 0.41\\
L4.00T26E82 & 4.00 & 2600 & 8.20 & 524 & 0.98\\
L4.00T26E85 & 4.00 & 2600 & 8.50 & 524 & 0.82\\
L4.00T26E88 & 4.00 & 2600 & 8.80 & 524 & 1.17\\[0.5ex]
L3.70T28E88 & 3.70 & 2800 & 8.80 & 295 & 0.26\\
L3.85T28E85 & 3.85 & 2800 & 8.50 & 393 & 0.46\\
L3.85T28E88 & 3.85 & 2800 & 8.80 & 393 & 0.44\\
L4.00T28E85 & 4.00 & 2800 & 8.50 & 524 & 0.57\\
L4.00T28E88 & 4.00 & 2800 & 8.80 & 524 & 0.65\\
L4.00T28E91 & 4.00 & 2800 & 9.10 & 524 & 0.52\\[0.5ex]
L3.70T30E91 & 3.70 & 3000 & 9.10 & 295 & 0.14\\
L3.85T30E88 & 3.85 & 3000 & 8.80 & 393 & 0.25\\
L4.00T30E85 & 4.00 & 3000 & 8.50 & 524 & 0.46\\
L4.00T30E88 & 4.00 & 3000 & 8.80 & 524 & 0.31\\
L4.00T30E91 & 4.00 & 3000 & 9.10 & 524 & 0.21\\[0.5ex]
L4.00T32E88 & 4.00 & 3200 & 8.80 & 524 & 0.28\\
L4.00T32E91 & 4.00 & 3200 & 9.10 & 524 & 0.23\\[1.0ex]\hline
\end{tabular}
\end{table}

The model calculations are computationally demanding, and we do not calculate a model grid at this point. The model setups are taken from {\rMWH} (table~2). We show our complete set of model parameters in Table~\ref{tab:resmodpar}. We collect our physical and numerical assumptions in Table~\ref{app:tabassumptions} for easy reference.

\subsection{Results}\label{sec:results}
We selected the models in this paper with the intention of studying relative changes compared to previously calculated wind models. This approach allows a quantitative and qualitative estimate of the importance of the adopted differences in modelling and physics.

Our main objective here is to introduce effects of drift, using one mean dust velocity. The drift models are compared to (non-drift) PC models that in all other aspects are the same as the drift models.

In addition to a modelled drift velocity {\vD} in drift models, we calculate the equilibrium drift velocity {\vDeq} for both PC and drift models assuming complete momentum coupling. That is, we assume that the drag force is equal to the radiative pressure on dust grains after subtracting the gravitational pull on the same dust grains (this term is negligible, but we keep it for completeness of the argument). Assuming specular collisions, equations~(\ref{eq:dmot}) and (\ref{eq:drag}) give
\begin{eqnarray}
\pi r^2_0K_{2}\rhog\vDeq\left(\vDeq^2+\vzeta^2\right)^{\frac{1}{2}}=\frac{4\pi}{c}\chi_{\text{H}}H-\frac{Gm_r}{r^2}\rhod,\nonumber
\end{eqnarray}
and this is written in terms of {\vDeq} as
\begin{eqnarray}
\vDeq^2=-\frac{\vzeta^2}{2}+\left\{\left(\frac{\vzeta^2}{2}\right)^2+\left(\frac{\displaystyle\frac{4\pi}{c}\chi_{\text{H}}H-\frac{Gm_r}{r^2}\rhod}{\pi r_0^2K_{2}\rhog}\right)^2\right\}^{\frac{1}{2}}.\label{eq:vdeq}
\end{eqnarray}
If we ignore the gravitational term as well as the thermal velocity term $\vzeta$ and use $\mdot=4\pi r^2\rhog u$, we get a simpler expression valid at supersonic velocities
\begin{eqnarray}
\vDeq^2=\frac{4}{r_0^2c}\frac{\chiH H}{K_{2}\rhog}=\frac{16\pi\chi_{\text{H}}Hr^2u}{r_0^2K_{2}\mdot c}.\label{eq:svdeq}
\end{eqnarray}
Notably, equation~\ref{eq:svdeq} can also be applied to results of PC models, but the resulting values of {\vDeq} are calculated with models that disregard dilution of the dust and are therefore not comparable with {\vD} of drift models.

Wind models are characterised with properties temporally averaged at the outer boundary. Two properties characterize the gas: the gas mass-loss rate {\mmdot} and the gas terminal velocity {\muinf}. The dust is characterized with four properties: the degree of condensation {\mfcond}, the dust-to-gas mass-loss ratio {\mmdmdot}, the mean grain radius {\mradd}, and the terminal drift velocity {\mvdrinf} (only for drift models). The dust-to-mass mass loss ratio is
\begin{eqnarray}
\frac{\dot{M}_{\text{d}}}{\dot{M}}=\frac{\rhod}{\rhog}\frac{\vinf}{\uinf}=\ddg\FD,\label{eq:dmddm}
\end{eqnarray}
where $\ddg=\rhod/\rhog$ and $\FD=\vinf/\uinf=1+\vDinf/\uinf$ is the drift factor. Moreover, since the dust component is diluted by the drift factor, we define the degree of condensation based on fluxes as
\begin{eqnarray}
\fcond=\frac{\rhod\FD}{\rho_{\text{c}}^{\text{tot}}(\FD)}\approx\frac{\FD K_{3}}{\FD K_{3}+n_{\text {C}}}=\frac{\vinf K_3}{\vinf K_3+\uinf n_{\text {C}}},
\end{eqnarray}
where $\rho_{\text{c}}^{\text{tot}}$ is the total density of condensable matter (present in both the gas and the dust phases). The value can, with this definition, become larger than unity when either velocity is negative, which is unphysical; this could occasionally be the case in the wind formation region when the gas velocity is negative (see Fig.~\ref{fig:tdiscwind}c at $r\simeq2\,\Rs$ for an example of how this appears). Assuming PC, the expression reduces to
\begin{eqnarray}
\fcondPC\approx\frac{K_3}{K_3+n_{\text{C}}}.
\end{eqnarray}

Providing a measure for the variability of the model structure, each outflow property ($\mathcal{Q}$) is accompanied by a relative fluctuation amplitude $\rfluc=\sistd/\mathcal{Q}$, where {\sistd} is the (sample) standard deviation of the property $\mathcal{Q}$ in the measured time interval.

\begin{figure}
\includegraphics{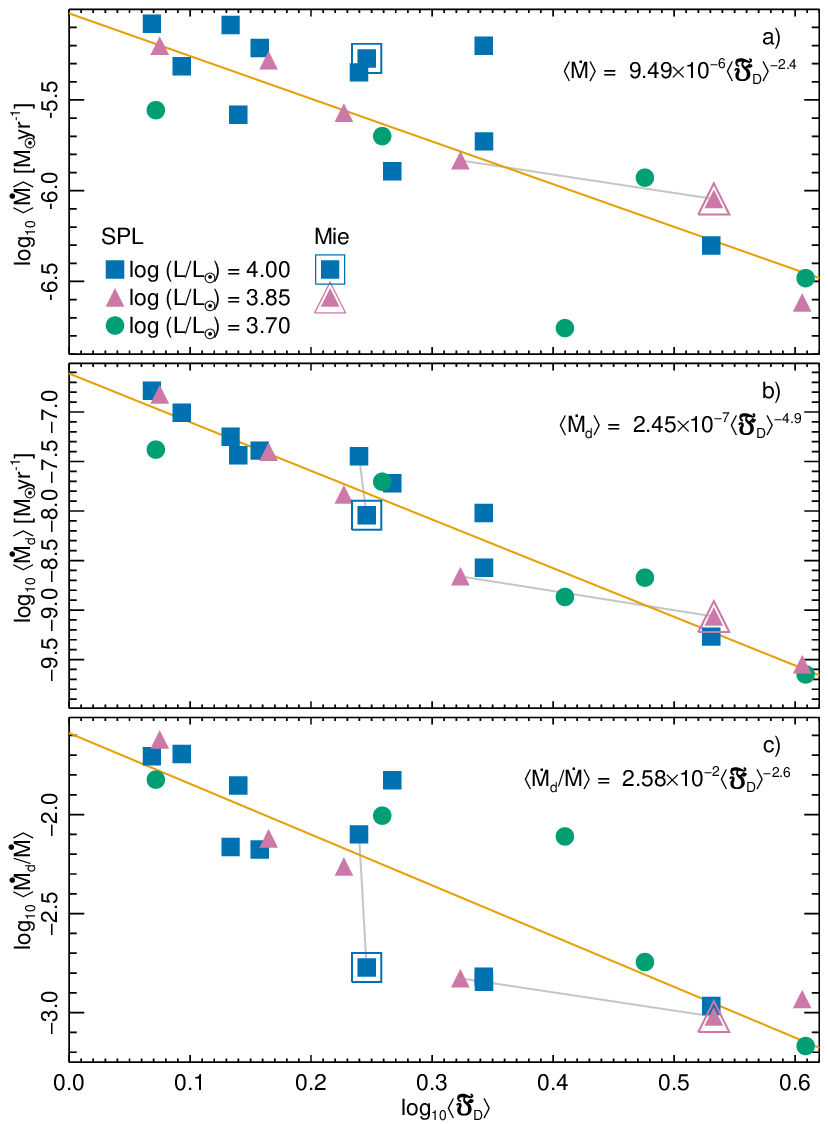}
\caption{Average properties of drift models versus the drift factor $\mFD$. The three panels show: (a) mass-loss rate, {\mmdot}; (b) mass-loss rate of the dust, {\mdmdot}; and (c) mass-loss ratio, $\mmdmdot=\langle\drhog\rangle$. All ordinates are logarithmic. The coefficients of a linear fit to the data is shown in the upper right corner of each panel and the orange line shows the fit. Grey lines connect the SPL and Mie values of models L3.85T28E88 and L4.00T28E88.\label{fig:discdriftfactorm}}
\end{figure}

\begin{figure}
\includegraphics{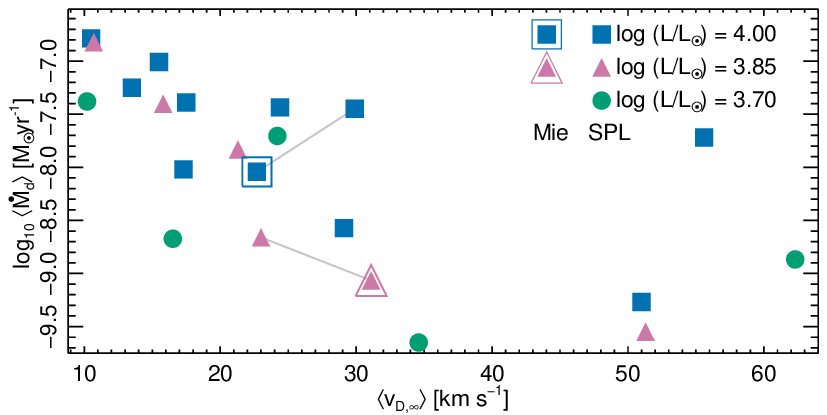}
\caption{Dust mass loss rate {\mdmdot} (log) plotted versus the average terminal drift velocity $\mvDinf$.\label{fig:discvDr}}
\end{figure}

\begin{figure}
\includegraphics{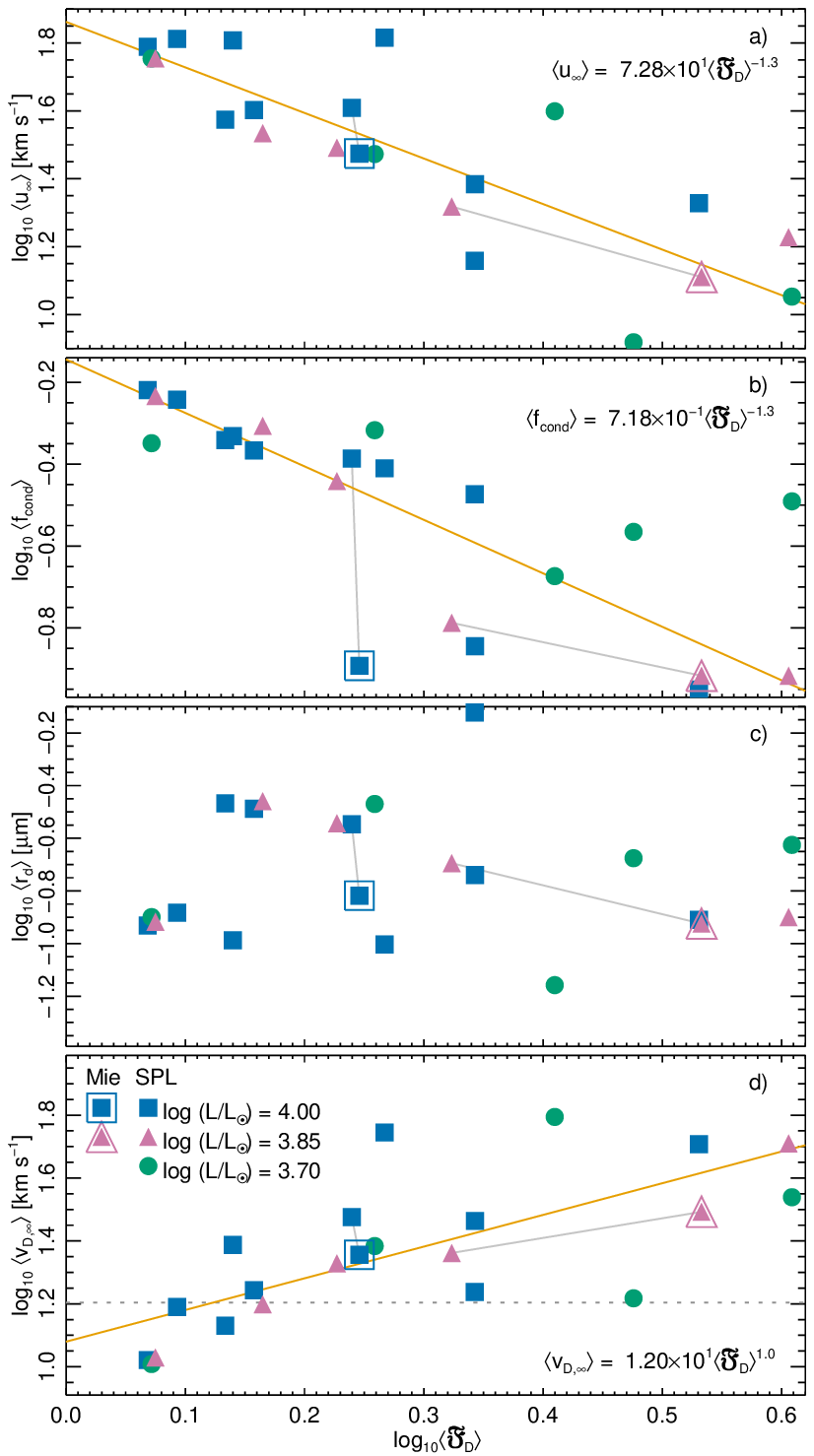}
\caption{Average properties of drift models versus the drift factor $\mFD$. The four panels show: (a) terminal velocity, {\muinf}; (b) degree of condensation, {\mfcond}; (c) grain radius, {\mradd}, and (d) terminal drift velocity, {\mvDinf}. All ordinates are logarithmic. The coefficients of a linear fit to the data is shown in the upper right corner of the top two panels. Grey lines connect the SPL and Mie values of models L3.85T28E88 and L4.00T28E88. Horizontal dotted lines indicate the values $\mvDinf=5.0$ and $16\kms$ (see Section~\ref{sec:discdriftfactor}).\label{fig:discdriftfactoro}}
\end{figure}

\begin{figure}
\includegraphics{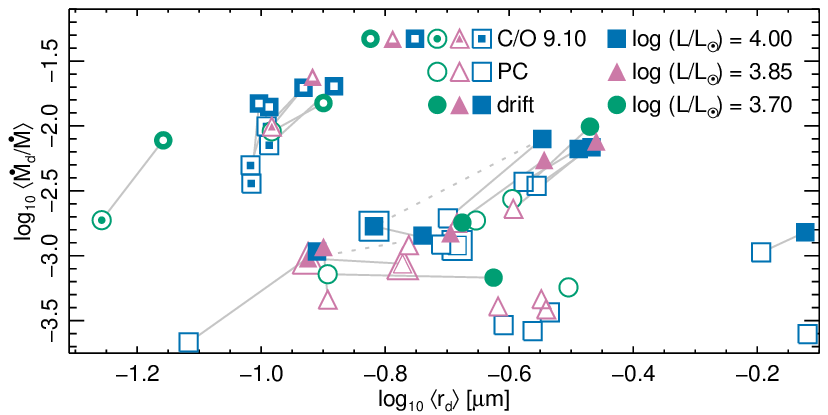}
\caption{Mass loss ratio {\mmdmdot} (log) versus the average grain radius $\mradd$ (log). From the top left corner, the diagonally appearing groups of models have $\CtO=9.1$, $8.8$, $8.5$ (discontinuous), and the single model in the bottom right corner is the PC model where $\CtO=8.2$.\label{fig:discddg}}
\end{figure}

We show results of all our model calculations using $\Xi_{0.34}$ in Table~\ref{tab:resall}. The average mass loss properties of the drift models are plotted against {\FD} in Fig.~\ref{fig:discdriftfactorm}. Moreover, the average dust mass loss rate is plotted against the drift velocity in Fig.~\ref{fig:discvDr}. The four remaining averaged properties -- terminal velocity, degree of condensation, grain radius, and terminal drift velocity -- are plotted against {\FD} in Fig.~\ref{fig:discdriftfactoro}. Finally, the average mass loss ratio is plotted against the average grain radius in Fig.~\ref{fig:discddg}.

\begin{table*}
\caption{Temporally averaged quantities at the outer boundary; see Section~\ref{sec:results}. From the left, the first two columns specify the model name (see Table~\ref{tab:resmodpar}) and if PC or drift is used (P/D). Six column pairs show the averaged: mass loss rate {\mmdot}, terminal velocity {\muinf}, degree of condensation {\mfcond}, dust-to-gas mass-loss rate {\mdrhog}, dust radius {\mradd}, and terminal drift velocity {\mvdrinf} (only for drift models). A relative fluctuation amplitude {\rfluc} is prvided for each property; a subscript m (c, w, $\mu$) indicates that the shown value was multiplied with a factor $10^3$ ($10^2$, $10^4$, $10^6$). The final columns show the outflow classification class: irregular (i), periodic ($l\times$p), and quasi-periodic ($l$q); $l$ indicates the (multi-)periodicity of the gas/dust outflow in the unit of the piston period $P$, and if the dust extinction is calculated using Mie scattering instead of the default SPL. Rows of drift models are shown in boldface.}
\label{tab:resall}
\begin{tabular}{$l@{\quad}^c@{\:\:}^r%
                @{\ }^l@{\quad}^ll%
                @{\ }^l@{\quad}^ll%
                @{\ }^l@{\quad}^ll%
                @{\ }^l@{\quad}^ll%
                @{\ }^l@{\quad}^ll%
                @{\ }^l@{\quad}^l^r^r}%
              \hline\hline\\[-1.8ex]
   \multicolumn{1}{l}{model name} & P/D &&
          \multicolumn{2}{c}{$10^7$\,\mmdot} &&
          \multicolumn{2}{c}{\muinf}  &&
          \multicolumn{2}{c}{\mfcond} &&
          \multicolumn{2}{c}{$10^4$\mdrhogf} &&
          \multicolumn{2}{c}{$10^2$\mradd}  &&
          \multicolumn{2}{c}{\mvdrinf}  & class & [SPL] / Mie\\
      & && \multicolumn{2}{c}{$[\mdotu]$} &&
          \multicolumn{2}{c}{$[\kms]$}    &&
          \multicolumn{2}{c}{}      &&
          \multicolumn{2}{c}{} &&
          \multicolumn{2}{c}{$[\mu\text{m}]$}      &&
          \multicolumn{2}{c}{$[\kms]$}\\
      & &&& \multicolumn{1}{c}{\rfluc}&&
        &\multicolumn{1}{c}{\rfluc} &&
        &\multicolumn{1}{c}{\rfluc} &&
        &\multicolumn{1}{c}{\rfluc} &&
        &\multicolumn{1}{c}{\rfluc} &&
        &\multicolumn{1}{c}{\rfluc}\\[1.0ex]\hline\\[-1.0ex]
L3.70T24E88          & P && 31.8 & \textit{$\phantom{0}$21} && 30.8 & \textit{0.98} && 0.440 & \textit{0.080} && $\phantom{0}$27.3 & \textit{$\phantom{00}$5.3} && 25.4 & \textit{$\phantom{0}$2.5}  &&      &             & 1.6q\\[0.15ex]
\rowstyle{\bfseries} & D && 20.0 & \textit{$\phantom{0}$27} && 29.7 & \textit{1.6}  && 0.482 & \textit{0.38}  && $\phantom{0}$98.7 & \textit{150}               && 33.9 & \textit{$\phantom{0}$3.8}  && 24.2 & \textit{13} & i\\[0.15ex]
L3.70T24E91          & P && 30.0 & \textit{$\phantom{0}$25} && 57.8 & \textit{1.9}  && 0.763 & \textit{0.048} && $\phantom{0}$90.9 & \textit{$\phantom{00}$5.5} && 10.4 & \textit{$\phantom{0}$0.26} &&      &             & 1q\\[0.15ex]
\rowstyle{\bfseries} & D && 27.8 & \textit{$\phantom{0}$25} && 56.8 & \textit{1.5}  && 0.448 & \textit{0.40}  && 150               & \textit{324}               && 12.6 & \textit{$\phantom{0}$0.47} && 10.2 & \textit{$\phantom{0}$5.8} & q\\[0.15ex]
L3.85T24E85          & P && 10.8 & \textit{510\tsm}         && $\phantom{0}$3.90 & \textit{4.7\tsc} && 0.151 & \textit{0.12\tsc} && $\phantom{00}$4.63 & \textit{$\phantom{0}$37\tsm} && 28.3 & \textit{$\phantom{0}$0.65} && && 1p\\[0.15ex]
L3.85T24E88          & P && 52.7 & \textit{150}             && 28.5 & \textit{3.9} && 0.371 & \textit{0.13} && $\phantom{0}$23.0 & \textit{$\phantom{00}$8.3} && 25.5 & \textit{$\phantom{0}$3.2} && && i\\[0.15ex]
\rowstyle{\bfseries} & D && 52.3 & \textit{$\phantom{0}$46} && 34.2 & \textit{0.69} && 0.494 & \textit{0.36}  && $\phantom{0}$75.8 & \textit{110} && 34.7 & \textit{$\phantom{0}$3.4} && 15.8 & \textit{$\phantom{0}$7.4} & i\\[0.15ex]
L3.85T24E91          & P && 62.1 & \textit{$\phantom{0}$52} && 57.5 & \textit{2.6}  && 0.827 & \textit{0.11}  && $\phantom{0}$98.9 & \textit{$\phantom{0}$11}   && 10.4 & \textit{$\phantom{0}$0.53} && && 1q\\[0.15ex]
\rowstyle{\bfseries} & D && 62.9 & \textit{$\phantom{0}$50} && 56.8 & \textit{1.8}  && 0.584 & \textit{0.43}  && 239 & \textit{360} && 12.1 & \textit{$\phantom{0}$0.59} && 10.7 & \textit{$\phantom{0}$3.3} & q\\[0.15ex]
L4.00T24E85          & P && 77.4 & \textit{$\phantom{0}$56} && 18.9 & \textit{1.1}  && 0.339 & \textit{0.12}  && $\phantom{0}$10.6 & \textit{$\phantom{00}$3.7} && 64.1 & \textit{12} && && i\\[0.15ex]
\rowstyle{\bfseries} & D && 62.9 & \textit{$\phantom{0}$12} && 14.4 & \textit{0.11} && 0.336 & \textit{0.27}  && $\phantom{0}$15.2 & \textit{$\phantom{0}$17}   && 75.4 & \textit{$\phantom{0}$5.0} && 17.3 & \textit{$\phantom{0}$1.7} & 1p\\[0.15ex]
L4.00T24E88          & P && 88.1 & \textit{120}             && 37.2 & \textit{2.2}  && 0.554 & \textit{0.12}  && $\phantom{0}$34.5 & \textit{$\phantom{00}$7.4} && 27.8 & \textit{$\phantom{0}$2.4} &&      &                           & q\\[0.15ex]
\rowstyle{\bfseries} & D && 81.9 & \textit{$\phantom{0}$63} && 37.5 & \textit{2.1}  && 0.456 & \textit{0.37}  && $\phantom{0}$68.6 & \textit{110}               && 34.1 & \textit{$\phantom{0}$4.1} && 13.5 & \textit{$\phantom{0}$8.1} & i\\[0.15ex]
L4.00T24E91          & P && 86.1 & \textit{100}             && 60.6 & \textit{3.3}  && 0.831 & \textit{0.13}  && $\phantom{0}$99.1 & \textit{$\phantom{0}$15}   && 10.2 & \textit{$\phantom{0}$0.30} && && q\\[0.15ex]
\rowstyle{\bfseries} & D && 83.2 & \textit{$\phantom{0}$66} && 61.5 & \textit{1.7}  && 0.604 & \textit{0.41}  && 197               & \textit{300}               && 11.7 & \textit{$\phantom{0}$0.40} && 10.5 & \textit{$\phantom{0}$4.0} & i\\[1.0ex]
L3.70T26E85          & P && $\phantom{0}$1.64 & \textit{$\phantom{0}$39\tsm} && $\phantom{0}$2.63 & \textit{2.7\tsc} && 0.199 & \textit{4.8\tsm} && $\phantom{00}$5.71 & \textit{$\phantom{0}$14\tsm} && 31.3 & \textit{$\phantom{0}$5.3\tsc} && && 1p\\[0.15ex]
L3.70T26E88          & P && 20.3 & \textit{$\phantom{0}$16} && 25.1 & \textit{1.6} && 0.302 & \textit{0.077} && $\phantom{0}$18.7 & \textit{$\phantom{00}$5.0} && 22.2 & \textit{$\phantom{0}$3.3} && && i\\[0.15ex]
\rowstyle{\bfseries} & D && 11.8 & \textit{180\tsm}         && $\phantom{0}$8.29 & \textit{6.6\tsc} && 0.272 & \textit{0.14} && $\phantom{0}$18.0 & \textit{$\phantom{0}$11} && 21.0 & \textit{$\phantom{0}$0.24} && 16.5 & \textit{$\phantom{0}$0.59} & 2q\\[0.15ex]
L3.85T26E85          & P && $\phantom{0}$6.00 & \textit{240\tsm} && $\phantom{0}$4.40 & \textit{7.8\tsm} && 0.134 & \textit{0.19\tsm} && $\phantom{00}$4.07 & \textit{$\phantom{0}$58\tsm} && 24.1 & \textit{$\phantom{0}$0.43} && && 1p\\[0.15ex]
L3.85T26E88          & P && 37.2 & \textit{$\phantom{0}$22} && 26.6 & \textit{0.75} && 0.290 & \textit{0.074} && $\phantom{0}$17.9 & \textit{$\phantom{00}$4.8} && 20.7 & \textit{$\phantom{0}$3.3} &&      &             & 3q\\[0.15ex]
\rowstyle{\bfseries} & D && 26.9 & \textit{$\phantom{0}$27} && 31.0 & \textit{0.95} && 0.362 & \textit{0.36}  && $\phantom{0}$54.6 & \textit{$\phantom{0}$86}   && 28.6 & \textit{$\phantom{0}$4.0} && 21.3 & \textit{11} & 4p\\[0.15ex]
L4.00T26E82          & P && $\phantom{0}$3.70 & \textit{$\phantom{0}$95\tsm} && $\phantom{0}$1.84 & \textit{1.4\tsc} && 0.163 & \textit{0.88\tsm} && $\phantom{00}$2.49 & \textit{$\phantom{0}$14\tsc} && 76.1 & \textit{$\phantom{0}$7.1} && && 1p\\[0.15ex]
L4.00T26E85          & P && 28.7 & \textit{$\phantom{00}$7.6} && $\phantom{0}$8.39 & \textit{0.23} && 0.120 & \textit{0.027} && $\phantom{00}$3.67 & \textit{$\phantom{0}$82\tsc} && 29.2 & \textit{$\phantom{0}$4.7} && && 4.6q\\[0.15ex]
L4.00T26E88          & P && 85.1 & \textit{$\phantom{0}$50} && 40.6 & \textit{1.7} && 0.603 & \textit{0.097} && $\phantom{0}$37.1 & \textit{$\phantom{00}$5.7} && 26.5 & \textit{$\phantom{0}$3.0} &&      &             & q\\[0.15ex]
\rowstyle{\bfseries} & D && 61.2 & \textit{$\phantom{0}$57} && 40.0 & \textit{2.0} && 0.430 & \textit{0.36}  && $\phantom{0}$66.6 & \textit{130}               && 32.5 & \textit{$\phantom{0}$3.3} && 17.5 & \textit{10} & i\\[1.0ex]
L3.70T28E88          & P && $\phantom{0}$7.88 & \textit{470\tsm} && 14.0 & \textit{5.2\tsc} && 0.118 & \textit{0.13\tsc} && $\phantom{00}$7.21 & \textit{$\phantom{0}$78\tsm} && 12.8 & \textit{$\phantom{0}$3.9\tsc} &&      &                           & 1p\\[0.15ex]
\rowstyle{\bfseries} & D && $\phantom{0}$3.30 & \textit{130\tsm} && 11.3 & \textit{2.0\tsc} && 0.323 & \textit{0.23}     && $\phantom{00}$6.78 & \textit{$\phantom{00}$7.1}   && 23.7 & \textit{$\phantom{0}$2.0}     && 34.6 & \textit{$\phantom{0}$1.8} & 1p\\[0.15ex]
                     & P && $\phantom{0}$7.19 & \textit{$\phantom{00}$1.0} && 20.6 & \textit{0.16} && 7.97\tsc & \textit{0.37\tsc} && $\phantom{00}$4.89 & \textit{$\phantom{0}$23\tsc} && 11.1 & \textit{$\phantom{0}$0.35} && && 1p & Mie\\[0.15ex]
L3.85T28E85          & P && $\phantom{0}$4.67 & \textit{$\phantom{0}$89\tsm} && $\phantom{0}$4.79 & \textit{3.1\tsc} && 0.129 & \textit{0.43\tsm} && $\phantom{00}$3.87 & \textit{$\phantom{0}$15\tsm} && 28.8 & \textit{$\phantom{0}$5.1\tsc} && && 1p\\[0.15ex]
L3.85T28E88          & P && 27.4              & \textit{$\phantom{00}$8.5} && 26.4 & \textit{0.40} && 0.197 & \textit{0.034}   && $\phantom{0}$12.1  & \textit{$\phantom{00}$2.1} && 17.3 & \textit{$\phantom{0}$2.0}  &&      &                           & 1p\\[0.15ex]
\rowstyle{\bfseries} & D && 14.7              & \textit{$\phantom{00}$2.7} && 20.8 & \textit{0.18} && 0.163 & \textit{0.22}    && $\phantom{0}$14.9  & \textit{$\phantom{0}$23}   && 20.2 & \textit{$\phantom{0}$1.0}  && 23.0 & \textit{$\phantom{0}$1.6} & 1p\\[0.15ex]
                     & P && 22.4              & \textit{$\phantom{0}$76}   && 34.8 & \textit{5.0}  && 0.141 & \textit{0.042}   && $\phantom{00}$8.68 & \textit{$\phantom{00}$2.6} && 16.9 & \textit{$\phantom{0}$2.0}  &&      &                           & i  & Mie\\[0.15ex]
\rowstyle{\bfseries} & D && $\phantom{0}$9.01 & \textit{220\tsm}           && 12.9 & \textit{2.5\tsc} && 0.121 & \textit{0.15} && $\phantom{00}$9.55 & \textit{$\phantom{0}$14}   && 11.9 & \textit{$\phantom{0}$0.72} && 31.1 & \textit{$\phantom{0}$2.3} & 1p & Mie\\[0.15ex]
L4.00T28E85          & P && 16.0              & \textit{$\phantom{0}$47\tsm} && $\phantom{0}$6.18 & \textit{3.3\tsm} && 9.56\tsc & \textit{0.29\tsm} && $\phantom{00}$2.92 & \textit{$\phantom{0}$89\tsw} && 24.6 & \textit{$\phantom{0}$5.6\tsc} && && 1p\\[0.15ex]
L4.00T28E88          & P && 51.3              & \textit{$\phantom{0}$34}   && 34.1 & \textit{1.2}  && 0.315 & \textit{0.046} && $\phantom{0}$19.4 & \textit{$\phantom{00}$2.8} && 20.0 & \textit{$\phantom{0}$1.9} &&      &             & q\\[0.15ex]
\rowstyle{\bfseries} & D && 44.9              & \textit{$\phantom{0}$45}   && 40.6 & \textit{0.55} && 0.411 & \textit{0.39}  && $\phantom{0}$79.4 & \textit{130}               && 28.4 & \textit{$\phantom{0}$0.96} && 29.9 & \textit{$\phantom{0}$5.6} & 1p\\[0.15ex]
                     & P && 36.4              & \textit{120}               && 50.9 & \textit{9.4}  && 0.196 & \textit{0.098} && $\phantom{0}$12.0 & \textit{$\phantom{00}$6.0} && 20.7 & \textit{$\phantom{0}$3.2} &&      &             &  q & Mie\\[0.15ex]
\rowstyle{\bfseries} & D && 53.6              & \textit{$\phantom{0}$65}   && 29.8 & \textit{2.0}  && 0.128 & \textit{0.21}  && $\phantom{0}$16.9 & \textit{$\phantom{0}$41}   && 15.2 & \textit{$\phantom{0}$2.0} && 22.7 & \textit{19} &  i & Mie\\[0.15ex]
L4.00T28E91          & P && 60.1              & \textit{150}               && 62.2 & \textit{6.4}  && 0.589 & \textit{0.22}  && $\phantom{0}$71.1 & \textit{$\phantom{0}$26}   && 10.3 & \textit{$\phantom{0}$0.94} &&      &             & i\\[0.15ex]
\rowstyle{\bfseries} & D && 48.5              & \textit{$\phantom{0}$89}   && 64.8 & \textit{5.0}  && 0.572 & \textit{0.36}  && 202               & \textit{340}               && 13.1 & \textit{$\phantom{0}$2.1}  && 15.5 & \textit{12} & i\\[1.0ex]
L3.70T30E91          & P && $\phantom{0}$3.68 & \textit{$\phantom{00}$3.4} && 40.6 & \textit{1.2}  && 0.158 & \textit{0.012} && $\phantom{0}$18.8 & \textit{$\phantom{00}$1.5} && $\phantom{0}$5.53 & \textit{$\phantom{0}$0.39} &&      &             & 1p\\[0.15ex]
\rowstyle{\bfseries} & D && $\phantom{0}$1.75 & \textit{$\phantom{00}$1.8} && 39.7 & \textit{1.2}  && 0.212 & \textit{0.22}  && $\phantom{0}$77.5 & \textit{200}               && $\phantom{0}$6.95 & \textit{$\phantom{0}$1.1}  && 62.3 & \textit{18} & 1q\\[0.15ex]
L3.85T30E88          & P && $\phantom{0}$1.90 & \textit{910\tsm} && 11.4 & \textit{0.51} && 7.53\tsc & \textit{0.015} && $\phantom{00}$4.59 & \textit{$\phantom{0}$89\tsc} && 12.8 & \textit{$\phantom{0}$2.1} &&      &                           & q\\[0.15ex]
\rowstyle{\bfseries} & D && $\phantom{0}$2.42 & \textit{580\tsm} && 16.9 & \textit{0.27} && 0.121    & \textit{0.18}  && $\phantom{0}$11.7  & \textit{$\phantom{0}$22}     && 12.6 & \textit{$\phantom{0}$1.9} && 51.3 & \textit{$\phantom{0}$4.5} & 1p\\[0.15ex]
L4.00T30E85          & P && $\phantom{0}$8.60 & \textit{150\tsm} && $\phantom{0}$6.20 & \textit{1.3\tsc} && 8.62\tsc & \textit{0.15\tsm} && $\phantom{00}$2.63 & \textit{$\phantom{0}$43\tsw} && 27.4 & \textit{$\phantom{0}$3.2\tsc} && && 1p\\[0.15ex]
L4.00T30E88          & P && 26.0              & \textit{$\phantom{0}$19}   && 33.2 & \textit{1.2}  && 0.199 & \textit{0.025} && $\phantom{0}$12.2 & \textit{$\phantom{00}$1.5} && 19.5 & \textit{$\phantom{0}$1.3} &&      &                           & 1p\\[0.15ex]
\rowstyle{\bfseries} & D && 18.7              & \textit{$\phantom{00}$2.5} && 24.2 & \textit{0.42} && 0.143 & \textit{0.21}  && $\phantom{0}$14.3 & \textit{$\phantom{0}$24}   && 18.2 & \textit{$\phantom{0}$2.5} && 29.1 & \textit{$\phantom{0}$2.8} & 1p\\[0.15ex]
L4.00T30E91          & P && 24.6              & \textit{$\phantom{0}$49}   && 67.7 & \textit{6.1}  && 0.411 & \textit{0.19}  && $\phantom{0}$49.6 & \textit{$\phantom{0}$23}   && $\phantom{0}$9.61 & \textit{$\phantom{0}$1.3} &&      &             & 2p\\[0.15ex]
\rowstyle{\bfseries} & D && 26.2              & \textit{$\phantom{0}$47}   && 64.2 & \textit{5.3}  && 0.466 & \textit{0.34}  && 140               & \textit{220}               && 10.3              & \textit{$\phantom{0}$1.8} && 24.4 & \textit{17} & i\\[1.0ex]
L4.00T32E88          & P && 42.2\tsc          & \textit{$\phantom{0}$19\tsm} && $\phantom{0}$4.48 & \textit{2.1\tsc} && 3.52\tsc & \textit{7.8} && $\phantom{00}$2.15 & \textit{$\phantom{0}$47\tsw} && $\phantom{0}$7.64 & \textit{$\phantom{0}$2.3\tsc} && && 1p\\[0.15ex]
\rowstyle{\bfseries} & D && $\phantom{0}$4.99 & \textit{$\phantom{00}$2.8}  && 21.3 & \textit{0.34} && 0.112 & \textit{0.18} && $\phantom{0}$10.8 & \textit{$\phantom{0}$21} && 12.3 & \textit{$\phantom{0}$2.0} && 51.0 & \textit{$\phantom{0}$9.0} & 1p\\[0.15ex]
L4.00T32E91          & P && $\phantom{0}$9.29 & \textit{$\phantom{0}$28} && 72.4 & \textit{7.8} && 0.299 & \textit{0.11} && $\phantom{0}$36.0 & \textit{$\phantom{0}$14} && $\phantom{0}$9.64 & \textit{$\phantom{0}$0.45} &&      &             & 1p\\[0.15ex]
\rowstyle{\bfseries} & D && 12.8              & \textit{$\phantom{0}$32} && 65.4 & \textit{3.3} && 0.389 & \textit{0.30} && 149               & \textit{270}             && $\phantom{0}$9.92 & \textit{$\phantom{0}$1.5}  && 55.6 & \textit{25} & 1p\\[1.0ex]\hline\\[-1.0ex]
\end{tabular}
\end{table*}

\section{Discussion}\label{sec:discussion}
We discuss the following topics in the next six subsections: the role of the spatial resolution (Section~\ref{sec:resolution}), wind formation when models include drift (\ref{sec:discwind}), temporal variability and averaged properties (\ref{sec:temporal}), the importance of using Mie scattering in place of SPL (\ref{sec:Mie}), a comparison with observations (\ref{sec:disccomp}), and a comparison with the theory of {\rEI} (\ref{sec:discEI}).

\subsection{How results depend on the used spatial resolution}\label{sec:resolution}
A strong argument in favour of using an adaptive grid equation is that such models are able to resolve shocks; this is shown by, for example, \citet{DoFe:91} and \citet{FeDo:94} for single shocks and pulsations.

We show in {\rSa} (where $\ngrid=500$ and $700$) that winds are better modelled without resolving shocks. The adaptive grid equation is still used to move the inner boundary to simulate stellar pulsations. Models become smoother and also often periodic than when the adaptive grid equation is used to resolve shocks, contrary to what is claimed in the literature.

\begin{figure*}
\includegraphics{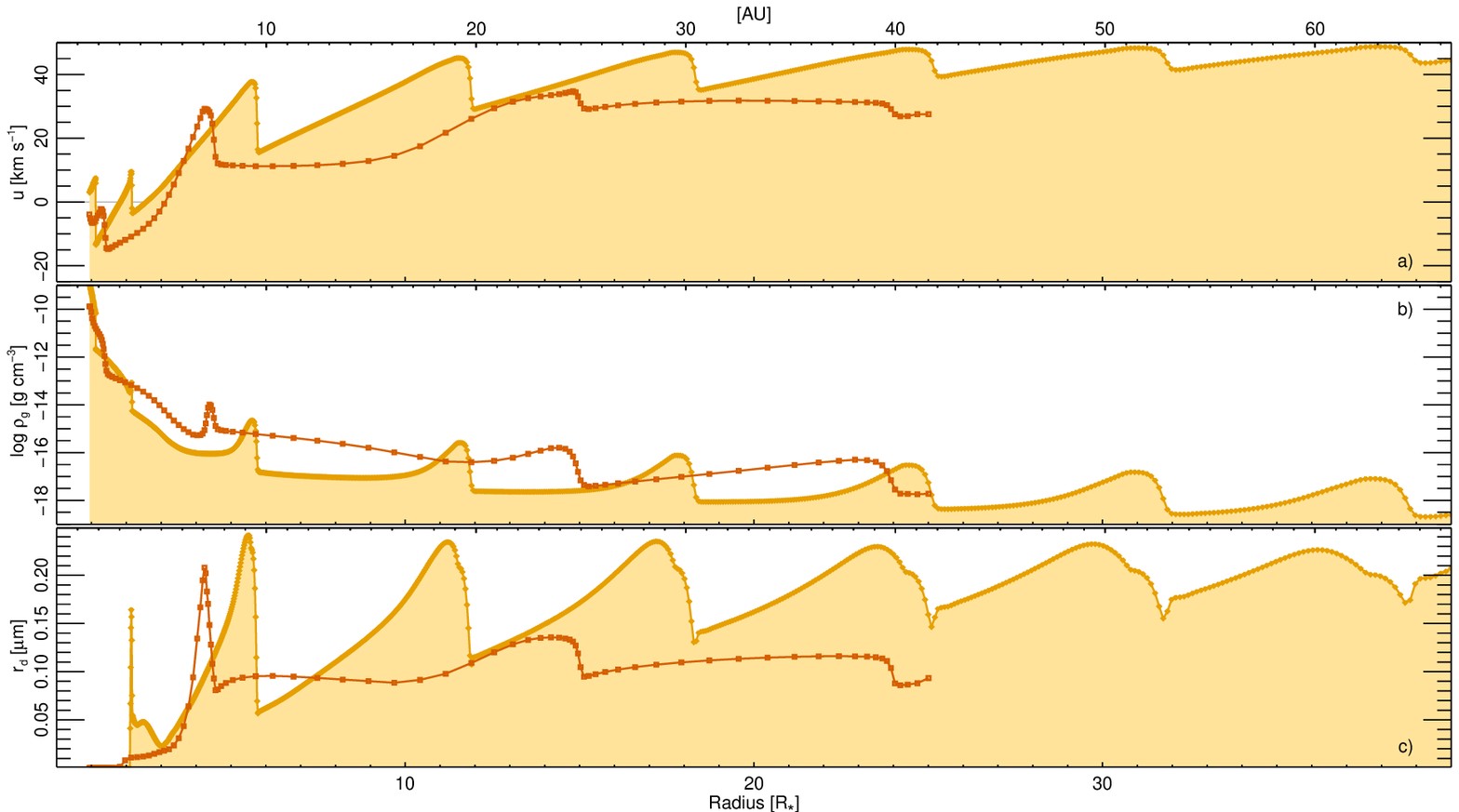}
\caption{Radial structure of an arbitrary snapshot of two PC models of setup L3.85T28E88 for the full modelled region, using $\Xi_{1.00}$. The model using a fixed grid and $\ngrid=1024$ (adaptive grid and $\ngrid=100$) is showed with an orange solid line on a light orange background and diamond symbols (red solid line and squares), each symbol shows the location of a gridpoint. The three panels show the: (a) gas velocity $u$, (b) gas density {\rhog} (log), and (c) average grain radius {\radd}. \label{fig:resolution}}
\end{figure*}

In this paper, we took the approach of {\rSa} further and kept the grid fixed at all radii $r>2\Rs$, which alleviates the numerical advection of dust in drift models; the spikes seen in plots of the drift velocity in {\rSaHoa} and {\rSaHob} are thereby avoided as the relative velocity between grid points and the dust is always greater than zero (Appendix~\ref{app:vDtech}). We also increased the number of gridpoints to $\ngrid=1024$, which is more than a factor 10 higher than what is currently used with extant results of the {\darwin} stellar wind model \citep[e.g., {\rHGAJ}; {\rMWH}; {\rENHAW};][hereafter \rBEM]{BlErMaLi.:19}, who use $\ngrid=100$. We also used the higher-accuracy advection scheme of {\rSa}. All shocks of the gas and discontinuities of the dust are resolved with at least one gridpoint across a shock front or a discontinuity with this approach; we illustrate this in Fig.~\ref{fig:resolution} where we show the physical structure of the PC model L3.85T28E88. And they are resolved at all times as the gridpoints are fixed. Shocks are less steep in the outer parts owing to the artificial viscosity, which length scale {\lav} $\propto r$, see equation~(\ref{eq:avisc}). A large fraction of the PC models (mainly) are periodic, mostly with a periodicity $1P$ (see Section~\ref{sec:periodic}).

In comparison, models that use the adaptive-grid equation and $\ngrid=100$ track one or two shocks in the outer parts of the wind, and show much higher variability; such models appear to be unable to adequately resolve multiple shocks. The reason is that most gridpoints gather about one or two shocks leaving remaining regions unresolved, see Fig.~\ref{fig:resolution}. Thereby, it is hardly possible to follow shocks that develop with each pulsation period; there are no gridpoints available to resolve them. Our conclusion could perhaps have been different if {\ngrid} was higher in these models. Notably, our comparison of benchmark test results shows reasonable agreement between results of {\teh} and {\darwin} that use the same physics setup (Appendix~\ref{app:benchmark}). Our results suggest that wind models are more accurate overall when the adaptive grid equation is not used and all shocks and all regions are resolved. This is particularly important in drift models where dust is not affixed to gas shocks.

\subsection{Wind formation in our new models}\label{sec:discwind}
\subsubsection{The radial structure of model L3.70T28E88}\label{disc:L370T28E88}
\begin{figure*}
\includegraphics{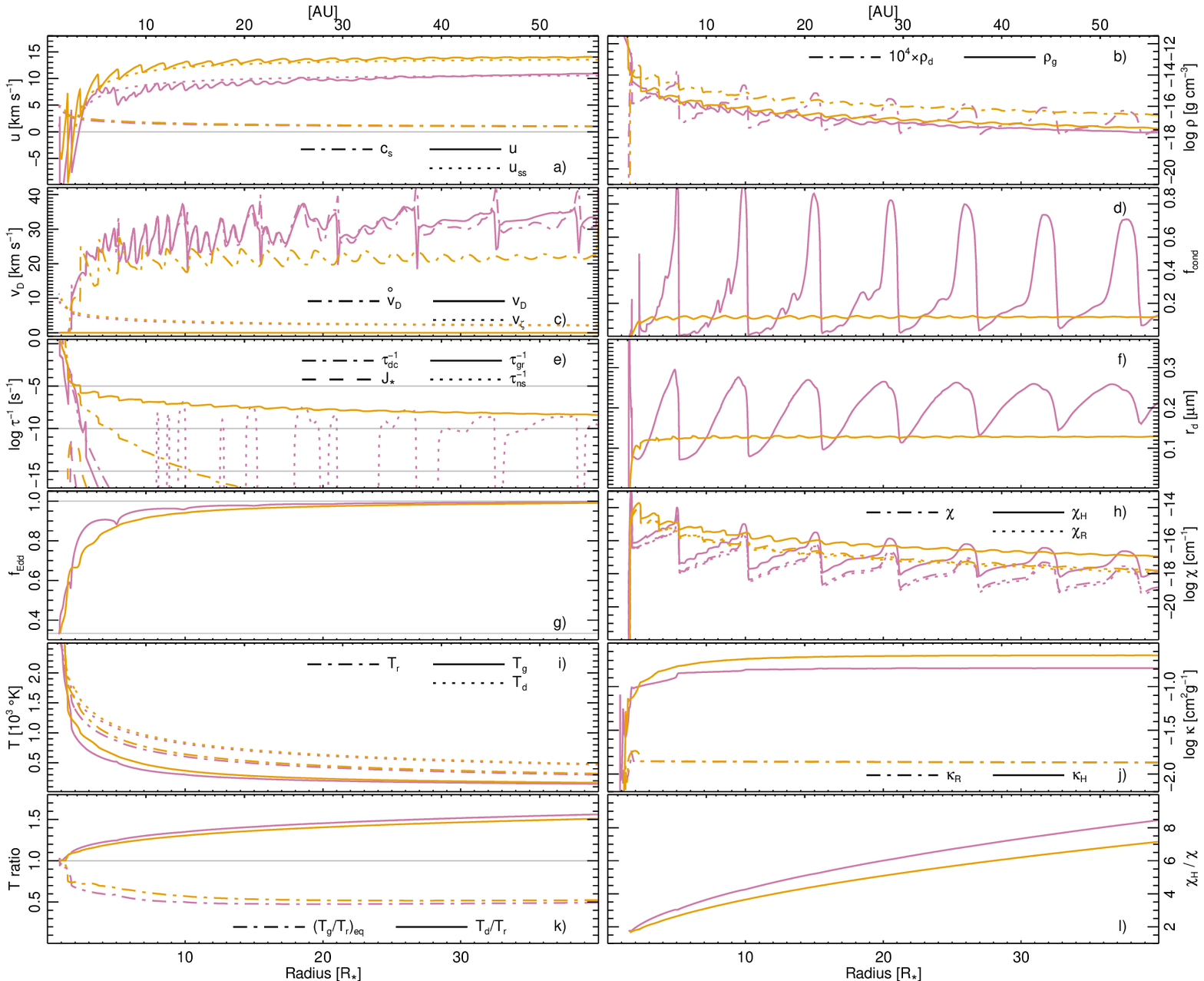}
\caption{Radial structure of a snapshot of setup L3.70T28E88 for the full modelled region, using the sticking-coefficients setup $\Xi_{0.34}$. The drift (PC) model is shown with purple (orange) lines. From the top left, the 12 panels show: (a) gas velocity $u$, sound speed \cs, and self-similar gas velocity \uss; (b) gas density $\rhog$, dust density $10^4\times\rhod$ (log); (c) drift velocity {\vD}, equilibrium drift velocity {\vDeq}, and thermal velocity \vzeta; (d) degree of condensation {\fcond}; (e) net growth rate {\tauGi}, net decay rate {\taudc}, nucleation rate $J_{\star}$, and non-thermal sputtering rate {\tausp} (log); (f) average grain radius {\radd}; (g) Eddington factor {\fedd}; (h) extinction coefficient {\chiH}, grey extinction coefficient {\chigrey}; (i) gas temperature {\teg}, radiative temperature {\ter}, and dust temperature {\ted}; (j) opacity {\kappaH} and Rosseland mean opacity {\kappaR} (log); (k) temperature ratios $\ted/\ter$ and $(\teg/\ter)_{\text{eq}}$; and (l) extinction coefficient ratio $\chiH/\chigrey$. All properties are drawn as function of the stellar radius {\Rs} (lower axis) and astronomical units (AU; upper axis). Grey horizontal lines are guides.\label{fig:discwind}}
\end{figure*}

We illustrate the physical structure for the PC and drift models of setup L3.70T28E88 in Fig.~\ref{fig:discwind}; this drift model shows the smallest dust mass-loss rate of all our models. (We show a similar figure for the PC and drift models L3.85T28E88 using Mie scattering instead of the SPL in Fig.~\ref{fig:discMie}). The PC model shows a periodic structure with small-amplitude variations in both the gas and the dust. Dust formation takes place throughout the envelope (Fig.~\ref{fig:discwind}e), although at much reduced rates in the outer parts and the rate of grain growth at, say, radii $r\ga10\,\Rs$ is usually negligible. Neither the degree of condensation (Fig.~\ref{fig:discwind}d) nor the grain radius (Fig.~\ref{fig:discwind}f) increase by much for $r\ga7\Rs$. The degree of condensation is rather low compared to denser grey opacity models (cf. figs.~2 and 4 in {\rSa}). We calculated the equilibrium drift velocity {\vDeq} (equation~\ref{eq:vdeq}), which reaches about 20\,\kms already at small radii (Fig.~\ref{fig:discwind}c); this velocity is supersonic for $r\ga2.5\,\Rs$ as $\vDeq\gg\vzeta$.

The drift model also shows a periodic structure. The mass loss rate is less than half of the corresponding PC model. The drift velocity (Fig.~\ref{fig:discwind}c) attains values of $\vD=20\kms$ already at $r\simeq2.5\Rs$, and increases to about $30$--$35\,\kms$ at larger radii; this is in sharp contrast to earlier grey models where drift velocities were always small ($\vD<16\kms$ in {\rSaHoa}--{\rSa}; cf.~figs.~2c and 4b in {\rSa}). The equilibrium drift velocity {\vDeq} of the drift model is very similar to {\vD} for $r\lsim10\,\Rs$ and shows larger deviations of up to $5\,\kms$ in less dense regions at larger radii. Our own tests show that the momentum coupling is nearly complete (Appendix~\ref{app:cmc}). There is -- in this case -- some similarity between {\vDeq} of the drift model and that of the PC model (see Section~\ref{disc:fullsample}). The drift velocity is supersonic at large Mach numbers throughout the radial domain ($\vD\gg\vzeta$).

The variations in the drift velocity and the dust density (Fig.~\ref{fig:discwind}b) show discontinuities that are moving outwards at speeds more than twice as high as the gas. The dust appears to accumulate in separated shells, where there is little dust between the shells. The high variability of the drift model makes a direct comparison of amounts of dust between the PC and drift models difficult (Fig.~\ref{fig:discwind}d).

In the shown snapshot, dust formation mostly occurs in a narrow region, where $1.5\lsim r\lsim2.2\Rs$ (Fig.~\ref{fig:discwind}e). At larger radii, gas particles do not stick to dust particles any more if drift velocities are too high (cf., e.g., fig.~1 in {\rSa}). In such locations, grain decay dominates, and in particular grains are ablated by gas particles (\tausp). However, the grain decay is inefficient.

\begin{figure*}
\includegraphics{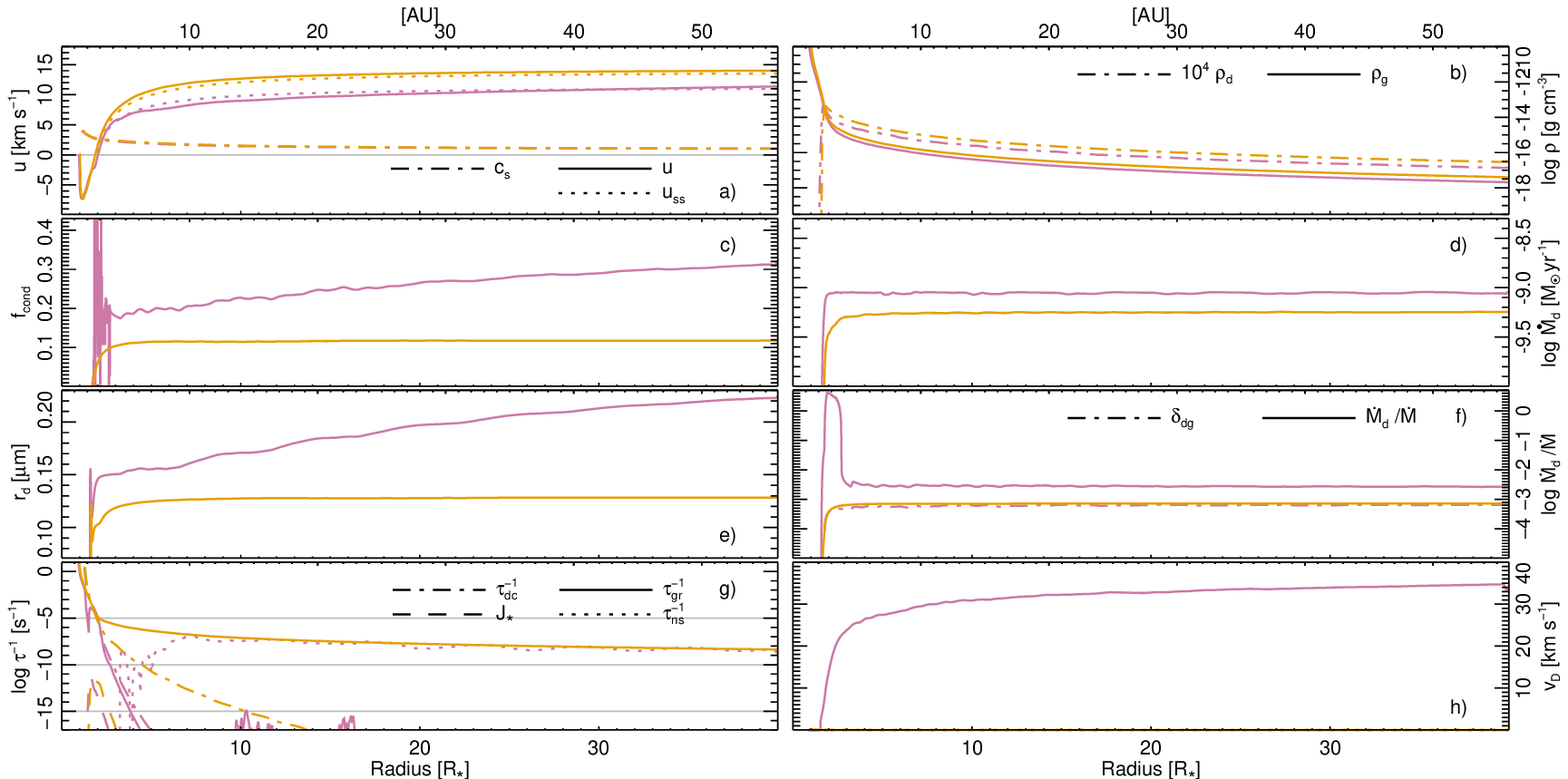}
\caption{Temporally averaged radial structure of setup L3.70T28E88 for the full modelled region. The drift (PC) model is shown with purple (orange) lines. From the top left, the eight panels show the: (a) gas velocity $u$, sound speed \cs, and self-similar velocity {\uss}; (b) gas density {\rhog}, dust density $10^4\times\rhod$ (log); (c) degree of condensation {\fcond}; (d) dust mass loss rate {\dmdot} (log); (e) average grain radius {\radd}; (f) dust-to-mass loss rate ratio $\dmdot/\mdot$, dust-to-gas density ratio {\ddg} (log); (g) net growth rate {\tauGi}, net decay rate {\taudc}, nucleation rate $J_{\star}$, and total non-thermal sputtering rate {\tausp} (log); and (h) drift velocity {\vD}. All properties are drawn versus the stellar radius {\Rs} (lower axis) and astronomical units (AU; upper axis). Grey horizontal lines are guides.\label{fig:tdiscwind}}
\end{figure*}

It is easier to scrutinise the wind formation in a temporal average of the radial structure in models of pulsating atmospheres. We show such an average plot for the same model in Fig.~\ref{fig:tdiscwind}. The figure reveals smooth structures in all shown properties of both models. In particular, the gas velocity structure is -- in this case -- very similar to the self-similar structure presented by {\rIE} (equation 1, at small optical depth)
\begin{eqnarray}
\uss=\muinf\left(1-r_{\text{c}}/r\right)^{\frac{2}{3}},\label{eq:uss}
\end{eqnarray}
where we set the dust condensation radius $r_{\text{c}}=2\Rs$.

The drift model shows structures that are as smooth as those of the PC model. The mass loss rates of both the gas and dust are constant for $r\ga5\Rs$ (Figs.~\ref{fig:tdiscwind}d and \ref{fig:tdiscwind}f). Meanwhile, both the degree of condensation (Fig.~\ref{fig:tdiscwind}c) and average grain radius (Fig.~\ref{fig:tdiscwind}e) show a significant increase throughout the model domain; this cannot be explained by dust formation as both grain growth and decay are negligible (Fig.~\ref{fig:tdiscwind}g). Instead, the slope appears when dust leaks \citep[cf. ``free streams'',][]{HoLe:16} into more dust free regions between dust fronts (such dust fronts are seen in Figs.~\ref{fig:discwind}b and \ref{fig:discwind}d); more dust has moved into regions between fronts as the wind reaches larger radii. The larger the separation between dust fronts, the steeper the slope. Finally, also the drift velocity increases with the radius (Fig.~\ref{fig:tdiscwind}h).

\subsubsection{Radial structure properties of the full model sample}\label{disc:fullsample}
In our analysis of radial structures of the full model sample, we consider four properties that show some change: outflow velocity, drift velocity, dust formation, and amount of formed dust. We first discuss our PC model results and thereafter our drift models.

Outflow velocities of PC models typically increase by 10--70 per cent in the radial interval 10--{40\,\Rs}. Values are usually higher with higher model luminosity (\Ls) and [initial] carbon-to-oxygen ratio ($C/O$). The increase is lower in the radial interval 20--{40\,\Rs}, up to 20 per cent. The terminal outflow velocity is reached already at {20\Rs} in model L3.85T30E88. The self-similar solution for the outflow velocity {\uss} (equation~\ref{eq:uss}) provides a good description of the velocity structure in five models: L3.70T26E85, L3.70T28E88, L3.85T24E85, L3.85T28E85, and L3.85T28E85, where all but one model have a low C/O ratio; the velocity structure is somewhat to much less steep than {\uss} in the remaining models.

Grain growth occurs at some rate throughout the model domain, but is balanced by grain decay through evaporation and chemical sputtering for $r\lsim2\Rs$. In our examination of the radial dust mass loss rate structure, we see that dust formation is complete (within a few per cent) at $r\approx5$--$10\,\Rs$, typically, but in a few models it appears that a larger radial interval is needed, $r\approx20$--$30\,\Rs$. In models with a higher mass loss rate, the dust mass loss rate structure increases more or less monotonically with the radius, but in all models with a lower mass loss rate, a peak is reached at about $r=2\,\Rs$, where grain decay causes decreased values out to, say, $r\approx3\,\Rs$ (L3.8T24E85, L3.70T26E85, L3.85T26E85, L4.00T26E82, L3.85T28E85, and L4.00T32E88).

In drift models, the terminal velocity is typically reached at shorter radii. The self-similar solution describes average outflow velocities well in three models: L3.70T24E88, L3.70T28E88, and L3.85T28E88; the velocity structure is somewhat to much less steep than {\uss} in the remaining models. There is one exception, the velocity structure of M3.85T30E88 increases more than {\uss}; also here, the optically thin exponent of {\uss} (2/3) provides a better fit than when using the optically thick exponent (2/5, see \rIE). The outflow velocity increases by up to 40 per cent in the radial interval 10--{40\,\Rs} (up to 12 per cent for 20--{40\,\Rs}).

The drift velocity varies with the model setup and the radius. Dust grains accelerate fast from, say $\vD\simeq2$--$10\,\kms$ when they form at $r\lsim2\,\Rs$, to higher velocities $\vD\simeq20$--$100\,\kms$ at $3\Rs\lsim r\lsim5$--$10\,\Rs$. Values are typically higher in regions between more dense shells of dust where $5\lsim\vD\lsim30\,\kms$. Such differences are not seen in a temporal average of the radial structure. The equilibrium drift velocity overlaps the drift velocity well at lower radii $r\lsim10\,\kms$, and mostly shows a somewhat larger deviation at larger radii where the difference is $5$--$50\kms$. Differences are larger in regions between dust shells. The equilibrium drift velocity of the PC models is sometimes similar to the drift velocity throughout the model domain (L3.85T24E91 and L3.70T30E91), but is more often similar in denser dust shells. For most models, the equilibrium drift velocity bears little resemblence to the actual drift velocity, which is not strange considering that the two values are calculated using different physical structures!

Grain formation turns into ablation when drift velocities are high enough. Typically, such higher values are reached where $r\ga4\Rs$, and then in regions where there is less dust between dust shells. We find that ablation has a minor effect on the dust formation and is unimportant in models where $\mvdrinf\lsim30\,\kms$. As a comparative remark, in his stationary wind models, \citet{Kw:75} finds that all dust is ablated when $\vD\ge20\,\kms$.

Our scrutiny shows that it is necessary to model a larger region that extends out to, say, $40\,\Rs$ to calculate a more accurate terminal velocity. The full region of dust formation should mostly be covered in a model that extends to $10\Rs$. Currently, our models use one average dust velocity for grains of all sizes. It is possible that our results would be different if the models would include grain size-dependent dust velocities, which could affect the drift velocity and thereby the rate of ablation for grains of different size.

\subsubsection{Reasons for higher drift velocities than in grey models}\label{sec:greydiff}
Our results using grey and constant opacities show minor effects of drift ({\rSaHoa}--{\rSa}), where the average drift velocity is typically $\mvD\approx5\,\kms$; for the full sample of grey models, $\mvD\lsim16\,\kms$. The results we present here show larger effects. In our analysis of the reason behind this discrepancy, we here examine the terms that are different in the two approaches.

The dust temperature is calculated differently in the grey and frequency-dependent approach. In grey models, we have used $\tedg=\ter$, whilst in frequency-dependent models, the radiative temperature is weighted with the extinction coefficient ratio $\left(\chiJ/\chiS\right)^{1/4}$ (equation~\ref{eq:tdust}). Furthermore, in radiative equilibrium,
\begin{eqnarray}
\kappaJ J=\kappaS S(\teg^{\text{eq}}),\quad\mbox{and so}\quad \left(\frac{\teg}{\ter}\right)^{\text{eq}}=\left(\frac{\kappaJ}{\kappaS}\right)^{\frac{1}{4}}\,\greq\,1.
\end{eqnarray}
Temperature ratios $\left(\teg/\ter\right)^{\text{eq}}$ that deviate from $1$ indicate the importance of non-grey RT (cf. section~3.1 in {\rHGAJ}).
We show the gas, dust, and radiative temperatures as well as the temperature ratios $\ted/\ter$ and $\left(\teg/\ter\right)^{\text{eq}}$ in Figs.~\ref{fig:discwind}i and \ref{fig:discwind}k (cf. figs.~3c, 5b, and 5c in \rHGAJ). In this context, our PC model ratio $\ted/\ter$ shows a good agreement with what {\rHGAJ} present for a model with different parameters. Meanwhile, our model shows a ratio that is lower than theirs; the ratio decreases towards $\left(\teg/\ter\right)^{\text{eq}}\simeq0.5$ for larger radii whilst their value is $\left(\teg/\ter\right)^{\text{eq}}>0.8$ for $2\lsim r<7.7\Rs$. The drift model ratios are similar, with a somewhat steeper temperature gradient of the gas, which might indicate an even stronger importance of non-grey RT when drift is included.

A fundamental difference between grey and frequency-dependent models is how the extinction is calculated. The ratio between {\chiH} (equation~\ref{eq:chi}) and {\chigrey} (equation~\ref{eq:chigrey}) for L3.70T28E88 is about 1.8 when dust first forms and increases to larger values with the radius, see Fig.~\ref{fig:discwind}l (cf. fig.~5d in \rHGAJ). The ratio increases somewhat faster with radius in the drift model.

\begin{figure}
\includegraphics{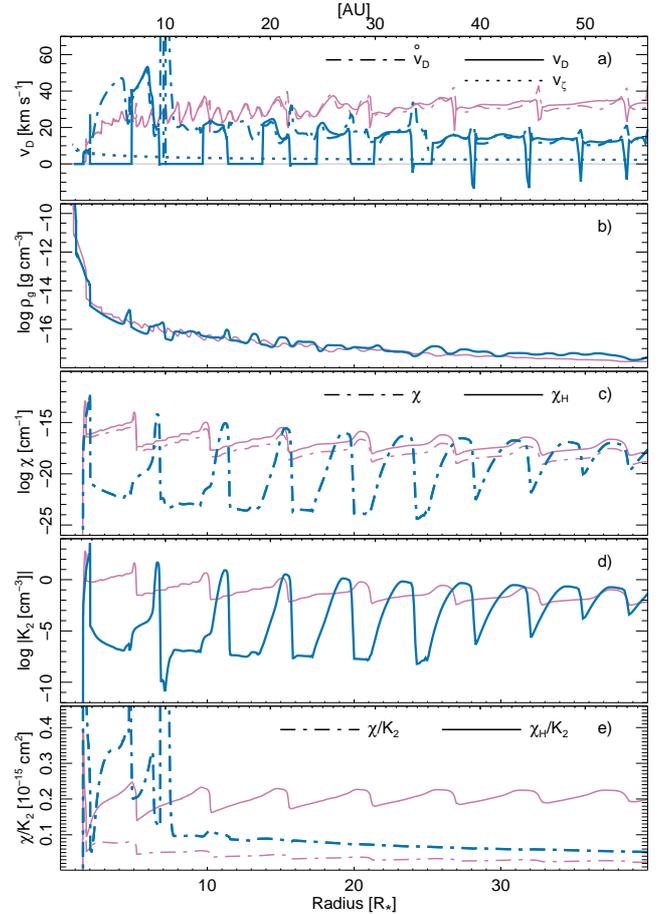}
\caption{Comparison of two drift models using {\chiH} (purple lines) versus {\chigrey} (thick blue lines) using radial structures of a snapshot of setup L3.70T28E88 for the full modelled region. The five panels show: (a) drift velocity {\vD}, equilibrium drift velocity {\vDeq}, and thermal velocity {\vzeta}; (b) gas density {\rhog} (log); (c) extinctions {\chiH} and {\chigrey}; (d) second dust moment $K_{2}$ (log); and (e) the ratios $\chiH/K_{2}$ and $\chi/K_{2}$. See Fig.~\ref{fig:discwind} for more details.\label{fig:discchigrey}}
\end{figure}

To examine consequences owing to this discrepancy more closely, we calculated a drift model of setup L3.70T28E88 where we used the grey extinction {\chigrey} instead of the frequency-dependent weighted average extinction {\chiH}. We show the resulting radial structure of the two drift models (that use {\chiH} and {\chigrey}) in Fig.~\ref{fig:discchigrey}; the figure shows all variables in the equilibrium drift velocity {\vDeq} (equation~\ref{eq:svdeq}) except the radiative flux $H$, which is nearly identical in the selected snapshots of the two models.

The drift velocity {\vD} is significantly lower in the model that uses the grey extinction {\chigrey}, $10\la\vD(\chigrey)\la15\kms$ instead of {\chiH}, $20\la\vD(\chiH)\la40\kms$, see Fig.~\ref{fig:discchigrey}a. Dust grains collect in shells with nearly dust free regions between shells (Figs.~\ref{fig:discchigrey}c and \ref{fig:discchigrey}d); the shells are smeared out when moving outwards. The gas density {\rhog} is similar throughout the radial domain, Fig.~\ref{fig:discchigrey}b. The dust extinction {\chigrey} (the dust moment $K_{2}$) is, moreover, a factor $10^8$ ($10^7$--$10^9$) higher at the front of the dust shells than between them; the ratio is smaller towards the outer boundary. The ratio between the dust extinction and the dust moment $K_{2}$ (Fig.~\ref{fig:discchigrey}e) illustrates more clearly that these two properties are responsible for the higher drift velocity when using {\chiH} instead of {\chigrey} (see equation~\ref{eq:vdeq}). However, because the physical structure changes with the different physical conditions (whence all three variables are modulated), we have not proven that the lower grey extinction is the only reason behind the differences.

\subsubsection{Parameter combinations that fail to form a wind}\label{sec:nowind}
Models fail to form a wind when too little dust forms. Such conditions are characterized by smaller amounts of carbon $C/O$ and low luminosity {\Ls}, as well as high effective temperature {\teff}.

In five PC models with low outflow velocities (L3.85T24E85, L4.00T26E82, L3.85T28E85, L4.00T28E85, and L4.00T30E85), the dust stays in place and accumulates until the combined radiative pressure on larger amounts of dust is able to form a wind. In the corresponding drift models, the drift velocity reaches high values near where dust is first formed. The high drift velocities ablate dust grains, and instead of accumulating, dust grains move outwards through the gas and leave the gas without dragging it along. Our models are not setup to handle such cases where the outer boundary falls back towards the photosphere. Instead, the increased amounts of dust around the star heat up and affect the RT and overall physical structure in the enclosed star.

Six models show a situation where the mass loss rate and outflow velocity in the PC model become higher in the drift model (L3.85T24E91, L3.85T30E88, L4.00T30E91, L4.00T32E88, L4.00T32E91, and L4.00T28E88 that is calculated using Mie scattering). The more efficient dust formation in the drift model is able to form more dust that is able to drive the wind more efficiently than in the PC case.

The aim of our study is not to set the limits of wind formation, but we find that the dust mass loss rate $\mdmdot\la2.0$--$5\times10^{-10}\,\mdotu$ in setups that fail to form a wind. It is tricky to make a better determination as the amount of formed dust also depends on if the drift velocity is high enough that grains are ablated by non-thermal sputtering. Also, a smaller value of $C/O\la8.50$ appears to make wind formation difficult; in our current set of models, only one such drift model forms a wind, at high luminosity and low effective temperature (L4.00T24E85).

\subsection{Evolution of temporally averaged properties}\label{sec:temporal}

\subsubsection{Classification of periodic and irregular variations}\label{sec:periodic}
The temporally averaged properties in Table~\ref{tab:resall} were calculated using different time intervals. Some models show periodic variations relatively quickly after calculations begin, whilst other never turn periodic. We use three variability classes: irregular (i), periodic ($l\times$p), and nearly periodic or quasi-periodic ($x\times$q). Variations of quasi-periodic winds are not perfectly periodic in both gas and dust properties or show a multiplicity that deviates from an integer factor of the pulsation period. Winds with a lower outflow velocity show a perfectly periodic radial structure with very low-amplitude variations that do not appear as periodic at the outer boundary -- we still classify these winds as periodic.

\begin{figure*}
\includegraphics{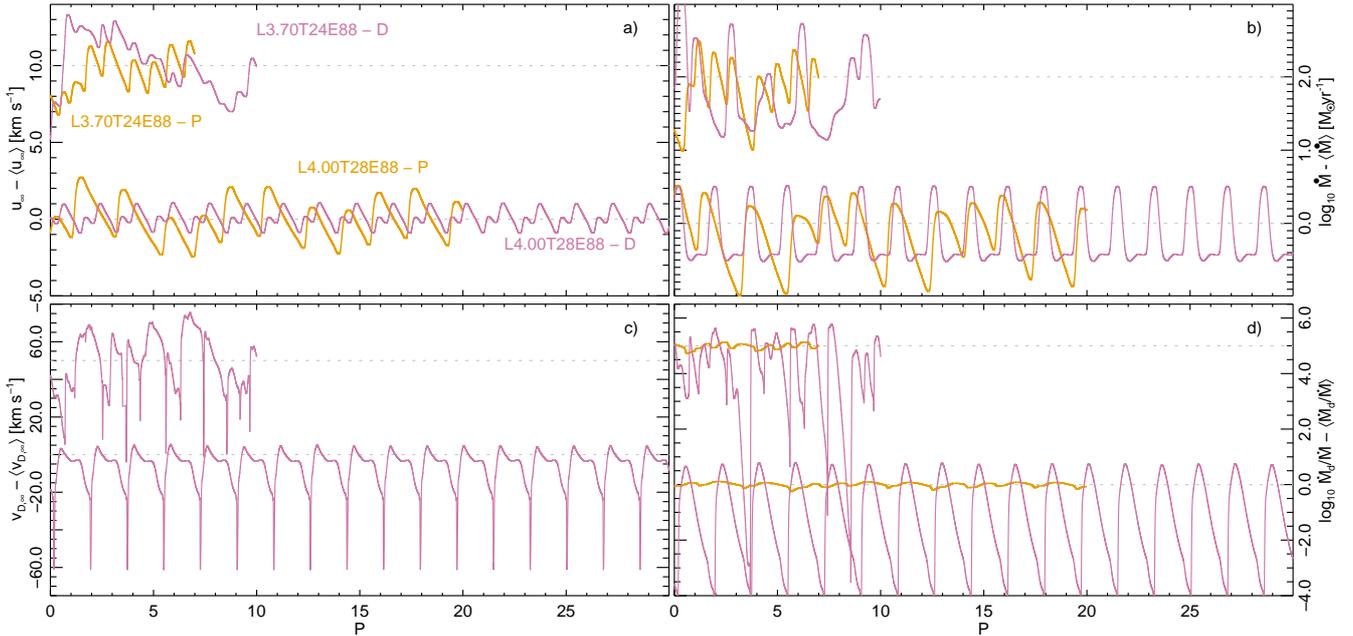}
\caption{Temporal structures for the last set of periods of models L4.00T28E88 (lower pair of lines) and L3.70T24E88 (upper pair of lines). The two drift (PC) models are shown with purple (orange) lines. From the top left, the four panels show: (a) terminal valocity {\uinf}, (b) mass loss rate {\mdot} (log), (c) terminal drift velocity {\vDinf}, and (d) dust-to-gas mass loss ratio $\dmdot/\mdot$ (log). All properties are drawn using the value at the outer boundary, subtracting the respective mean value shown in Table~\ref{tab:resall}, versus the model age in pulsation periods $P$. Grey dotted lines are guides that show the offset of the values of model L3.70T24E88.\label{fig:disctemp}}
\end{figure*}

Out of the 31 PC model winds in Table~\ref{tab:resall}, 26 models are classified as either periodic (15) or quasi-periodic (11). The remaining five models are classified as irregular. The classification of the 22 drift models are different with 8 irregular structures, and 14 structures that are either periodic (10) or quasi-periodic (4). We show four temporal structures of both drift and PC models to illustrate the three classifications in Fig.~\ref{fig:disctemp}, the two model setups are L4.00T28E88 and L3.70T24E88. The PC model of setup L3.70T24E88 is only evolved for a couple of periods after the structure turns quasi-periodic, of period $1.6P$. Both PC models show a variability in the terminal velocity and mass-loss rate that change in amplitude and are also not integer multiples of the respective pulsation period, which is why both model structures are classified as quasi-periodic instead of periodic. The drift model L3.70T24E88 shows a clear irregular structure, and finally the drift model L4.00T28E88 shows a clear periodic structure.

The classification is occasionally a bit uncertain between the periodic and quasi-periodic classes. The classification might also change with longer modelling intervals. However, we believe that it is more important that the models are further developed with more physics, which could change the structure completely, before the assessments on this level of detail are attempted anew. We note, however, that the classification of all more numerically accurate models in {\rSa} are stationary or periodic; this result is not reproduced here with our new models, currently.

\begin{figure*}
\includegraphics{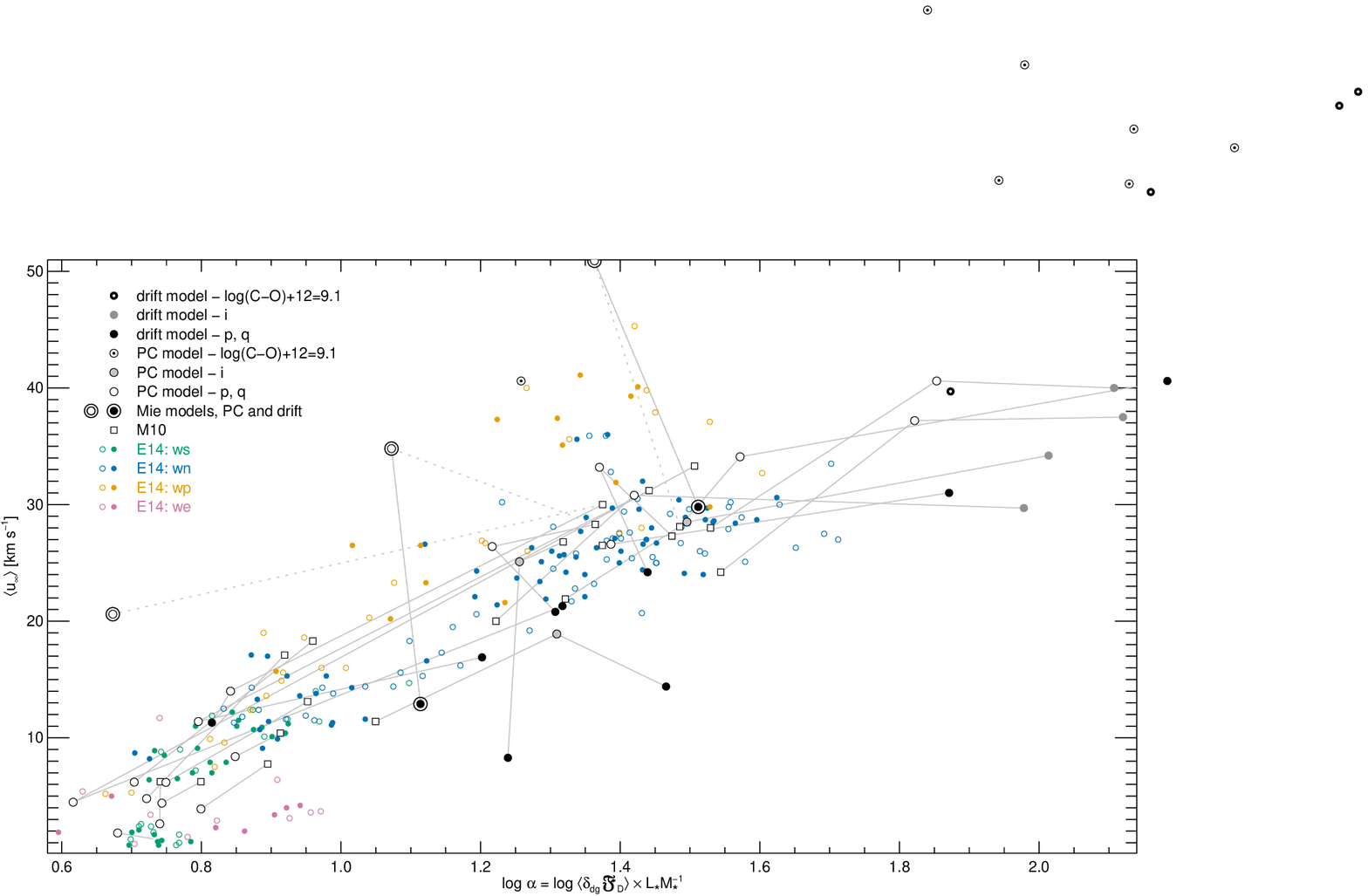}
\caption{Terminal velocity versus $\alpha$ (equation~\ref{eq:alpha}). The figure is largely a reproduction of fig.~5 in {\rENHAW}, whose model values are shown with coloured symbols where the colour indicates properties of the wind classification; the authors vary the inner boundary luminosity using $f_{\text{L}}$ that is set to 1 (circles $\circ$) and 2 (bullets $\bullet$). Original model values of {\rMWH} are shown with a square $\square$ and values of our PC (drift) models are shown with circles $\circ$ (bullets $\bullet$). Drift, PC, and original model values using the same input parameters are connected with light grey solid lines. Mie models are connected with grey dotted lines.\label{fig:discalpha}}
\end{figure*}

\subsubsection{Outflow velocity versus radiative acceleration}\label{sec:acceleration}
The outflow velocity (and mass-loss rate) is often related to the ratio of radiative to gravitational acceleration \citep[of the dust; e.g.][chapter~7]{LaCa:99}
\begin{eqnarray}
  \Gamma=\frac{\chiH\Ls}{4\pi cG\Ms}.
\end{eqnarray}
In {\rMWH} (equation~7), we define the wind-formation efficiency parameter $\alpha=\ddg\Ls\Ms^{-1}$ that is, in principle, proportional to $\Gamma$. Here, we adjust this parameter to account for the dilution of the dust component owing to drift (see equation~\ref{eq:dmddm})
\begin{eqnarray}
  \alpha=\ddg\FD\frac{\Ls}{\Ms}\label{eq:alpha}.
\end{eqnarray}
{\rMWH} (fig.~3) plot $\uinf(\alpha)$ for all models, as do {\rENHAW} (fig.~5). We plot our results on top of the values of {\rENHAW} (who use SPL to calculate dust opacities), see Fig.~\ref{fig:discalpha}.

The wind-driving mechanism is stronger at higher outflow velocities. Our PC model values, as well as the associated values of {\rMWH}, overlap the values of {\rENHAW} well. Some of our PC model $\alpha$ values are higher than what both {\rMWH} and {\rENHAW} find. It is more interesting to see that the drift-model values are shifted towards higher to drastically higher values of $\alpha$. Whilst the relation shows a high correlation for PC models, the correlation is lower with drift models. The figure illustrates the drastically higher amounts of dust formed in drift models at increasing values of $\alpha$.

In comparison to our study, {\rENHAW} define four different pulsation classes for PC models that form winds: \textit{steady} winds with small temporal variations (\textit{ws}), winds with \textit{periodic} variations in properties (\textit{wp}), winds with more irregular \textit{non-periodic} variations (\textit{wn}), and winds that show an intermittent \textit{episodic} outflow (\textit{we}). Most models are non-periodic, slightly fewer models are steady or episodic (at lower outflow velocities ($\uinf\la13\,\kms$) and the remaining models are periodic (see their figure~5). The authors show a fact sheet for one model that they classify as periodic (\textit{wp}; fig.~C.1), which reveals a temporally variable structure that we would here classify as irregular. We believe our models show more accurate and periodic structures than {\darwin} (Appendix~\ref{sec:numdiff}). We find periodic variations across a larger region of the plot than {\rENHAW}; including all models with $\muinf<17\,\kms$ and all drift models with $\muinf<25\,\kms$. Notably, our PC models that were calculated using Mie scattering are found at higher terminal velocities than all our other models (at the same $\alpha$ value). Also noteworthy, already in our work leading up to {\rMWH}, we find that an intermittent nature of many time series of {\darwin} prevents a meaningful comparison of uncertainties in outflow velocities (see Appendix~\ref{sec:numdiff}).

\subsubsection{Characterizing mean wind properties with {\FD}}\label{sec:discdriftfactor}
Figures~\ref{fig:discdriftfactorm} and \ref{fig:discdriftfactoro} illustrate well defined ranges of physical values that result with our current set of wind models. Additional sets of models could likely extend the relations further towards both less and more massive winds.

The figures reveal an exponentially decreasing dependence with {\FD} in the mass loss rate {\mmdot} (Fig.~\ref{fig:discdriftfactorm}a), dust mass loss rate {\mdmdot} (Fig.~\ref{fig:discdriftfactorm}b), dust-to-gas mass loss ratio $\mmdmdot=\ddg\FD$ (Fig.~\ref{fig:discdriftfactorm}c), terminal velocity {\muinf} (Fig.~\ref{fig:discdriftfactoro}a), and degree of condensation {\mfcond} (Fig.~\ref{fig:discdriftfactoro}b); fits are shown in the respective figure panel. The mean terminal drift velocity {\mvDinf} instead increases with {\FD} (Fig.~\ref{fig:discdriftfactoro}d). There is little correlation between the luminosity of each fit with the central star luminosity, except that the models using $\log\Ls=3.85$ seem to result in the best exponential fits. It is unclear that any similar relation exists for the mean grain radius {\mradd} (Fig.~\ref{fig:discdriftfactoro}c).

Our current models reveal a maximum mass loss rate of $\mmdot(\FD=1)\approx10^{-5}\mdotu$, where $\mvD=0\kms$. It appears that drift models do not form higher mass loss rates; optically dense models might play a role to understand the occurrence of such high mass-loss rates (see the discussion in \rIE). And the mass loss rate decreases with increasing drift factor, down to $\mdot(\FD\simeq4.1))\approx2\times10^{-7}\,\mdotu$. Our models show a lack of lower mass loss rates, which is likely a result of the model parameters we have chosen for our calculations. But the results are also a function of the assumptions in form of grain properties and the interaction between the gas and dust.

The fit to the dust mass loss rate is most strongly correlated and it is also shows the steepest decrease with the drift factor. The range of values is $2.4\times10^{-10}(\FD=4.1)\la\mdmdot\la2.5\times10^{-7}(\FD=1.0)\,\mdotu$. Whilst dust formation in the form of grain growth increases with lower values of the drift velocity (cf. fig.~1 in \rSaHoc), the grain growth quickly becomes less efficient at higher drift velocities where grains are also ablated by non-thermal sputtering; compare Fig.~\ref{fig:discvDr}, which shows low dust mass loss rates in five out of six models where $\mvD>30\,\kms$. The exception is the high carbon-content model $L4.00T32E91$. At some point where $\vD\ga40\,\kms$, there is a cutoff where dust is unable to drive a wind (see Section~\ref{sec:nowind}).

All values of the dust-to-gass mass loss ratio lie in the range $0.66\la10^3\ddg\FD\la2.6$, which is higher than most of the non-drift dust-to-gas density ratios {\ddg} that are reported in earlier studies (see figure~4 in {\rMWH}, figure~5 in \rENHAW, and figure~8 in \rBEM), also see Section~\ref{sec:disccomp}. Non-drift models show too low values except with the highest mass loss rates when $\FD=1$.

The terminal velocity decreases with the drift factor, which indicates increasing difficulties at forming high outflow velocities when the drift velocity increases. The highest terminal velocity in our models is about $65\,\kms$ at $\FD=1$; notably, observations do not show such high values (see Section~\ref{sec:disccomp}). All terminal velocities $\muinf>39\,\kms$ are found in the high carbon-to-oxygen-models, which are also not seen in observations. Moreover, the degree of condensation decreases with the power 1.3 of the drift factor and all values are found in the range $0.11\la\mfcond\la0.60$.

The mean grain radius shows no evident relation with the drift factor. Instead, we show the dust-to-gass mass loss ratio {\mmdmdot} versus the mean grain radius {\mradd} in Fig.~\ref{fig:discddg}. The figure shows that the mean grain radius increases with the carbon-to-oxygen ratio.

The drift velocity reveals a trend of increasing values with the drift factor. The minimum value of our fit is $\mvD(\FD=1)\simeq12\,\kms$, whilst the models show $\mvD>10\,\kms$. Drift velocities are with few exceptions higher than in earlier higher accuracy Planck-mean models where $3\le\mvD\le5\kms$ (\rSa), and lower-accuracy Planck-mean models where $4\le\mvD\le16\kms$ and constant-opacity models where $2\le\mvD\le13\kms$ (\rSaHob). The upper limit on the drift velocity in these older papers is indicated with a horizontal guide in fig.~\ref{fig:discdriftfactoro}d. Notably, the older results are calculated for a different region in the parameter space.

\subsection{Replacing SPL with Mie scattering in the dust extinction}\label{sec:Mie}
We show in Section~\ref{sec:greydiff} that the higher dust extinction of frequency-dependent models makes a difference compared to when a lower grey extinction is used. For the models presented in this paper, which have a mean grain radius in the range $0.070\la\mradd\la0.75\,\mu$m, the radiative pressure efficiency factor {\Qnuabspr} is up to about a factor 5 higher when Mie scattering is used instead of SPL (see Fig.~\ref{fig:mie}); this implies that effects can be even stronger compared to when using grey extinction values, and drift velocities can be even higher. We calculated three sets of models for setups L3.70T28E88, L3.85T28E88, and L4.00T28E88 using Mie scattering to see how the outcome is affected.

Compared to the results of the respective SPL model, the PC model values are 8.8--38 per cent lower (\mmdot, \mfcond, \mmdmdot). Differences are smaller in the mean grain radius {\mradd}, $13$ per cent lower to $3.5$ per cent higher. However, the terminal velocity {\muinf} is 32--49 per cent higher. Compared to the values of the model of {\rMWH}, the grain radius {\mradd} is $55$ lower to $56$ per cent higher, the terminal velocity {\muinf} 13 per cent lower to 81 per cent higher, and the remaining properties 17--82 per cent lower.

The changes are high in the two drift models as well where the terminal drift velocity {\mvDinf} is 35 per cent higher in model L3.85T28E88 and the mass loss rate {\mmdot} is 19 per cent higher in model L4.00T28E88 than in the corresponding SPL model. The remaining properties are 24--79 per cent lower; additionally, the periodic structure of the SPL model L4.00T28E88 turns irregular using Mie scattering. Compared to the values of the PC model of {\rMWH}, the values of L3.85T28E88 are 7--67 per cent lower. The differences are seemingly lower for model L4.00T28E88 where the degree of condensation {\mfcond} is 56 per cent lower whilst the other properties are 6.9 per cent lower to 14 per cent higher. No wind forms using Mie scattering with setup L3.70T28E88.

\begin{figure*}
\includegraphics{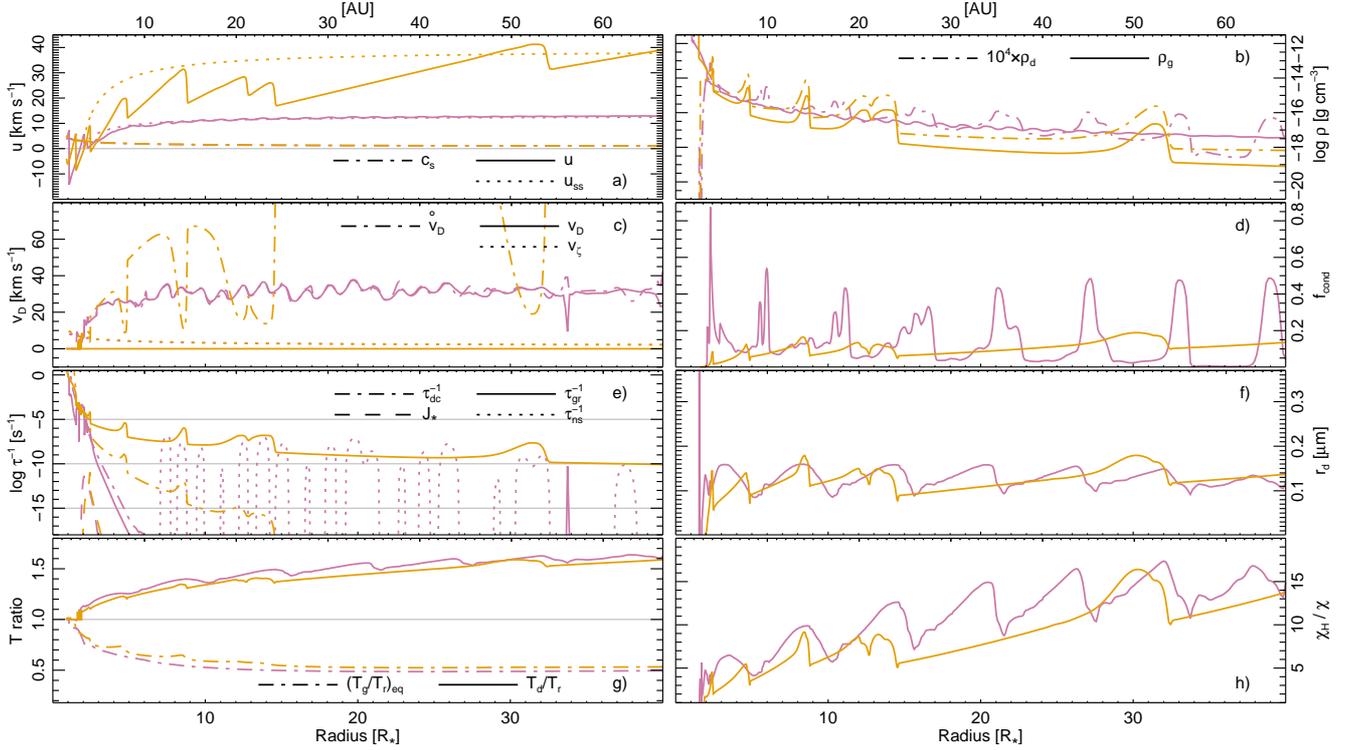}
\caption{Radial structure of a snapshot of setup L3.85T28E88 using Mie scattering. The drift (PC) model is shown with purple (orange) lines. From the top left, the eight panels show: (a) gas velocity $u$, sound speed \cs, and self-similar gas velocity \uss; (b) gas density $\rhog$, dust density $10^4\times\rhod$ (log); (c) drift velocity {\vD}, equilibrium drift velocity {\vDeq}, and thermal velocity \vzeta; (d) degree of condensation {\fcond}; (e) net growth rate {\tauGi}, net decay rate {\taudc}, nucleation rate $J_{\star}$, and non-thermal sputtering rate {\tausp} (log); (f) average grain radius {\radd}; (g) temperature ratios $\ted/\ter$ and $(\teg/\ter)_{\text{eq}}$; and (h) extinction coefficient ratio $\chiH/\chigrey$. All properties are drawn as function of the stellar radius {\Rs} (lower axis) and astronomical units (AU; upper axis). Grey horizontal lines are guides.\label{fig:discMie}}
\end{figure*}

We show the radial structure of the PC and drift models of setup L3.85T28E88 that use Mie scattering in Fig.~\ref{fig:discMie}. The drift model shows a more stationary appearing structure in the gas density and velocity structures (Figs.~\ref{fig:discMie}a--\ref{fig:discMie}c) than the corresponding SPL model (not shown, but compare with the similarly appearing drift-model structure in Fig.~\ref{fig:discwind}). The moderately high and nearly constant value of the drift velocity results in little ablation (Fig.~\ref{fig:discMie}c). Whilst the temperature ratios are about the same as for model L3.70T28E88 (Fig.~\ref{fig:discwind}k), higher values of the extinction are seen in Fig.~\ref{fig:discMie}h; the extinction ratio is about twice as high at the outer boundary and also shows a radial structure.

Considering the amount of formed dust and comparing with the values of the SPL models (Fig.~\ref{fig:discalpha}), the three PC models form less dust and attain drastically increased outflow velocities, which places the models above and away from all other model values. The drift models also form less dust than when SPL is used, but in combination with the lower outflow velocities the new values are closer to the other model values.

In our comparison of results of Mie and SPL models in {\rMH}, we find for two sets of models that effects are rather small; model values are both lower and higher than when SPL is used, although we see a tendency towards smaller amounts of formed dust in all models that use Mie scattering. Here, we have modelled three setups using Mie scattering -- which is why it is too early to generalise the results to all other models. However, the lower terminal velocity achieved with Mie scattering places the two drift models in a region that is closer to values of observations (Fig.~\ref{fig:discmdot}). The large changes, which are larger than we find in the PC models of {\rMH}, indicate that, in models that use a high spatial resolution, Mie scattering is an important process that needs to be used in place of SPL in all models of carbon-rich mass loss rates.

Models that are calculated using Mie scattering result in lower outflow velocities, when models include drift, which also implies higher drift velocities. The models using Mie scattering follow the same trends in properties as the SPL models when the properties are related to the drift factor.

\subsection{Comparison with observations}\label{sec:disccomp}

\subsubsection{Radio and infrared observations}\label{sec:coir}
One approach is to measure mass loss rates (of the gas) using radio observations of emission lines of CO and apply the stationary wind approach of \citet{Mo:80} and \citet{KnMo:85}, see, for example, \citet{OlErGuCa:93}, \citet{KnYoLeJo:98}, \citet[who derive their values using an improved RT model]{ScOl:01}, \citet{ScRyOl:02}, \citet{GrSeSpPe:02,GrSeSpPe:02b}, \citet{RaOl:14}, \citet{DaTeJu.:15}, and \citet{RaVlDo.:20}.

A more common approach, currently, is to measure spectra in the infrared wavelength range for both C-rich and O-rich stars at a known distance [typically the Magellanic Clouds (MC)], fit a spectral energy distribution (SED) to the spectrum and use the dust optical depth to extract mass loss rates of the gas \citep[e.g.,][]{vLoGrKo.:99,GrWoSl.:07,GrSlSoPe:09} or dust \citep[e.g.,][]{Ju:86,ZiLoWa.:96,SrMeLe:09,SaSrMe.:10,SaSrMe:11,SrSaMe:11,JoKeSa.:12,JoKeSr.:14,RiSrSaMe:12}. The chosen approach is to -- amongst other assumptions -- use a fixed terminal velocity of the dust $\langle\vinf\rangle=10\,\kms$ and fix the dust-to-gas-ratio $\ddg=0.005$; the terminal drift velocity is also assumed to be zero, so $\muinf=\langle\vinf\rangle$. \citet{NaMaGi.:18}, and their more recent publications, instead fit a SED by solving a set of differential equations based on a stationary wind; these authors also ignore drift.

The parameter selection is important. The results of our drift models -- that we calculated using solar metallicity opacities -- indicate that it is very hard to reach both the highest mass loss rates and low terminal velocities (cf. Fig.~\ref{fig:discdriftfactorm}a and \ref{fig:discdriftfactoro}a). For example, $\mmdot=1.0\times10^{-5}\,\mdotu$ requires $\muinf=75\,\kms$, which is a high value that is not observed. When we instead use $\muinf=45\,\kms$, $\mvDinf=17\,\kms$ (Fig.~\ref{fig:discdriftfactoro}c), $\mmdot=3.9\times10^{-6}\,\mdotu$ (Fig.~\ref{fig:discdriftfactorm}a), and $\ddg=6.8\times10^{-3}$ (Fig.~\ref{fig:discdriftfactorm}c). Please note, however, that error bars are significant! Considering the conditions where $\muinf=10\,\kms$ -- as in the lower metallicity MC-specific parameters mentioned above -- the terminal drift velocity $\mvDinf=55\,\kms$, the mass loss ratio $\mmdot=4.9\times10^{-6}$ and the dust-to-gas ratio $\ddg=1.1\times10^{-4}$ (Fig.~\ref{fig:discdriftfactorm}c). These values are naturally not directly applicable to the MCs; instead one should use relations based on models that are calculated using MC-specific metallicities). Differences are large, the dust-to-gas ratio assumed by the authors is 50 times higher than this value; in good agreement with the range of values {\rBEM} present for the MCs. Drift velocities are, moreover, never negligible, and it appears plausible that they are even higher in low-metallicity environments.

The gas mass loss rate can be derived in the SED approach without knowledge of the drift velocity, assuming it is possible to estimate the dust density {\rhod} and dust-to-gas density ratio {\ddg}. However, to measure the dust mass loss rate {\mdmdot}, it is necessary to estimate the drift velocity {\mvdrinf}. Nearly all mass-loss estimates based on SED data ignore drift altogether. \citet{NaGrAr.:19} mention drift, but ignore it based on the results of the grey and stationary model of \citet{KrSe:97}, who include grain growth and drift and find that drift velocities are very small ($\mvDinf\la5\,\kms$). The systematically higher mass loss rates found by \citet{GrSeSpPe:02b} in their CO-based study imply lower drift velocities, $\mvDinf\simeq3\,\kms$; \citet{ScRyOl:02} find the same value. \citeauthor{GrSeSpPe:02b} also point out that their observations of gas-to-dust ratios are in good agreement with theory, but note that the cited theoretical studies are all based on grey gas opacities and as their own study, none considers effects of drift. There appear to be two exceptions to low drift velocities: based on stationary models, \citet{PaPe:86} and \citet{OlErGuCa:93} estimate higher drift velocities that are more similar to our model values.

Average drift velocities in stellar winds can be estimated with time-dependent models such as those that are calculated here. Alternatively, as a first approach, one can assume equilibrium between the radiative pressure and the drag force, as in equation~(\ref{eq:vdeq}). Physical assumptions regarding both the gas and the dust influence the result. The dust velocity is required for accurate estimates of yields of dust, which are often provided as dust-to-gas (equation~\ref{eq:dmddm}; or gas-to-dust) ratios.

\subsubsection{Model values of {\rMWH}, {\rENHAW}, and observations}\label{sec:obscomp}
\begin{figure*}
\includegraphics{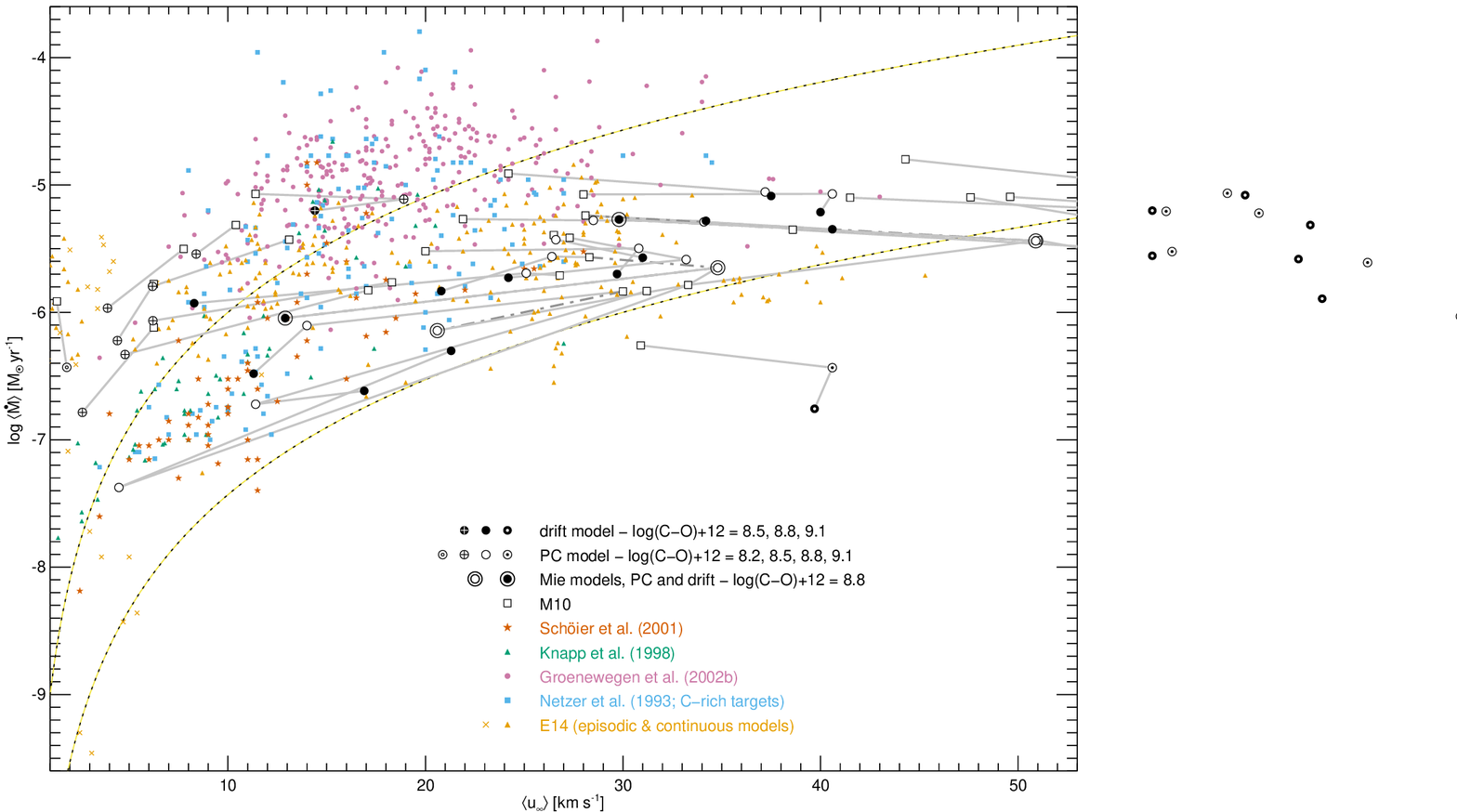}
\caption{Mass-loss rate versus the outflow velocity. The figure shows both observations of different sources as indicated and model values (orange and black symbols). Original model values of {\rMWH} are shown with a square $\square$ and values of our PC (drift) models are shown with circles $\circ$ (bullets $\bullet$). Drift, PC, and original model values using the same input parameters are connected with light grey solid lines. The yellow and black dashed lines delimit the drift-dominated region according to {\rIE}.\label{fig:discmdot}}
\end{figure*}

In their comparison of results of C-rich models using {\darwin}, {\rENHAW} plot observed values along with model values of the respective mass-loss rate and outflow velocity. We re-create this figure with Fig.~\ref{fig:discmdot}, and consequently include the C-star observations of \citet[who gather values of 18 references]{NeEl:93}, \citet{KnYoLeJo:98}, \citet{ScOl:01}, and \citet{GrSeSpPe:02b}. We also overplot our new values as well as the delimited region of lower optical depth drift-dominated mass loss according to {\rIE} (equation~10); we use their fig.~3 to set the delimiting values $10\le A\le30$ and they take the data from \citet{OlErGuCa:93}. Reddening dominates in the region with higher mass-loss rates (above the upper line), and here drift is of less importance. In this plot, we omit our models using $\CtO=9.1$, with one exception, as they achieve very high outflow velocities.

Observations of \citet{KnYoLeJo:98} and \citet{ScOl:01} mostly lie within the limits of drift-dominated models, whilst more values of \citet{GrSeSpPe:02b} lie in the region that is classified as reddening-dominated based on the data of \citet{OlErGuCa:93}; mass-loss rates of six objects that overlap in these two studies agreee to within a factor two when using the same object distance (M.~Groenewegen, priv.comm. 2020) and three (two) of these six objects are shifted out to the redshift-dominated (in to the drift-dominated) region when using the distances of the more recent study.
The mixed-origin values of \citet{NeEl:93} are spread out over most of the plot where there are values of observations.

A majority of the PC models of {\rENHAW} are found in the drift-dominated region, with the exception of models with very low outflow velocities, which are also in the reddening-dominated region. Except low-velocity ``episodic'' model values, remaining model values are found in the strip where $-6.5\la\log_{10}\mmdot\la-5.0\,\mdotu$. Few values are found in the region $-7.5\la\log_{10}\mmdot\la-6.3\,\mdotu$, $3\la\muinf\la14\,\kms$, which combines lower outflow velocities and mass loss rates.

Our new PC model values, which use more gridpoints, more frequency points in the RT, and improved numerical features, differ from the original values of {\rMWH} throughout the plot. Differences are the largest in the drift dominated region, where outflow velocities change by $-87$ to $+81$ per cent and mass loss rates by $-97$ to $+5.3$ per cent. Few of our new values change so much that they enter the region of values of \citet{KnYoLeJo:98} and \citet{ScOl:01}. All our models that use $\CtO\le8.5$ are found in the reddening-dominated region at velocities higher than in the drift-dominated region.

Values of our drift models are both higher and lower compared to the respective PC model; changes in the outflow velocities and mass loss rates are with one exception $-67$ to $+48$ and $-58$ to $+38$ per cent, respectively. The changes are the largest in model L4.00T32E88, $+380$ and $+1100$ per cent, respectively. The correspoinding changes relative to the PC model values of {\rMWH} are $-69$ to $+56$ per cent and $-83$ to $-3.2$ per cent. The PC models with the lowest terminal velocities ($\muinf\la8\,\kms$) and where $\log(C-O)+12=8.2$ do not form a wind. The drift models L3.70T26E88 and L4.00T24E85 lie in the reddening-dominated region. Model setups L3.85T30E88 and L4.00T32E88 are two cases near the lower limit of the drift dominated region where both the terminal velocity and the mass-loss rate of the drift models are higher than in the corresponding PC model. Most of the observations of \citet{GrSeSpPe:02b} show higher values that are not covered in our parameter space.

There is a deficiency of drift models with a lower outflow velocity. Except the three models in the reddening dominated region and model L3.70T28E88, there are no (SPL) drift models with $\mvD<17\,\kms$. Notably, when model L3.85T28E88 uses Mie scattering instead of SPL, the outflow velocity decreases by 39 per cent to $\muinf=12.7\kms$, and the mass loss rate decreases by 46 per cent (also see Section~\ref{sec:Mie}). It appears that a study is valuable where all models use Mie scattering to get model values that agree better with observations.

We have found that it is difficult to model high mass loss rate values in the optically thick region. Meanwhile, effects of drift dominate in the optically thin region, as predicted by \rEI, and drift will be an important component when finding model setups that match the observations in this region.

\subsection{Comparison with the theory of \rEI}\label{sec:discEI}
Instead of solving the formidable wind formation problem described here, {\rEI} present a solution that is much less numerically demanding and yet yields general properties of dust driven stationary winds and grain drift. A comparison with our results is justified.

The main idea of the work of {\rEI} is that properties of the wind at large distances are expressed using attributes in the wind formation region. The assumptions are the following. All dust forms promptly at the grain condensation radius {\rcond} and grains neither grow nor are destroyed outside of this radius. Grains are described with their type, assuming a constant size, condensation temperature, and absorption and scattering efficiencies. The pressure gradient is ignored as winds of interest are assumed to be highly supersonic. The star is losing mass at specified mass-loss rate at the condensation radius $\rcond=2\Rs$. The authors present results in the form of radial density and velocity structures, structures plotted versus a drift-effect parameter that is the ratio of radiation pressure to drift effects ($P$), and reddening correction factors plotted versus the optical depth {\tauV} (at the visual wavelength 550nm; \rEI, equation~17).

\begin{figure}
\includegraphics{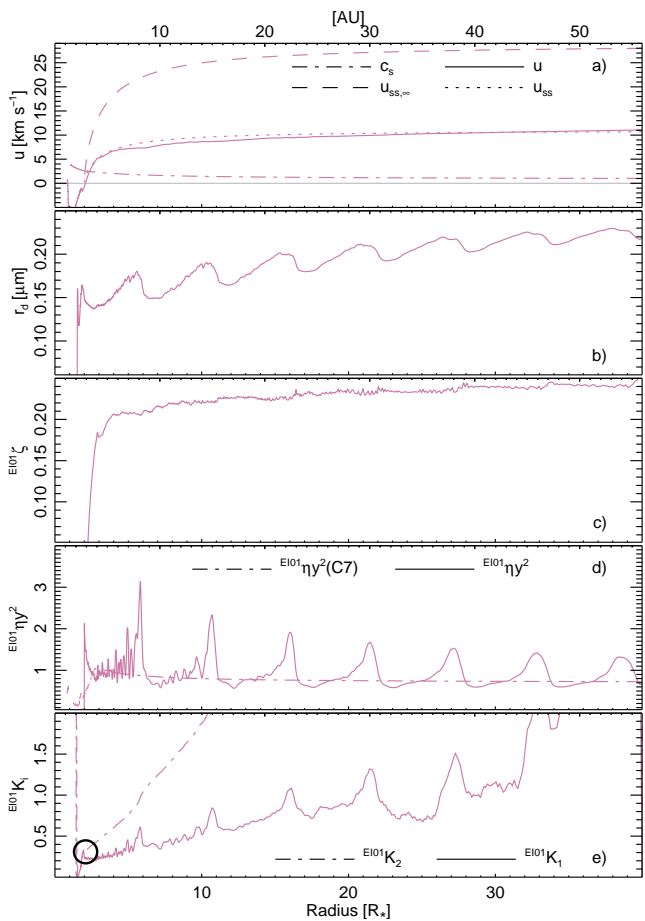}
\caption{Temporally averaged radial structure of a set of snapshots of the drift model L3.70T28E88 showing calculated properties of {\rEI}. From the top, the five panels show: (a) gas velocity $u$, sound speed \cs, self-similar gas velocities {\uss} and \ussi; (b) average grain radius {\radd}; (c) drift profile {\EIzeta}; (d) calculated (\rEI, equation~18) and parameterised (\rEI, equation~C7) dust density profiles $\EIeta y^2={\EIeta}(r/\rcond)^2$; and (e) reddening correction factors {\EIKone} and {\EIKtwo} that are used in the circled region at the condensation radius \rcond. See Fig.~\ref{fig:discwind} for more details.\label{fig:discEI}}
\end{figure}

The winds of our study with the lowest (L3.70T28E88) and highest (L4.00T24E91) dust mass loss rate yield optical depths in the range $0.65\la\tauV\la15$. For the discussion here, we show temporally averaged radial structures of the drift model L3.70T28E88 in Fig.~\ref{fig:discEI}. In comparison to Fig.~\ref{fig:discwind}, the properties show larger variations here owing to a shorter time interval ($5P$) and low numbers of models (54) used to calculate the averages. Moreover, we only consider values where $r\ge\rcond$.

The velocity structure (Fig.~\ref{fig:discEI}a) agrees well with the self-similar velocity structure {\uss} of {\rIE} (their equation~1 and our equation~\ref{eq:uss}) when we use the terminal gas velocity at the outer boundary as {\muinf}. The agreement is less optimal when we instead use equation~35 in {\rEI} to calculate {\ussi} and their expressions for {\muinf}\footnote{Using the equations and variable names of {\rEI}, we calculate ${\vinf}={w_{\infty}}v_{\text{m}}$ using equations~23, 29, 10, 4, and 9.}. The resulting velocity structure {\ussi} is about $2.5$ times higher than {\uss}. Furthermore, the drift profile $\EIzeta$ ($u/v=\FD^{-1}$, equation~12 and figure~2 in \rEI) shows the same trend as in {\rEI} (Fig.~\ref{fig:discEI}c). Our model value lies closer to the $\tauV=0.08$ line of \rEI. This property is always less than 1 when the model accounts for gas-to-dust drift.

The density profile {\EIeta} (see Fig.~\ref{fig:discEI}d; cf. equation~18 and figure~2 in \rEI) is very similar to what {\rEI} show. Admittedly, our line is shows rather large variations owing to the small number of models used in the shown average (see above). The trend of the parameterised density ($\EIeta$ calculated using equation~C7 in \rEI) is very similar to {\EIeta}. The agreement of the two profiles supports the statement that the density profile does not depend on drift (\rEI).

We also show the reddening correction factors {\EIKone} and {\EIKtwo} (equations~48 and 49 and the top two panels in figure~4 in \rEI) in Fig.~\ref{fig:discEI}e; these corrections are scalar properties that apply to the wind formation point {\rcond}. We show radial profiles of these structures to emphasize that as the wind formation point is not a fixed spatial coordinate in our models. According to its definition, the reddening correction $\EIKone(\tauV=0.7)\simeq1.1$ and $\EIKone\ge1\,\forall\,\tauV$. We find a somewhat lower value, $\EIKone\simeq0.3$. Finally, we find $\EIKtwo\simeq0.3$, which is also lower than the value according to its definition, $\EIKtwo(\tauV=0.7)\simeq0.95$ and $\EIKtwo<1\,\forall\tauV$. Both values are, however, in qualitative agreement with \rEI, and in particular $\EIKtwo$ indicates a stronger need for significant reddening corrections for the lower optical depth value of our model.

The authors show that the mass-loss rate depends on the optical depth as $\tauV^{3/4}$ (equation~55) owing to drift. Considering how influential we find that the inclusion of drift is in this study, it would be interesting to see how mass-loss rates derived in infrared observations are affected if they would use this relation instead of a usual relation that is linear with $\tauV$. Moreover, according to the theory of {\rEI} and accounting for effects of drift, the mass-loss rate is proportional to the outflow velocity as $\mmdot\propto\muinf^3$ and it is not related to the stellar luminosity. When drift is ignored, the outflow velocity is instead proportional to the luminosity as $\muinf^4\propto\Ls$. With a limited number of exceptions, the mass-loss rate of the PC models of {\rMWH} (figure~1), {\rENHAW} (figure~4), {\rBEM} (figure~6) show a non-existant to very weak dependence on the outflow velocity (all model values appear in a wide band), seemingly in agreement with the finding of {\rEI}. Our drift models follow the expected behavior in the sense that they, with two exceptions, lie inside the drift-dominated region indicated in Fig.~\ref{fig:discmdot}. Notably, we find that models calculated using Mie scattering show a trend of both lower outflow velocities and mass-loss rates. It seems plausible that all models calculated using such scattering lie inside this region. Our three PC models calculated using Mie scattering show increased drift velocities and appear to follow the relation $\muinf\propto\Ls$.

Our models are time-dependent and not stationary, and drifting dust initiates the stellar wind and dust grains form wherever conditions are suitable in a pulsating atmosphere. The two approaches show a qualitative agreement, but the physical differences result in quantitative differences as shown.

\section{Conclusions}\label{sec:conclusions}
We have extended our grey dust-driven wind models of Papers I--IV to include frequency-dependent RT in both the gas and dust components. We have also rewritten the RT solver and included our improvements to the numerical description of advection terms and the discretisation scheme. With our new model code \mbox{\teh}, we are able to model more realistic configurations of stellar winds of carbon-rich stars than was possible in our earlier papers; we can use the same physics as the other stellar wind model code that is currently used, {\darwin} \citep{HoBlArAh:16} -- with the added advantage that we can also model effects of gas-to-dust drift (two-fluid flow) and use high spatial resolution. To our knowledge, this is the first time a cool stellar wind is modelled at this high level of physical detail. And it appears that effects of drift become stronger with the level of detail.

Based on the set of models of {\rMWH}, we have calculated both PC (non-drift) and drift models to clarify differences owing to drift. As in {\rSa} and in contrast to the approach of {\darwin}, we have skipped the adaptive grid equation to resolve shocks in the gas. Instead of the commonly used 100 gridpoints, our models use 1024 gridpoints that are fixed in space where they resolve shock fronts in the gas and fronts in the dust simultaneously, at all times. Here, we have mostly calculated dust extinction rates using the SPL, but we used Mie scattering in a few cases for comparison. Benchmark results of {\teh} and {\darwin} using the same PM model setups shows reasonable agreement.

We have found periodic variations in a majority of the calculated structures of both PC and drift models, whilst earlier studies have reported irregular variations; we attribute this difference to the higher spatial resolution and improved numerical accuracy in our models as we also achieve irregular structures when we use the same numerical and physical setup as those studies. The results reveal intermediate to large changes of 50--1000 per cent in properties such as terminal velocities and mass-loss rates when we compare with our PC models in {\rMWH}. Outflow velocities, in particular, appear to increase greatly when PC models use Mie scattering instead of SPL; changes are greater than we find in {\rMH}.

In comparison to our earlier work on drift models, drift velocities are significantly higher than before -- we find mean terminal drift velocities in the range $10\la\mvdrinf\la64\,\kms$. Moreover, we have found that six out of seven model properties are correlated with the dust-to-gas velocity ratio, which we refer to as the drift factor, \FD; the exception is the mean grain radius. Five properties show an exponential dependence with the drift factor: the mass-loss rate ($\mmdot\propto\FD^{-2.4}$), dust mass-loss rate ($\mdmdot\propto\FD^{-4.9}$), dust-to-gas mass loss ratio ($\mmdmdot\propto\FD^{-2.6}$), terminal velocity ($\muinf\propto\FD^{-1.3}$), and degree of condensation ($\mfcond\propto\FD^{-1.3}$). The terminal drift velocity instead shows a linear increase. The strongest correlation is seen in the dust mass-loss rate. Our set of 20 drift models yields average values in the following set of ranges (Figs.~\ref{fig:discdriftfactorm} and \ref{fig:discdriftfactoro}):
\noindent\begin{tabular}{lll}\\[-1.8ex]
Drift factor           & \mFD & $1.2$--$4.2$\\
Mass-loss rate         & \mmdot & $0.03$--$1\times10^{-5}\,\mdotu$\\
Dust mass-loss rate    & \mdmdot & $0.0022$--$2.4\times10^{-7}\,\mdotu$\\
Density ratio          & \mddg & $0.015$--$2.6\times10^{-2}$\\
Mass loss ratio        & $\langle\ddg\FD\rangle$ & $0.063$--$2.6\times10^{-2}$\\
Terminal velocity       & \muinf & $10$--$70\,\kms$\\
Degree of condens.     & \mfcond & $0.11$--$0.72$.
\end{tabular}

Calculating dust yields, the dust-to-gas density ratio is multiplied with the drift factor. Consequently, our drift models yield 1.2--4.2 times as much dust as a corresponding PC model that ignores drift, and where the dust-to-gas density ratio is unchanged; this result corroborates our find in {\rSaHoc} and \citet{CSa:03} that dust formation is more efficient owing to drift. It is impossible to get correct yields when drift is ignored. Furthermore, our results show that ablation is unimportant, grain decay rates are too small at the large radii where drift velocities are high enough to activate ablation by non-thermal sputtering.

With two exceptions, our drift models lie in the interval of drift-dominated outflows as discussed by {\rEI} assuming $\mmdot\propto\muinf^3$ (Fig.~\ref{fig:discmdot}); this interval is in turn based on the CO-based observational data of \citet{OlErGuCa:93}. We could not calculate corresponding drift models for many PC models that lie outside of this interval. And consequently, most of the high-mass loss observations of, for example, \citet{GrSeSpPe:02b} also lie in a different parameter space than our models. Nearly all observational studies disregard drift and in view of our results therefore achieve too small yields of dust, in particular for lower mass loss rates where $\mdot\la1.0\times10^{-5}\,\mdotu$.

A comparison between the results of our time-dependent models and the simplified theory developed by {\rEI} shows a qualitative agreement, but the simplifications prevent a quantitative agreement. We agree with {\rEI} that drift is always present and an important factor of the wind when calculating models that result in realistic outflow velocities that match observations.

Our current dust-driven wind models do not reproduce the combination of higher mass loss rates -- $\mdot\ga1.0\times10^{-5}\,\mdotu$ -- at low expansion velocities. The models can, of course, be improved further with more physical detail. We could add a set of additional equations of motion of the dust to describe binned size-dependent drift velocities. Considering the already high complexity of our models, it is difficult to predict what the effects of such a treatment would be. Plausibly, particles of some sizes would move faster through the gas, which might increase ablation rates. Moreover, the current set of models need to be calculated using Mie scattering only, SPL is too inaccurate in the grain size interval we model. It would also be of interest to calculate models at lower metallicities to see how drift affects results in such circumstances (\citealt{MaArAn:15}; \rBEM). In a longer perspective, it would be highly valuable to extend {\teh} with oxygen-rich chemistry including silicates to include drift in models of M star winds.

\section*{Acknowledgements}
We thank {\v{Z}}. Ivezi\'c (U. Washington) and M. Elitzur (U. Kentucky) for a valuable discussion on their work on analytic dust-driven wind models. We thank K. Eriksson, S. Bladh, and S. H{\"o}fner (Uppsala Univ.) for providing information for the benchmark test between {\teh} and {\darwin}. We thank Y. Yasuda (Univ. Hokkaido) for stimulating discussions on the creation of stellar wind models. We also thank B. Villarroel for inspiring discussions on the scientific presentation. L. Mattsson acknowledges support by the Swedish Research Council (Vetenskapsr{\aa}det), grant no. 2015-04505. This work was in part supported by the Knut and Alice Wallenberg Foundation through the grant Dnr.\ KAW 2014.0048 on ``Bottlenecks for particle growth in turbulent aerosols.'' The simulations were performed on resources at Center for High Performance Computing (PDC), Chalmers Centre for Computational Science and Engineering (C3SE), and High Performance Computing Center North (HPC2N), which are all provided by SNIC. In particular, we thank Tor Kjellsson Lindblom at PDC for assistance concerning technical aspects in making {\teh} run on the PDC resources.

\section*{Data availability}
All model parameter and log files are available for download at Zenodo, where we also provide all data of models L3.70T28E88, L3.85T28E88, and L4.00T28E88 \citep{dSaMa:20}. Additionally, we provide a description of the binary file format and tools that read the same files. Remaining (binary) data files will be shared on reasonable request to the corresponding author.


\bibliographystyle{mnras}
\bibliography{AGB_Refs}


\appendix

\section{Glossary}
We collect all abbreviations used in this paper in Table~\ref{app:tababbreviations}, and all symbols in Tables~\ref{app:tabglossary} and \ref{app:tabcglossary}, where the second table contains model input parameters and properties only calculated at the outer boundary.

\begin{table}
\caption{Glossary of used abbreviations}
\label{app:tababbreviations}
\begin{tabular}{ll}\hline\hline\\[-1.8ex]
Term & Description\\\hline
AGB  & asymptotic giant branch\\
amC  & amorphous carbon\\
CO   & carbon monoxide\\
MC   & Magellanic Clouds\\
ODE  & ordinary differential equation\\
PAH  & polyaromatic hydrocarbons\\
PC   & position coupling (non-drift)\\
PDE  & partial differential equation\\
PPM  & piecewise parabolic method, advection scheme (\rSa)\\
RHD  & radiation hydrodynamics\\
RT   & radiative transfer\\
SED  & spectral energy distribution\\
SPL  & small particle limit\\\hline
\end{tabular}
\end{table}

\begin{table}
\caption{Glossary of used symbols}
\label{app:tabglossary}
\begin{tabular}{lll}\hline\hline\\[-1.8ex]
Symbol  & Unit & Description\\[1.0ex]
$A$ & & atomic weight; $A_{\text{C}}=12.01115$\\
$\alpha$ & & wind-formation efficiency property\\
\agr & $\text{cm}$ & grain radius\\
$B$  & $\text{erg}\,\text{cm}^{-2}\,\text{s}^{-1}$ & Planck function; $B=T^4\StefanB/\pi$\\
\Bnu & $\text{erg}\,\text{cm}^{-2}\,\text{s}^{-1}$ & Planck function;\\
     & $\,\,\text{Hz}^{-1}\,\text{ster}^{-1}$ & $\Bnu=2h\nu^3/c^2\left[\exp\left(h\nu/\left(\kB T\right)\right)-1\right]^{-1}$\\
$c$     & $\text{cm}\,\text{s}^{-1}$ & speed of light\\
$c_{\text{s}}$ & $\text{cm}\,\text{s}^{-1}$ & speed of sound\\
$C_{\text{D}}^{\text{LA}}$ & & limits approximation drag coefficient\\
\cost && Mite theory average scattering angle\\
$\delta_{\text{dg}}$ & & $\delta_{\text{dg}}=\rhod/\rhog$\\
$e$     & $\text{erg}\,\text{g}^{-1}$  & specific internal energy of the gas\\
$\varepsilon$ & & fraction of specular collisions\\
\EIeta  & & $\EIeta=n_{\text{d}}/\int_{1}^{\infty}n_{\text{d}}\text{d}y$, $y=r/\rcond$\\
$\zeta$ & $\text{erg}\,\text{K}^{-1}\,\text{g}^{-1}$ & $\zeta=128\kB/(9\pi\mu \mH)$\\
\EIzeta & & $\EIzeta=u/v=\FD^{-1}$\\
\FD     &                             & drift factor, $\FD=1+\vD/u$\\
\fcond  &                             & degree of condensation,\\
                                     && $\fcond\simeq\FD K_{3}/(\FD K_{3}+\nC)$\\
\fdrag  & $\text{g}\,\text{cm}^{-2}\,\text{s}^{-2}$ & drag force\\
\fedd   & & Eddington factor\\
\fgravd & $\text{g}\,\text{cm}^{-2}\,\text{s}^{-2}$ & dust gravitational term\\
\fin    & $\text{g}\,\text{cm}^{-2}\,\text{s}^{-2}$ & inertial term\\
\fradd  & $\text{g}\,\text{cm}^{-2}\,\text{s}^{-2}$ & dust radition pressure term\\
$G$     & $\text{dyn}\,\text{g}^{-2}\,\text{cm}^2$ & gravitational constant\\
$\Gamma$ & & radiative to grav. acceleration\\
$\gamma$ && ratio of specific heats\\
$h$ & $\text{erg}\,\text{s}$ & Planck constant\\
$H$ & $\text{erg}\,\text{cm}^{-2}\,\text{s}^{-1}$  & 1st moment of the radiation field\\
$H^{\text{int}}$ & $\text{erg}\,\text{cm}^{-2}\,\text{s}^{-1}$  & $H$ at the inner boundary\\
$J$ & $\text{erg}\,\text{cm}^{-2}\,\text{s}^{-1}$  & 0th moment of the radiation field\\
$J_{\star}$ & $\text{s}^{-1}\,\text{cm}^{-3}$ & net grain nucleation rate per volume\\
$K$ & $\text{erg}\,\text{cm}^{-2}\,\text{s}^{-1}$  & 2nd moment of the radiation field\\
{\dissc} & $\text{dyn}\,\text{cm}^{-2}$ & dissociation constants\\
$K_{j}$ & $\text{cm}^{-3}$           & moments of the grain size distribution;\\
        &                            & $0\le j\le3$\\
\EIKone & & reddening correction -- {\rEI}, equation~48\\
\EIKtwo & & reddening correction -- {\rEI}, equation~49\\
\kB & $\text{erg}\,\text{K}^{-1}$ & Boltzmann constant\\
$k_{\nu}$ &                          & refractive index, extinction coefficient\\
$k_{\text{d,X}}$ &  & extincion coefficient ratio of X\\
$k_{\text{g,X}}$ &  & gas opacity ratio of X\\
\kappanu & $\text{cm}^2\text{g}^{-1}\,\text{Hz}^{-1}$ & gas mass absorption coefficient\\
\kappanugd & $\text{cm}^2\text{g}^{-1}\,\text{Hz}^{-1}$ & total mass absorption coefficient\\
$\kappa_{\text{d,P}}$ & $\text{cm}^2\,\text{g}^{-1}$ & Planck mean dust absorption coefficient\\
\kappaJHS & $\text{cm}^2\,\text{g}^{-1}$ & frequency average of \kappanu, weighted\\
                                       && by $J$, $H$, and \Sourcef, respectively\\
$\kappa_{\text{P}}$ & $\text{cm}^2\,\text{g}^{-1}$ & Planck mean gas opacity\\
$\kappa_{\text{R}}$ & $\text{cm}^2\,\text{g}^{-1}$ & Rosseland mean gas opacity\\
\lav     & & artificial viscosity length scale\\
$m_1$   & g                           & monomer mass\\
\mdot  & \mdotu                    & mass loss rate\\
$m_{\text{bol}}$ & & apparent bolometric magnitude\\
$M_{\text{bol},\sun}$ & & absolute bolometric magnitude of the sun;\\
 &&$M_{\text{bol},\sun}=4.74$\\
\dmdot & \mdotu                    & dust mass loss rate\\
$\mH$ & g                             & mass of a hydrogen atom\\
$m_{\nu}$ &                            & complex refractive index\\
\massp & g                            & mass of a proton\\
$m_r$   & g                           & integrated mass at radius $r$\\
$\mu$   &                             & mean molecular weight\\
$\overline{\mu}$ & &                  $H/J$ at the outer boundary\\
$\mu_{\text{D}}$   &                   & distance modulus\\
$\langle N\rangle$ &                 & mean grain size, $\langle N\rangle=K_{3}/K_{0}$\\\hline
\end{tabular}
\end{table}

\begin{table}
\contcaption{Glossary of used symbols}
\begin{tabular}{lll}\hline\hline\\[-1.8ex]
Symbol  & Unit                        & Description\\[1.0ex]
\NA     &  $\text{mol}^{-1}$           & Avogadros constant\\
\ncore  &                             & number of rays inside the model\\
\nC     & $\text{cm}^{-3}$             & gas phase total number density of\\
        &                             & condensible material\\
\ngrid  &                             & number of gridpoints\\
\nnu    &                             & number of frequencies used in the RT\\
$n_{\nu}$ &                            & refractive index, phase velocity\\
$n_{\text{d}}$ & $\text{cm}^{-3}$   & dust number density, $n_{\text{d}}\equiv K_{0}$\\
$N_l$ &                           & lower size limit of macroscopic grains\\
$\nu$ & $\text{s}^{-1}$ & frequency\\
\Pg     & $\text{dyn}\,\text{cm}^{-2}$ & gas pressure\\
$q$     &                             & sphericality\\
$Q_{\text{abs},\nu}$ &                  & \chiJ, \kappaS: $\Qnuabs=\Qnuext-\Qnusca$,\\
                                     && \chiH: $\Qnuabs=\Qnuext-\cost_{\nu}\Qnusca$\\
                                     && \chiR: $\Qnuabs=\Qnuext$, as $\Qnusca=0$\\
\Qnuext &                             & extinction efficiency\\
\Qnusca &                             & scattering efficiency\\
\Qpnu & $\text{cm}^{-1}$             & absorption / extinction efficiency,\\
&& $\Qpnu=\Qnu/\agr$\\
$r$     & cm                          & radius\\
$r_0$   & cm                          & monomer radius\\
\Rs     & cm                          & stellar photosphere radius\\
\rcond  & cm                          & dust condensation radius\\
\mradd & $\text{cm}$ & mean grain radius, $\mradd=r_0K_{1}/K_{0}$\\
$R$     & $\text{erg}\,\text{K}^{-1}\,\text{mol}^{-1}$ & ideal gas constant\\
\rhod   & \dens  & dust density, $\rhod=m_1K_{3}$\\
\rhog   & \dens  & gas density\\
\rhoint & \dens  & grain intrinsic density\\
$\mathcal{S}$ & $\text{cm}^{-3}\,\text{s}^{-1}$ & amount of condensed material\\
\Sourcef & $\text{erg}\,\text{cm}^{-2}\,\text{s}^{-1}$  & gas source function\\
$\langle\sigma\rangle$ & $\text{cm}^2$ & mean grain cross section,\\
                                      && $\langle\sigma\rangle=\pi r_0^2K_{2}/K_{0}$\\
\surft & \ergcm & grain surface tension\\
{\StefanB} & $\text{erg}\,\text{cm}^{-2}\,\text{s}^{-1}\,\text{K}^{-4}$ & Stefan-Boltzmann constant\\
$t$     & s                           & time\\
\ted  & K                  & dust temperature, $\ted^4=\ter^4\chiJ/(\kappadS\rhog)$\\
\tedg & K                  & grey dust temperature, $\tedg=\ter$\\
\teg  & K                  & gas temperature\\
\ter  & K                  & radiation temperature, $\ter^4=J\pi/\StefanB$\\
$T_{\text{r,ext}}$ & K        & radiative temp. at the outer boundary\\
\taui  & $\text{s}^{-1}$             & net grain growth rate,\\
 &&$\taui=\tauGi-\taudc-\tausp$\\
\tauGi & $\text{s}^{-1}$             & total grain growth rate\\
\taudc & $\text{s}^{-1}$             & total grain decay rate; by evaporation\\
                                    && and chemical sputtering\\
\tausp & $\text{s}^{-1}$             & total non-thermal sputtering rate\\
\tauV  &                             & optical depth of the dust, at 550nm\\
$u$     & $\text{cm}\,\text{s}^{-1}$ & gas velocity\\
\uss & $\text{cm}\,\text{s}^{-1}$ & self-similar gas velocity using\\
 && equation~1 in {\rIE}\\
$v$     & $\text{cm}\,\text{s}^{-1}$ & dust velocity\\
\vD     & $\text{cm}\,\text{s}^{-1}$ & drift velocity, $\vD=v-u$\\
\vDeq   & $\text{cm}\,\text{s}^{-1}$ & equilibrium drift velocity (equation~\ref{eq:vdeq})\\
\vzeta  & $\text{cm}\,\text{s}^{-1}$ & thermal speed in $C_{\text{D}}^{\text{LA}}$\\
$\Xi$   &                           & sticking coefficient setup\\
$\xi$   &                           & atom and molecule-specific\\
        &                           & sticking coefficient \\
\chigrey & $\text{cm}^{-1}$          & grey extinction coeffient,\\
        &                           & $\chigrey=\pi r_0^3K_{3}\times4.4\tedg$\\
\chinu  & $\text{cm}^{-1}\,\text{Hz}^{-1}$ & dust extinction coefficient\\
\chiJHS & $\text{cm}^{-1}$ & frequency average of {\chinu}, weighted\\
                         && by $J$, $H$, and \Sourcef, respectively\\
\chiR & $\text{cm}^{-1}$ & Rosseland mean extinction coefficient\\[1.0ex]\hline
\end{tabular}
\end{table}

\begin{table}
\caption{Glossary of model parameters}
\label{app:tabcglossary}
\begin{tabular}{lll}\hline\hline\\[-1.8ex]
Symbol  & Unit & Description\\[1.0ex]
$\epsilon_{\text{C}}$ &                & elemental abundance of carbon\\
$\epsilon_{\text{O}}$ &                & elemental abundance of oxygen\\
\Ls & $\text{erg}\,\text{s}^{-1}$ & stellar luminosity\\
\Ms     & g                         & stellar mass\\
$P$     & s                           & stellar pulsations: piston period\\
\Rs     & cm                          & stellar radius\\
\rint   & cm                          & radial location of the inner boundary\\
\rext   & cm                          & radial location of the outer boundary\\
\teff   & K                           & effective temperature\\
\deltauvelp & $\text{km}\,\text{s}^{-1}$ & stellar pulsations: piston amplitude\\[1.5ex]
\multicolumn{3}{l}{\hspace*{0.3cm}Properties only calculated at the outer boundary}\\
$\alpha$ & $\text{erg}\,\text{s}^{-1}\,\text{g}^{-1}$ & $\alpha=\left(\dot{M}_{\text{d}}/\dot{M}\right)\Ls/\Ms=$\\
         & & $\phantom{\alpha}=\ddg\mathfrak{F}_{\text{D}}\Ls/\Ms$\\
\rfluc  &                             & relative fluctuation ampl., $\rfluc=\sistd/\mathcal{Q}$\\
$\langle\mathcal{Q}\rangle$ &         & temporal mean of the property $\mathcal{Q}$\\
\sistd  &                             & standard deviation\\
\uinf   & $\text{km}\,\text{s}^{-1}$  & terminal gas velocity\\
\vinf   & $\text{km}\,\text{s}^{-1}$  & terminal dust velocity\\[1.0ex]\hline
\end{tabular}
\end{table}


\section{Hydrostatic initial models}\label{app:initial}
The system of equations is -- assuming a dust-free hydrostatic equilibrium -- reduced to a set of four equations. The equation of specific inner energy of the gas (equation~\ref{eq:gene}) reduces to the relation $\kappaJ J=\kappaS S_{\text{g}}$, and therefore $J=\Sourcef\kappaS/\kappaJ$. The equation of integrated mass (equation~\ref{eq:integratedmass}), the equation of motion of the gas (equation~\ref{eq:gmot}), the equation of radiative energy (equation~\ref{eq:radenergy}), and the equation of radiative flux (equation~\ref{eq:radflux}) can be rearranged to the following set of four ordinary differential equations (ODEs),
\begin{eqnarray}
\frac{\text{d}m_{r}}{\text{d}r}&=&4\pi r^2\rhog\label{eq:ODEm}\\
\frac{\text{d}\Pg}{\text{d}r}&=&-\frac{Gm_r}{r^2}\rhog+\frac{4\pi}{c}\kappa_{\text{H}}\rhog H\label{eq:ODEP}\\
\frac{\text{d}H}{\text{d}r}&=&-\frac{2}{r}H\label{eq:ODEH}\\
\frac{\text{d}\teg}{\text{d}r}&=&-\frac{q\kappaH\rhog H}{\teg^3}\frac{\pi}{4\StefanB q\fedd}\frac{\kappaJ}{\kappaS} - \frac{\text{d}\left(q\fedd\right)}{\text{d}r}\frac{\teg}{4q\fedd},\label{eq:ODET}
\end{eqnarray}
where the four primary variables are: $m_r$, {\Pg }, $H$, and \teg.

The initial model code of {\teh}, {\johnc}, solves the four ODEs in equations~(\ref{eq:ODEm})--(\ref{eq:ODET}) together with four conditions that specify the stellar mass $M_{\star}$ , the effective temperature \teff, the luminosity {\Ls}, and the oxygen and carbon abundances $\epsilon_{\text{O}}$ and $\epsilon_{\text{C}}$ [$C/O=\CtO=\log\left(10^{\epsilon_{\text{C}}}-10^{\epsilon_{\text{O}}}\right)$], as well as the mean molecular weight $\mu$. Additional input is required in form of the radial extent of the model domain ($\left[r_{\text{int}},r_{\text{ext}}\right]$), frequency-dependent opacities, required accuracy of the calculations, etc.

{\johnc} calculates an initial model in the form of a hydrostatic stellar atmosphere of an AGB star in four steps:
\begin{enumerate}
\item Calculate a model using grey RT. The calculations begin at the photosphere radius {\Rs} with preset values of the mass, effective temperature, and luminosity. The photosphere radius is taken from the relation
\begin{equation}
\StefanB\teff^4=\frac{\Ls}{4\pi\Rs^2}.\label{eq:rstellar}
\end{equation}
Using an initial guess of the pressure, the equations are integrated outwards to the outer boundary. This integration is iterated by adjusting the photosphere pressure until the temperature at the outer boundary is lower than the temperature {\teg} in the equation
\begin{equation}
\frac{\StefanB}{\pi}\teg^4=\frac{H_{\text{ext}}}{\bar{\mu}}=\frac{\Ls}{\left(4\pi r_{\text{ext}}\right)^2}\frac{1}{\bar{\mu}},
\end{equation}
where a value at the outer boundary is denoted with the subscript ``ext'', and the angular intensity distribution of the radiation field $\bar{\mu}=\left(H/J\right)_{\text{ext}}$ is taken from the RT calculations; initially, $\bar{\mu}=0.5$. Thereafter, the ODEs are integrated inwards from the photosphere to the inner boundary, without solving the RT equation whilst assuming $\fedd=1/3$; this step provides initial values for all four primary variables at the inner boundary. So far, this procedure is very similar to the description given in \citet[chapter~4.10]{Do:98}.

Finally, the ODEs are integrated from the inner boundary to the outer boundary whilst calculating RT. The pressure at the inner boundary is again iterated to achieve the estimated temperature at the outer boundary. This step makes use of an ODE solver, such as the implicit predictor-corrector ODE solver \textsc{slga} \citep{RaScGl:79}, which handles stiff equations.
\item Calculate a model using frequency-dependent RT based on the primary variables at the inner boundary of the first step as a starting point. The ODE integration is anew iterated to achieve the estimated temperature at the outer boundary, which will be different compared to the grey calculations.
\item Calculate temperature corrections to achieve radiative equilibrium using the inner boundary values of the second step as starting point. The temperature corrections are calculated for the full radial extent using the Uns\"old-Lucy procedure and the description in {\rHM} (chapter 17.3). Consequently, three ODE:s are solved in this step, equations~(\ref{eq:ODEm})--(\ref{eq:ODEH}), making use of the ODE solver \textsc{dlsode}\footnote{The ``Livermore Solver for ODEs'', \textsc{dlsode}, is available at the web page \href{https://computation.llnl.gov/casc/odepack/}{https://computation.llnl.gov/casc/odepack/}.}. Temperature corrections are applied iteratively until the maximum temperature and luminosity corrections are both below a preset threshold. The pressure at the inner boundary is anew iterated until the temperature at the photosphere radius is close enough to the specified value \teff.
\item The resulting physical structure is relaxed on the adaptive grid, also when all grid weights are set to zero. Regardless of how many gridpoints the ODE solver has used, the solution is interpolated to use all or a part of the {\ngrid} required gridpoints, the latter option is the case when using a fixed grid with a larger extent than the initially modeled hydrostatic region. The gridpoint concentration in the centremost region is doubled relative to the remaining domain according to the description in Section~\ref{sec:grid}. The physical structure is stored in a binary file for use with {\teh}.
\end{enumerate}

\section{The role of sticking coefficients}\label{app:stick}
We wanted to compare our new results with those of {\rMWH} who use higher sticking coefficients. Here, we also calculated a set of PC models using the adaptive grid equation and $\ngrid=100$; these models use the RT of \citet{Yo:80} and 64 frequencies, as well as the same second order van Leer advection.

\begin{figure*}
\includegraphics{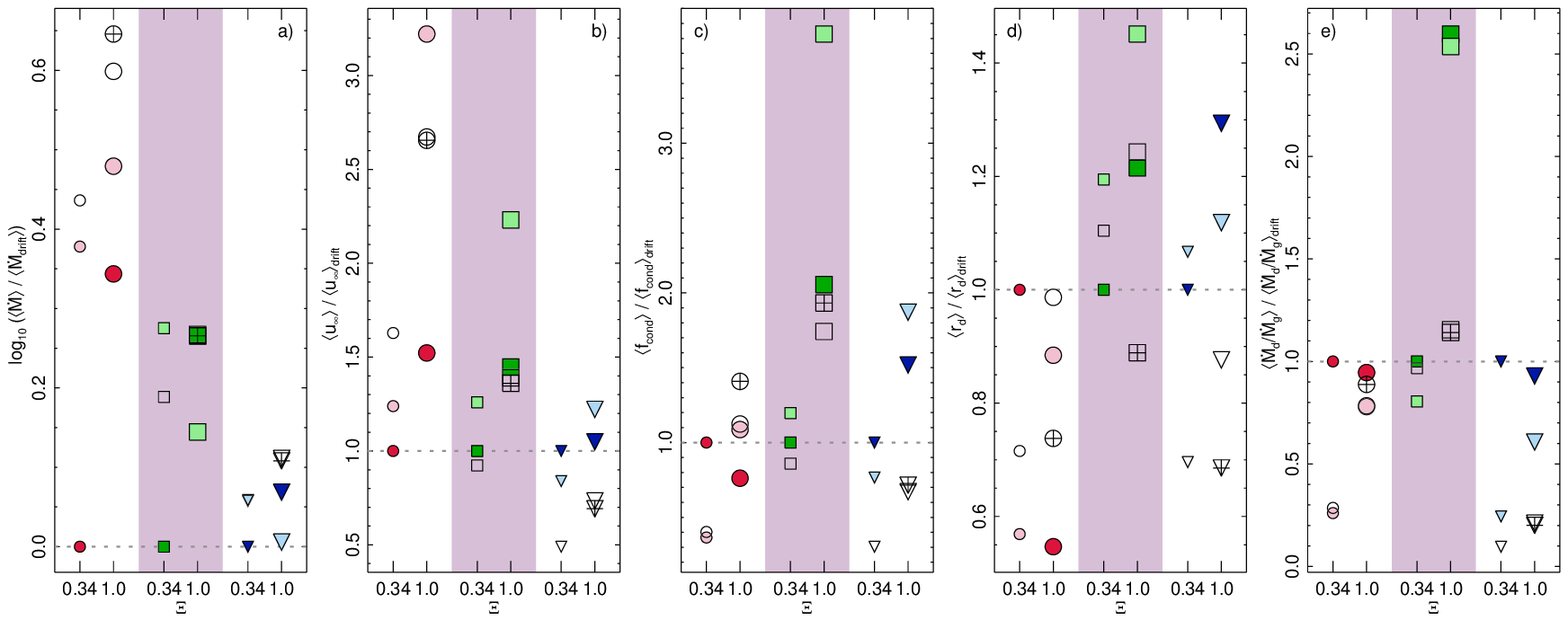}
\caption{Fractions of mean properties in Table~\ref{tab:scoeff} relative to the respective drift-model value. From the left, the five panels show ratios of the: (a) mass-loss rate \mmdot, (b) terminal velocity \muinf, (c) degree of condensation \mfcond, (d) mean grain radius \mradd, and (e) mass loss ratio $\mmdmdot$. Values of drift (PC) models where $\ngrid=1024$ are shown using symbols filled with a dark (light) colour, and values of PC models where $\ngrid=100$ are shown using open symbols. Reference values of {\rMWH} are shown with a plus symbol (+). Values of models using $\Xi_{1.0}$ ($\Xi_{0.34}$) are shown with larger (smaller) symbols. Values of model L3.70T28E88 (L3.85T28E88; L4.00T28E88) are shown to the left (centre; right) in each panel using red bullets $\bullet$ (green squares $\blacksquare$; blue triangles $\blacktriangledown$); the set of values in the centre of each panel are shown on a light purple backdrop to better separate the three sets of models. \label{fig:scoeff}}
\end{figure*}

We present our results of the three models that we calculated using the two sets of sticking coefficients $\Xi_{0.34}$ and $\Xi_{1.00}$ in Table~\ref{tab:scoeff}. We show the results in Fig.~\ref{fig:scoeff} where each property is plotted in relation to the respective drift-model value that uses $\Xi_{0.34}$. The table also contains the corresponding results of \citet[table~4]{AnHoGa:03} for their two models l10rdhou$\rho$185 ($\Ms,\,\Ls,\,\teff=1.0\Msun,\,10\times10^3\Lsun,\,2600\,$K) and l13drhou$\rho$185 ($\Ms,\,\Ls,\,\teff=1.0\Msun,\,13\times10^3\Lsun,\,2600\,$K); the authors do not specify a carbon-to-oxygen ratio for these two models.

Amongst other properties, \citet[section 4.3]{AnHoGa:03} study the role of sticking coefficients in frequency-dependent models of stellar winds. Their results show higher values when $\Xi_{1.00}$ is used compared to $\Xi_{0.34}$. The ratios of {\mmdot}, {\muinf}, and {\mfcond} between results of $\Xi_{1.00}$ and $\Xi_{0.34}$ are $3.0$, $3.3$, and $1.9$, respectively, for model l10drhou$\rho$185 ($1.4$, $2.3$, and $2.2$ for model l13drhou$\rho$185). The ratios are higher in model l10drhou$\rho$185.

We find smaller differences between PC models calculated using $\ngrid=100$ and the values of {\rMWH}. Considering all three models and $\Xi_{0.0}$, mass loss rates are 10 per cent lower to 0.69 per cent higher, terminal velocities 0.67--6.0 per cent higher, degree of condensations 6.1--20 per cent lower, mass loss ratios 12 per cent lower to 5.7 per cent higher, and mean grain radii 28--40 per cent higher. The agreement is very good, except that our grain radii are larger; this is likely owing to neglected differences in the dust parameters we used (Appendix~\ref{app:benchmark}). The corresponding values for $\Xi_{0.34}$ are 11--38 per cent lower (\mmdot), 29--39 per cent lower (\muinf), 56--71 per cent lower (\mfcond), 50--68 per cent lower (\mmdmdot), and 3.0 per cent lower to 24 per cent higher (\mradd). These values are lower than with $\Xi_{0.34}$. Our results agree with those of \citet{AnHoGa:03}.

Comparing our $\Xi_{0.34}$ PC models calculated using $\ngrid=1024$ to those using $\ngrid=100$, the values change as: {\mmdot} $-13$ to $+21$ per cent, {\muinf} $-24$ to $+71$ per cent, {\mfcond} $-9.2$ to $+50$ per cent, {\mmdmdot} $-8.8$ to $+150$ per cent, and {\mradd} $-20$ to $+53$ per cent. The same values of our $\Xi_{1.00}$ models are: {\mmdot} $-22$ to $-25$ per cent, {\muinf} $+21$ to $+66$ per cent, {\mfcond} $-3.3$ to $+180$ per cent, {\mmdmdot} $-0.4$ to $+190$ per cent, and {\mradd} $-10$ to $+28$ per cent. The differences are larger and values higher when $\ngrid=1024$ for the models that use $\Xi_{1.00}$, with the exception of mass loss rates that are all lower and the mean grain radius of model L4.00T28E88, which shows an increase using $\Xi_{0.34}$. Only considering the $\ngrid=1024$ models, the values are higher using $\Xi_{1.00}$, with two exceptions, the mass-loss rates are somewhat lower in models L3.85T28E88 and L4.00T28E88 that use $\Xi_{0.34}$. Relative fluctuation amplitudes are also somewhat smaller in the higher resolved models. We classify all three models that use $\Xi_{1.00}$ and $\ngrid=1024$ and two models that use $\Xi_{0.34}$ and $\ngrid=1024$ as periodic. We classify all six models calculated using $\ngrid=100$ as irregular.

Finally, we compare our drift models with the corresponding PC models where $\ngrid=1024$. The values of the $\Xi_{0.34}$ models change as: {\mmdot} $-(5.9$--$58)$ per cent, {\muinf} $-21$ to $+19$ per cent, {\mfcond} $-17$ to $+170$ per cent, {\mmdmdot} $23$--$310$ per cent, and {\mradd} $-17$ to $+76$ per cent. The same values of our $\Xi_{1.00}$ models are: {\mmdot} $-27$ to $+32$ per cent, {\muinf} $-14$ to $-53$ per cent, {\mfcond} $-19$ to $-45$ per cent, {\mmdmdot} $+2.4$ to $+53$ per cent, and {\mradd} $-38$ to $+16$ per cent. All but two drift models show a periodic structure, the exceptions are L3.85T28E88 and L4.00T28E88 using $\Xi_{1.0}$. The drift models that use $\Xi_{1.00}$ show larger values than the models that use $\Xi_{0.34}$, with three exceptions: the dust properties of model L3.70T28E88, the drift velocity of model L3.85T28E88, and the mass loss ratio of model L4.00T28E88.

Except that smaller amounts of dust forms in models that use $\Xi_{0.34}$, there is no simple relation between property values and sticking coefficient. In this paper, we use the set of sticking coefficients suggested by \citet{GaKeSe:84}, $\Xi_{0.34}$.

\begin{table*}
\caption{Properties temporally averaged at the outer boundary varying the sticking coefficients. From the left, the first four columns specify: the name; the model type, PC (P) or drift (D); sticking-coefficient setup ($\Xi$); and the number of gridpoints {\ngrid}. Remaining columns are the same as in Table~\ref{tab:resall}. All models have been calculated with the outer boundary fixed at $\rextf\!=\!40\,\Rs$ ($\ngrid=1024$) and $\rextf\!=\!25\,\Rs$ ($\ngrid=100$). Rows of drift models are shown in boldface. The four last lines show values of \citet[table 4]{AnHoGa:03}.\label{tab:scoeff}}
\begin{tabular}{$l@{\ \ }^r^r@{\ \ }^r^r@{\,}%
    ^l@{\,}^lc%
    ^l@{\,}^lc%
    ^l@{\,}^lc%
    ^l@{\,}^lc%
    ^l@{\,}^lc%
    ^l@{\,}^l^r}\hline\hline\\[-1.8ex]
   \multicolumn{1}{l}{model} & $\displaystyle\frac{\text{P}}{\text{D}}$ & \multicolumn{1}{c}{$\Xi$}& \multicolumn{1}{c}{$\ngrid$}&&
          \multicolumn{2}{c}{$10^7$\,\mmdot} &&
          \multicolumn{2}{c}{\muinf}  &&
          \multicolumn{2}{c}{\mfcond} &&
          \multicolumn{2}{c}{\mdrhogf} &&
          \multicolumn{2}{c}{$10^2$\,\mradd} &&
          \multicolumn{2}{c}{\mvdrinf}  & class\\
   \multicolumn{1}{r}{} &  &&&&\multicolumn{2}{c}{$[\mdotu]$} &&
          \multicolumn{2}{c}{$[\kms]$}    &&
          \multicolumn{2}{c}{}      &&
          \multicolumn{2}{c}{$[10^{-4}]$} &&
          \multicolumn{2}{c}{$[\mu\text{m}]$}  &&
          \multicolumn{2}{c}{$[\kms]$}\\
     &&&&&&\multicolumn{1}{c}{$\rfluc$} &&
          &\multicolumn{1}{c}{$\rfluc$} &&
          &\multicolumn{1}{c}{$\rfluc$} &&
          &\multicolumn{1}{c}{$\rfluc$} &&
          &\multicolumn{1}{c}{$\rfluc$} &&
          &\multicolumn{1}{c}{$\rfluc$}\\[1.0ex]\hline\\[-1.0ex]
L3.70T28E88 & P & 0.34 &  100 && $\phantom{0}$9.01 & \textit{$\phantom{00}$4.4} && 18.4 & \textit{$\phantom{0}$1.3} && 0.130 & \textit{1.0\tsc} && $\phantom{00}$7.91 & \textit{$\phantom{0}$62} && 16.1 & \textit{$\phantom{0}$1.1} && && i\\[0.15ex]
L3.70T28E88 & P & 1.00 &  100 && 13.1 & \textit{$\phantom{0}$16} && 30.2 & \textit{$\phantom{0}$4.1} && 0.363 & \textit{9.2\tsc} && $\phantom{0}$21.7 & \textit{$\phantom{00}$5.5} && 22.2 & \textit{$\phantom{0}$3.5} && && i\\[0.15ex]
\textit{\rMWH}&P& 1.00 &  100 && 14.6 &               && 30.0 &               && 0.455 &                 && $\phantom{0}$24.6 &               && 16.6 &      && & &  \\[1.0ex]
L3.70T28E88 & P & 0.34 & 1024 && $\phantom{0}$7.88 & \textit{$\phantom{00}$0.47} && 14.0 & \textit{$\phantom{0}$5.2\tsc} && 0.118 & \textit{1.2\tsm} && $\phantom{00}$7.21 & \textit{$\phantom{00}$7.8\tsc} && 12.8 & \textit{$\phantom{0}$4.0\tsc} && && 1p\\[0.15ex]
L3.70T28E88 & P & 1.00 & 1024 && $\phantom{0}$9.95 & \textit{$\phantom{00}$5.4}  && 36.4 & \textit{$\phantom{0}$0.93}  && 0.351 & \textit{3.1\tsc}    && $\phantom{0}$21.6 & \textit{$\phantom{00}$1.9}  && 19.9 & \textit{$\phantom{0}$1.0}  && & & 1p\\[1.0ex]
\rowstyle{\bfseries} L3.70T28E88 & D & 0.34 & 1024 && $\phantom{0}$3.30 & \textit{$\phantom{00}$0.13} && 11.3 & \textit{$\phantom{0}$0.20} && 0.323 & \textit{0.23} && $\phantom{0}$27.7 & \textit{$\phantom{0}$30} && 22.5 & \textit{$\phantom{0}$2.9} && 33.5 & \textit{$\phantom{0}$2.9} & 1p\\[0.15ex]
\rowstyle{\bfseries} L3.70T28E88 & D & 1.00 & 1024 && $\phantom{0}$7.28 & \textit{$\phantom{00}$0.33} && 17.2 & \textit{$\phantom{0}$3.4\tsc} && 0.246 & \textit{0.29} && $\phantom{0}$26.2 & \textit{$\phantom{0}$36} && 12.3 & \textit{$\phantom{0}$7.4} && 24.6 & \textit{$\phantom{0}$7.5} & 1p\\[1.0ex]\cline{1-5}\\
L3.85T28E88 & P & 0.34 &  100 && 22.7 & \textit{$\phantom{0}$25} && 19.2 & \textit{$\phantom{0}$4.3} && 0.140 & \textit{5.0\tsc} && $\phantom{00}$8.55 & \textit{$\phantom{00}$3.0} && 15.9 & \textit{$\phantom{0}$3.7} && && i\\[0.15ex]
L3.85T28E88 & P & 1.00 &  100 && 27.2 & \textit{$\phantom{0}$53} && 28.9 & \textit{$\phantom{0}$3.6} && 0.284 & \textit{8.2\tsc} && $\phantom{0}$17.2 & \textit{$\phantom{00}$5.0} && 17.9 & \textit{$\phantom{0}$2.9} && && i\\[0.15ex]
\textit{\rMWH}&P& 1.00 &  100 && 27.1 &               && 28.3 &               && 0.315 &                 && $\phantom{0}$17.0 &              && 12.8 &               && & &  \\[1.0ex]
L3.85T28E88 & P & 0.34 & 1024 && 27.4 & \textit{$\phantom{00}$8.5} && 26.4 & \textit{$\phantom{0}$0.40} && 0.197 & \textit{3.4\tsc} && $\phantom{0}$12.1 & \textit{$\phantom{00}$2.1} && 17.3 & \textit{$\phantom{0}$2.0} && && 1p\\[0.15ex]
L3.85T28E88 & P & 1.00 & 1024 && 20.5 & \textit{$\phantom{0}$23} && 46.4 & \textit{$\phantom{0}$1.7} && 0.608 & \textit{5.7\tsc} && $\phantom{0}$37.8 & \textit{$\phantom{00}$3.0} && 20.9 & \textit{$\phantom{0}$1.5} && && 1p\\[1.0ex]
\rowstyle{\bfseries}L3.85T28E88 & D & 0.34 & 1024 && 14.7 & \textit{$\phantom{00}$2.7} && 20.8 & \textit{$\phantom{0}$0.18} && 0.163 & \textit{0.22} && $\phantom{0}$14.9 & \textit{$\phantom{0}$23} && 14.4 & \textit{$\phantom{0}$6.0} && 21.9 & \textit{$\phantom{0}$6.3} & 1p\\[0.15ex]
\rowstyle{\bfseries}L3.85T28E88 & D & 1.00 & 1024 && 27.1 & \textit{$\phantom{0}$18} && 30.1 & \textit{$\phantom{0}$0.91} && 0.335 & \textit{0.29} && $\phantom{0}$38.7 & \textit{$\phantom{0}$62} && 17.5 & \textit{$\phantom{0}$7.2} && 18.5 & \textit{12} & i\\[1.0ex]\cline{1-5}\\
L4.00T28E88 & P & 0.34 &  100 && 51.4 & \textit{$\phantom{0}$79} && 19.9 & \textit{$\phantom{0}$2.3} && 0.125 & \textit{4.5\tsc} && $\phantom{00}$7.66 & \textit{$\phantom{00}$2.8} && 13.5 & \textit{$\phantom{0}$2.9} && && i\\[0.15ex]
L4.00T28E88 & P & 1.00 &  100 && 58.0 & \textit{120} && 29.8 & \textit{$\phantom{0}$3.8} && 0.276 & \textit{9.4\tsc} && $\phantom{0}$16.8 & \textit{$\phantom{00}$5.7} && 17.0 & \textit{$\phantom{0}$3.0} && && i\\[0.15ex]
\textit{\rMWH}&P& 1.00 &  100 && 57.6 &              && 28.1 &                && 0.294 &               && $\phantom{0}$15.9 &               && 13.3 &               && & &  \\[1.0ex]
L4.00T28E88 & P & 0.34 & 1024 && 47.7 & \textit{$\phantom{0}$34} && 34.1 & \textit{$\phantom{0}$1.2} && 0.315 & \textit{4.6\tsc}  && $\phantom{0}$19.4 & \textit{$\phantom{00}$2.8} && 20.7 & \textit{$\phantom{0}$2.0} && & & q\\[0.15ex]
L4.00T28E88 & P & 1.00 & 1024 && 45.5 & \textit{$\phantom{0}$54} && 49.6 & \textit{$\phantom{0}$2.5}   && 0.769 & \textit{9.6\tsc}   && $\phantom{0}$48.2 & \textit{$\phantom{00}$4.7}  && 21.7 & \textit{$\phantom{0}$0.86}  && & & 1p\\[1.0ex]
\rowstyle{\bfseries}L4.00T28E88 & D & 0.34 & 1024 && 44.9 & \textit{$\phantom{0}$45}  && 40.6 & \textit{$\phantom{0}$0.55}  && 0.411 & \textit{0.39}   && $\phantom{0}$79.4 & \textit{130}   && 19.4 & \textit{10}   && 22.8 & \textit{11} & 1p\\[0.15ex]
\rowstyle{\bfseries}L4.00T28E88 & D & 1.00 & 1024 && 52.6 & \textit{$\phantom{0}$61} && 42.6 & \textit{$\phantom{0}$1.6}  && 0.624 & \textit{0.30} && $\phantom{0}$73.8 & \textit{$\phantom{0}$87} && 25.1 & \textit{$\phantom{0}$8.4} && 26.5 & \textit{13} & i\\[1.0ex]\cline{1-5}\\
l13drhou$\rho$185 & P & 0.34 & 100 && 23$\phantom{.0}$ & && $\phantom{0}$3.6 & && 0.12$\phantom{.0}$\\[0.15ex]
l13drhou$\rho$185 & P & 1.00 & 100 && 70$\phantom{.0}$ & && 12$\phantom{.0}$ & && 0.23$\phantom{.0}$\\[1.0ex]
l13drhou$\rho$185 & P & 0.34 & 100 && 49$\phantom{.0}$ & && $\phantom{0}$7.4 & && 0.10$\phantom{.0}$\\[0.15ex]
l13drhou$\rho$185 & P & 1.00 & 100 && 70$\phantom{.0}$ & && 17$\phantom{.0}$ & && 0.22$\phantom{.0}$\\\hline\\[-1.0ex]
\end{tabular}
\end{table*}

\section{Physical and numerical setup}\label{app:assumptions}
We collect values and references of most physical and numerical parameters and assumptions in Table~\ref{app:tabassumptions}. Parameter values of the benchmark test models in Appendix~\ref{app:benchmark} are shown in column 3, separate from the parameter values used in the other parts of this paper in column 2.

\begin{table*}
\caption{Physical and numerical assumptions}
\label{app:tabassumptions}
\begin{tabular}{llll}\hline\hline\\[-1.8ex]
Parameter/Property  & New models & Benchmark test (Appendix~\ref{app:benchmark}) & Description\\[1.0ex]\hline
\textbf{Gas parameters:}\\
$\mu$ & 1.26 & 1.26 & Mean molecular weight.\\
$\gamma$ & $5/3$ & $5/3$ & Ratio of specific heats.\\[1ex]
\textbf{Dust parameters:}\\
Grain chemistry & H, H$_2$, C, C$_2$, C$_2$H, and C$_2$H$_2$ & H, H$_2$, C, C$_2$, C$_2$H, and C$_2$H$_2$ & Gas phase molecules considered in\\
&&& \quad equilib. chemistry of grain growth.\\
Type & amC & amC & Type of formed dust.\\
$\rhoint$ & $1.85\,\dens$ & $1.85\,\dens$ & Grain intrinsic density.\\
$\surft$ & $1400\,\ergcm$ & $1400\,\ergcm$ & Grain surface tension.\\
$\Xi$ & $\Xi_{0.34}=(0.37, 0.34, 0.34, 0.34)$, & $\Xi_{1.00}$ & Sticking coefficients:\\
      & $\Xi_{1.00}=(1.00, 1.00, 1.00, 1.00)$ & & \quad$\Xi=(\aC,\,\aCt,\,\aCtH,\,\aCtHt)$.\\
$\dissc\,[\text{dyn}\,\text{cm}^{-2}]$ & \citet{ShHu:90} & \citet{StPr:71} & Dissociation constants.\\
Dust velocity & mean grain-size {\vD} and $v=u$ & $v=u$ & Drag force calculated using one mean grain size.\\
$\varepsilon$ & $1.0$ & not applicable & Fraction of specular collisions between gas\\
&&& \quad particles and dust grains.\\
\multicolumn{4}{l}{Additional grain growth parameters are listed and discussed in, for example, {\rSaHoc}.}\\[1.0ex]
\textbf{Radiative transfer:}\\
Solver & Feautrier & \citet{Yo:80} & Radiative transfer solution approach.\\
$\nnu$, $\ncore$ & 319, 20 & 64, 5 & Number of frequencies and core rays.\\
$T_{\text{r,ext}}$ & $0\,\text{K}$ & $0\,\text{K}$ & Outer boundary radiative temperature.\\
$H_{\nu}^\text{int}$ & $\frac{\displaystyle\kappaR}{\displaystyle\kappanu}\frac{\displaystyle\partial B_\nu}{\displaystyle\partial\teg}\left(\frac{\displaystyle\partial B}{\displaystyle\partial\teg}\right)^{-1}H^{\text{int}}$ & $\frac{\displaystyle\Bnu}{\displaystyle B} H^\text{int}$ & Inner boundary rad. flux, see equation~(\ref{eq:Hint}).\\
$n_{\nu}$, $k_{\nu}$ ($m_{\nu}$) & \citet{RoMa:91} & \citet{RoMa:91} & Extinction data: refractive indices of dust grains.\\
&(104\,nm--300\,$\mu$m) & (104\,nm--300\,$\mu$m)\\
$Q_{\text{abs},\nu}$ & SPL, Mie & SPL & Approach to calculate the absorption efficiency.\\
$\kappanu\,[\text{cm}^2\text{g}^{-1}\,\text{Hz}^{-1}]$ & \citet{BAr:00,ArGiNo.:09} & \citet{BAr:00,ArGiNo.:09} & Gas opacities.\\
&(253\,nm--25\,$\mu$m or 39480--400\,cm$^{-1}$) & (253\,nm--25\,$\mu$m  or 39480--400\,cm$^{-1}$)\\
\kappaR, \kappaS, (\nnu) & Calculated $\forall\nu$ at & pre-calculated & Rosseland and Planck mean gas opacities\\
                         & model start                &                & \quad use all frequencies.\\
\chiR, \kappadS (\nnu)   & Calculated $\forall\nu$ at & pre-calculated & Rosseland and Planck mean dust extinctions\\
                         & model start                &                & \quad use all frequencies.\\[1.0ex]
\multicolumn{4}{l}{\textbf{Model domain, pulsations, temporal interval:}}\\
Model domain   & $\rint$ set as small as possible, & $\Pg=10^2\,\text{dyn}\,\text{cm}^{-2}\Rightarrow\rint$, & Definition of inner boundary location and\\
               & $\rhog(\rext)\approx10^{-6}\rhog(\rint)$ & $\rhog=10^{-16}\,\text{g}\,\text{cm}^{-3}\Rightarrow\rext$ & outer boundary locations.\\
\multicolumn{4}{l}{Simulated using a piston; the inner boundary moves in a radial sinusoidal motion of amplitude {\deltauvelp} and period $P$.}\\
Piston initialization & $2\,P$ & $15\,P$ & Starting at zero, the amplitude reaches full\\
&&&amplitude in this many pulsation periods.\\
Temporal interval & $11$--$200\,P$, or as short as possible to & $1000\,P$ or until 20 per cent of model & Models are calculated for this temporal interval.\\
& see periodic variations & domain mass remains\\[1.0ex]
\textbf{Numerical method:}\\
\multicolumn{3}{l}{implicit Henyey, staggered mesh, one dimensional} & Numerical approach.\\
grid                        & no property is resolved, & resolve {\rhog} and $e$, using the & Adaptive grid equation setup.\\
                            & fixed grid for $r>2\Rs$; & grid weight 1.0.\\
& twice the gridpoint density\\
&in the innermost 148 gridpoints\\
$\ngrid$ & 1024 & 100 & Number of gridpoints.\\
Discretisation & volume-weighted & arithmetic mean & Discretisation approach of properties on the\\
               &  && \quad staggered mesh (section~2.2 in \rSa).\\
Advection      & volume-weighted van Leer & arithmetic mean van Leer & Advection scheme (section~2.3 in \rSa).\\
$\lav$         & $ 3.5\times10^{-3}r\,\text{cm}$ & $r$ cm & Artificial viscosity length scale.\\[1.0ex]\hline

\end{tabular}
\end{table*}

\section{Numerical issues in drift models}\label{app:vDtech}

\subsection{Troughs in the drift velocity}
Our first generation of drift models often show high-value spikes in the drift velocity at the front of dust `shocks', see for example fig.~2 in {\rSaHoa}. Such spikes occur when the relative speed between a moving gridpoint (of the adaptive grid equation) and the dust is very close to zero. When this happens, the numerical diffusion may become too small to prevent spurious results to appear (see \rSaHoa, section~3.2). The problem is mitigated in the Planck-mean models presented in {\rSaHob}--{\rSa}, where both densities and drift velocities are lower than in the constant-opacity models of \rSaHoa.

Densities and drift velocities of our new frequency-dependent RT models presented here are higher than in the Planck-mean models. And if gridpoints are allowed to move about using the adaptive grid equation, conditions are anew favourable for the appearance of spurious spikes in the drift velocity. We decided to attempt to avoid these spurious spikes by keeping the grid fixed, which ought to work for as long as the dust velocity satisfies $v\ne0\,\kms$. As a compromise, we kept all gridpoints fixed where dust is present, $r>2\Rs$, while gridpoints at smaller radii were still allowed to stretch with the movements of the piston at the inner boundary.

\begin{figure}
\includegraphics{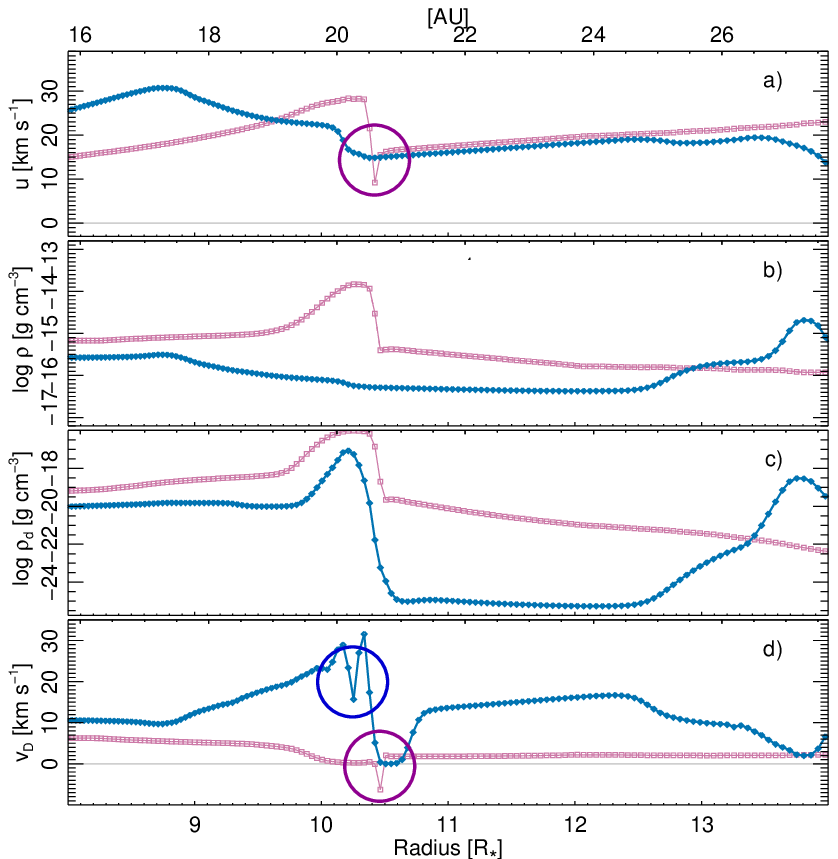}
\caption{Radial structures of snapshots of drift model setup L4.00T28E88 for a part of the radial region that shows a resolved shock and dust front; the purple line shows the model using non-compatible volume weighted van Leer advection with both the gas and dust velocities. Each square indicates the position of a gridpoint. The thick blue line (and diamonds) shows the model using non-compatible volume weighted van Leer advection using the integrated gas mass and compatible van Leer advection using the dust velocity. From the top, the four panels show: (a) gas velocity $u$, (b) gas density {\rhog}, (c) dust density {\rhod}, and (d) drift velocity {\vD}. All properties are drawn versus the stellar radius \Rs (lower axis) and astronomical units (AU, upper axis). Grey horizontal lines are guides. Circles indicate features discussed in the text.\label{fig:compat}}
\end{figure}

Our new drift models are free of spikes in the dust component owing to vanishing numerical diffusion. However, most models show `troughs' of lower -- and even negative -- drift velocities where there are strong negative radial gradients of the dust density (in dust fronts), for example see $r\simeq15,\,21,\,27,\,33,\,\text{and}\,38\Rs$ in Figs.~\ref{fig:discwind}b and \ref{fig:discwind}c, and Fig.~\ref{fig:compat}d (pink circle). The advection in the equation of motion of the dust (equation~\ref{eq:dmot}) is a term that is susceptible for the origin of these features.

\citet{WiNo:86} present an RHD model that is similar to ours in that it includes both an adaptive-grid equation and a staggered mesh. The authors advocate numerical accuracy as a reason to use the integrated mass instead of the velocity and density when advecting momentum (see their section~V.B); in particular the authors discuss an example where the ratio of the amount of mass in a gridpoint and the amount of mass that passes through the same gridpoint in a time step is near the machine precision. In fact, several of our models show similarly appearing `troughs' also in the gas velocity when we use the gas velocity and density in the advection terms instead of the integrated gas mass as is our standard procedure (an example is shown for model L4.00T28E88 in Fig.~\ref{fig:compat}a).\footnote{In comparison, the (volume-weighted van Leer-type) advection is found to work correctly using the velocity and density instead of the integrated mass in the grey RHD models presented by \citet{DoPiSt.:06} and the PC and drift models of {\rSa}.} However, the ratio of mass in the cell to the mass passing through a gridpoint is about $10^{-2}$, and the argument of \citet{WiNo:86} seems unessential.

To see if it could solve the problem and remove the troughs, we attempted a reverse situation where we add an equation of integrated mass of the dust (analogous to equation~\ref{eq:integratedmass} for the gas) and then use this integrated dust mass in the advection term of the dust equation of motion. Our test results show that the troughs disappear in the inner parts of the extended atmosphere, but further out, the dust velocity flutters between low and high values at short time intervals. It appears that the numerical accuracy of the integrated dust mass is insufficient to use in the advection term. Nevertheless, that it works partially in the inner parts is a clue that advection of momentum is easier using one variable (the integrated mass) instead of two (the velocity and density).

\citet{VaKa:98} present an improved van-Leer-type (`compatible') advection scheme that unlike previous ('non-compatible') schemes is designed to preserve the monotone character of, for example, both the density and momentum; this is achieved by delimiting the density and the velocity separately instead of just the momentum. Results of our tests using this advection scheme with the dust equation of motion and regular van Leer-type advection with the dust moment equations show that this approach is a promising solution as the troughs do not appear. A crucial point in this assessment is to observe that using a staggered mesh (as in {\teh}), the dust density is defined at gridpoint centres whilst the dust velocity and momentum are defined at gridpoint interfaces. Consequently, to allow the algorithm to identify non-monotonic density structures, it is necessary to replace the condition of a monotonic slope in the density at gridpoint interfaces ($\sign[\rhod[i]-\rhod[i+1]]=\sign[\rhod[i-1]-\rhod[i]]$, \citealt{VaKa:98}, equation~2.1.6) with the condition of a monotonic slope at gridpoint centres ($\sign[\rhod[i+1/2]-\rhod[i+3/2]]=\sign[\rhod[i-1/2]-\rhod[i+1/2]]=\sign[\rhod[i-3/2]-\rhod[i-1/2]]$). We show an example using compatible advection in Fig.~\ref{fig:compat} where a trough in front of the dust front with a negative drift velocity does not appear; meanwhile, another similar but non-negative trough occurs behind the dust front. Troughs such as these appear more frequently when first-order advection is used with more gridpoints when four gridpoints are used instead of three to identify non-monotone density slopes. It is consequently important to account for the additional gridpoint in the density slope ($i-3/2$) only in front of dust fronts.

Whilst this test study was made using regular van Leer-type advection, we find that it is necessary to use volume-weighted van Leer-type advection as in the other models of this paper. Regular van Leer-type advection is too inaccurate; using van Leer advection with the dust, at times and in some gridpoints, the dust velocity does not converge to an accurate value with the consequence that the time step decreases to a low value and the model evolution comes to a halt. To remove the occurrence of the troughs discussed here, we advocate the development of a both compatible and volume-weighted advection scheme. For now, we conclude that the impact of the velocity troughs on the model structure and evolution is minor, as they always appear in front of dust fronts where the dust density is some $10^2$--$10^6$ times lower than behind the front (see, e.g., Fig.~\ref{fig:compat}c). All terminal drift velocities are calculated by weighting values at individual times with the dust density.

\subsection{The role of numerics to properties of the model variability}\label{sec:numdiff}
A majority of both our PC and drift models presented here as well as in {\rSa} result in periodically varying structures. A vast majority of the models of {\darwin} meanwhile result in irregularly varying structures (S. H\"ofner, priv.comm. 2020). As we do in {\rSa}, we attribute the occurrence of periodic variations, as opposed to irregular variations, to our use of a high numerical accuracy in our models owing to the volume-weighted advection scheme and volume-weighted averaging on the staggared mesh. Additionally, we use a higher number of gridpoints ($\ngrid=1024$ instead of $\ngrid=100$) and the adaptive grid equation is kept fixed where $r>2\Rs$ and is not used to resolve shocks anywhere. We calculated a few additional PC setups of model L3.85T26E88 for comparison where we used the same approach as in {\rSa} and allowed all gridpoints to stretch with the movements of the piston at the inner boundary.

The model calculated using the PPM advection scheme and arithmetic averages instead of the volume-weighted van Leer scheme and volume-weighted averages does not become periodic in the first 77 periods; the models are in all other aspects identical. Likewise, the model where the adaptive grid is not fixed at $r=2\Rs$ also fails to develop periodic variations, as does a model that uses the regular van Leer advection scheme. Periodic variations put high demands on the numerical method, which is why we advocate using as many gridpoints as possible, volume-weighted averages, volume-weighted (van Leer) advection, and fixing the adaptive grid wherever possible to avoid the degradation of the solution when all regions are not equally resolved and gridpoints restructure.

\section{Benchmark test: {\darwin} vs. {\teh}}\label{app:benchmark}
We have compared the outcome of PC models of {\teh} and {\darwin} in a benchmark test where we, as far as we know, used the same physical and numerical setup as the models calculated using {\darwin}; the parameter-value setup is shown in Table~\ref{app:tabassumptions} on the right-hand side of the values used in the remaining parts of this study. We show the model parameters of this sample in Table~\ref{tab:resmodparb} together with the corresponding model values of {\rMWH} and {\rENHAW}; additionally, we used $M=1.0\Msun$ and $\deltauvelp=4\kms$. In these models, which use the adaptive grid equation to resolve shock fronts, the outer radius is at first allowed to move outwards from the location of the outer boundary of the hydrostatic inital model to $r=25\Rs$, before the wind is calculated. We evolved the models for a period of $200$--$1000P$. We followed the approach of {\rENHAW} and to the best of our abilities calculated temporally averaged values for the interval that begins when the initial transient has left the model domain and ends after 1000 periods or when 20 per cent of the mass in the envelope remains, and we use the same properties as with the other models in this study. We show the results of our comparison in Table~\ref{tab:resallb}.

\begin{table}
\caption{Model parameters of our benchmark test, cf. Table~\ref{tab:resmodpar}.}
\label{tab:resmodparb}
\begin{tabular}{lccccc}\hline\hline\\[-1.8ex]
model &$\log\left(\Ls\right)$       &{\teff}     & {\CtO} & \pP & $\displaystyle\frac{M_\text{e}}{M_\star}$\\
      &$[\Lsun]$&$[\mbox{K}]$& & $[\mbox{d}]$& [\%]\\[1.0ex]\hline\\[-1.8ex]
L3.70T26E85 & 3.70 & 2600 & 8.50 & 295 & 0.13\\
L3.70T26E88 & 3.70 & 2600 & 8.80 & 295 & 0.099\\
L3.85T26E85 & 3.85 & 2600 & 8.50 & 393 & 0.24\\
L3.85T26E88 & 3.85 & 2600 & 8.80 & 393 & 0.25\\
L3.70T28E88 & 3.70 & 2800 & 8.80 & 295 & 0.099\\
L3.85T28E85 & 3.85 & 2800 & 8.50 & 393 & 0.23\\
L3.85T28E88 & 3.85 & 2800 & 8.80 & 393 & 0.16\\[1.0ex]\hline
\end{tabular}
\end{table}

\begin{table*}
\caption{Temporally averaged quantities at the outer boundary; see Appendix~\ref{app:benchmark}. From the left, the first two columns specify the model name (see Table~\ref{tab:resmodparb}) and a reference for the values (when other than our own). Five columns show the averaged: mass loss rate {\mmdot}, terminal velocity {\muinf}, degree of condensation {\mfcond}, dust-to-gas ratio $\langle\ddg\rangle$, and dust radius {\mradd}. The final column shows our outflow classification class (irregular i) and the one of {\rENHAW}.}
\label{tab:resallb}
\begin{tabular}{$l^lc^lc^lc^lc^lc^l^l}\hline\hline\\[-1.8ex]
   \multicolumn{1}{l}{model name} & \multicolumn{1}{l}{ref.} &&
          $10^7$\,\mmdot && \muinf   && \mfcond && $\langle\ddg\rangle$ && $10^2$\mradd        & class\\
       &&& $[\mdotu]$    && $[\kms]$ &&         && $10^4$ && $[\mu\text{m}]$\\[1.0ex]\hline\\[-1.0ex]
L3.70T26E85 & {\rMWH}   && $\phantom{0}$7.60 && $\phantom{0}$6.24 && 0.211 && $\phantom{0}$5.72  &&          \\[0.15ex]
            & {\rENHAW} && $\phantom{0}$7.41 && $\phantom{0}$6.5  && 0.215 && $\phantom{0}$6.05  &&      & ws\\[0.15ex]
\rowstyle{\bfseries} &  && 11.8 && 12.2 && 0.210 && $\phantom{0}$6.26  && 23.7 &  i\\[0.15ex]
L3.70T26E88 & {\rMWH}   && 20.0 && 27.0 && 0.400 && 22.0 && 12.4     \\[0.15ex]
            & {\rENHAW} && 18.6 && 25.1 && 0.358 && 20.0 &&      & wn\\[0.15ex]
\rowstyle{\bfseries} &  && 27.3 && 22.5 && 0.384 && 23.0 && 12.3 &  i\\[0.15ex]
L3.85T26E85 & {\rMWH}   && 16.7 && $\phantom{0}$6.24 && 0.171 && $\phantom{0}$4.63 && 18.7     \\[0.15ex]
            & {\rENHAW} && 17.8 && $\phantom{0}$6.5  && 0.171 && $\phantom{0}$4.8  &&      & ws\\[0.15ex]
\rowstyle{\bfseries} &  && 17.2 && 14.0 && 0.196 && $\phantom{0}$5.88 && 22.3 &  i\\[0.15ex]
L3.85T26E88 & {\rMWH}   && 40.4 && 26.5 && 0.322 && 17.4 && 12.4 &   \\[0.15ex]
            & {\rENHAW} && 42.7 && 26.0 && 0.329 && 18.5 &&      & wn\\[0.15ex]
\rowstyle{\bfseries} &  && 49.3 && 30.3 && 0.385 && 22.5 && 11.8 &  i\\[0.15ex]
L3.70T28E88 & {\rMWH}   && 14.6 && 30.0 && 0.455 && 24.6 && 16.6     \\[0.15ex]
            & {\rENHAW} && 13.8 && 29.7 && 0.453 && 25.4 &&      & wn\\[0.15ex]
\rowstyle{\bfseries} &  && $\phantom{0}$9.81 && 27.5 && 0.326 && 19.9 && 12.2 &  i\\[0.15ex]
L3.85T28E85 & {\rMWH}   && 14.9 && 17.1 && 0.225 && $\phantom{0}$6.1  &&          \\[0.15ex]
            & {\rENHAW} && 14.5 && 15.7 && 0.211 && $\phantom{0}$5.93 &&      & wp\\[0.15ex]
\rowstyle{\bfseries} &  && 14.4 && 11.9 && 0.162 && $\phantom{0}$4.89 && 20.7 &  i\\[0.15ex]
L3.85T28E88 & {\rMWH}   && 27.1 && 28.3 && 0.315 && 17.0 && 12.8     \\[0.15ex]
            & {\rENHAW} && 26.9 && 26.7 && 0.371 && 20.8 &&      & wn\\[0.15ex]
\rowstyle{\bfseries} &  && 25.9 && 33.0 && 0.380 && 21.9 && 12.1 &  i\\[1.0ex]\hline
\end{tabular}
\end{table*}

Our model values mostly agree with the values of both {\rENHAW} and {\rMWH} within 30 per cent, and the agreement is typically better than that. Our terminal velocities for models L3.70T26E85 and L3.85T26E85 are 90 and 120 per cent higher than the other two studies, and our mass-loss rate of model L3.70T26E88 is about 60 per cent higher.

Whilst it has been our goal to use the same modelling approach, auxiliary data, and parameter input as {\darwin}, it is highly plausible that there still are differences that we have not accounted for. In absence of any published outcome of numerical tests of {\darwin}, we cannot asses to what extent numerical accuracy (of the numerical method -- not machine precision) plays a role when essentially identical PC models show significantly different mean flow properties. However, we note that output from {\darwin} sometimes shows considerable intermittency, which may be artificial and owing to numerical accuracy. We have noticed this phenomenon in {\rMWH}, but there we concluded that mean quantities were correct. But if the numerical accuracy is different, solutions of nonlinear PDEs can display different  pseudo-chaotic behaviour, in which case mean values may be altered. The relative numerical accuracy of individual iterations in {\teh} -- both when using the adaptive grid equation to resolve shocks and when it is deactivated -- is mostly $10^{-9}$ or better in 3 or 4 iterations, and the accuracy is mostly $10^{-12}$ or better after one more iteration. (The lowest accuracy is seen in the carbon number density, the dust moments and the dust velocity.) Next, we will attempt to asses all possible additional reasons that we could think of.

The initial model determines the amount of available mass in the model domain, mostly through the precise location of the inner boundary. The amount of mass in the model domain of these models is smaller than in our other models, compare the last column in Tables~\ref{tab:resmodparb} and \ref{tab:resmodpar}. Although we attempted to use the same criterion as {\rENHAW} when setting the boundary, the implication of a smaller amount of mass is that the model domain is emptied of mass faster, and differences are larger. The approach chosen by {\rENHAW} uses calculations that end after $1000\,P$ or when 20 per cent of the initial mass in the model domain remains. Under such circumstances, exact initial conditions are necessary for an accurate assessment in a comparison.

Starting model calculations, we initiate our dynamical models by turning on the piston from zero to full amplitude in $2\,P$; {\darwin} uses a longer initilialization period to handle the initial shockwave gracefully. Here, we have used $15\,P$ (to our experience). More mass can leave the model domain when the initialization takes longer. Additionally, in some cases, we find it difficult to estimate exactly when the transient of the starting model has left the model domain.

The discretisation of individual source and sink terms on a staggered mesh can be done in different ways. And the chosen approach may result in rather large differences in the outcome. To disentagle such differences one would have to make a minute study where the influence of each term is assessed and compared with {\darwin} individually. For example, for one of the benchmark models, L3.85T28E85, we got values that were off by 40--60 per cent from our other set of values when we changed the discretisation to subtract decay rates from growth rates before integrating the values over the volume of each grid cell (not shown in Table~\ref{tab:resallb}).

Other than that, differences owing to the nucleation rate appear to be small. Variations of results are much smaller with {\teh} when we do not resolve shocks using the adaptive grid equation. The larger differences in the outflow velocities plausibly originate in some other difference than listed here.

The agreement overall is decent to good, and we conclude that {\teh} is able to reproduce results of {\darwin}. Notably, {\teh} does not [yet] include a description of oxygen-rich chemistry, which is why we cannot reproduce such models.

\section{On complete momentum coupling}\label{app:cmc}
\citet{Gi:72} is the first to study the degree of momentum coupling between dust and gas. He finds that the momentum coupling is instantaneous and complete. Later, \citet{BeFr:83} show that the dust decouples from the gas at large radii ($r\ga1200\Rs$). \citet{McGSt:92} find that complete momentum coupling holds when grains are large ($\agr\ga0.1\,\mu$m) and that the gas and dust phases decouple with small grains ($\agr\la0.05\,\mu$m). Our models are dynamic instead of stationary, include both grain growth and ablation with dust grains that show different (mean) sizes, as well as RT. The models are more complex than before, and we think a comparison is justified.

We use the approach of \citet{McGSt:92} and compare the magnitudes of the following four terms in the dust equation of motion (equation~\ref{eq:dmot}): the inertial and temporal term {\fin}, the gravitational force {\fgravd}, the radiative pressure on the dust {\fradd}, and the drag force {\fdrag}. The inertial term is
\begin{eqnarray}
\fin=\frac{\partial}{\partial t}(\rhod v)+\nabla\cdot(\rhod v\,v),
\end{eqnarray}
where we discretised the advection term using the simpler arithmetic mean expression of van Leer.

\begin{figure}
\includegraphics{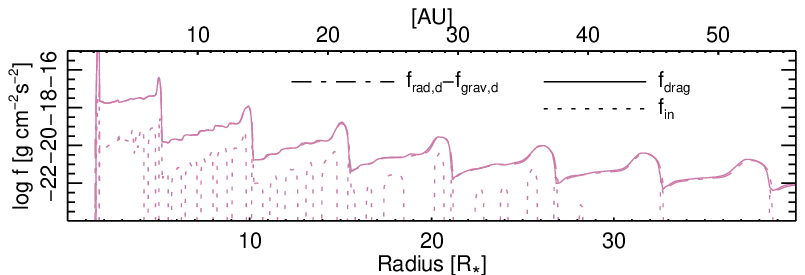}
\caption{Comparison of three terms: drag force {\fdrag}, radiative pressure and gravitational force $\fradd-\fgravd$, and inertial term {\fin} (log). This figure complements the panels in Fig.~\ref{fig:discwind}.\label{fig:discCMCone}}
\end{figure}

A comparison of the force terms for our set of models shows a mostly complete momentum coupling, where $\fdrag\simeq\fradd$ and $\fin<\fdrag$. However, we see some support for a relaxed coupling in the outer parts. We plot the radial structure of the force terms for the model with the lowest dust mass loss rate L3.70T28E88 (Section~\ref{disc:L370T28E88}) in Fig~\ref{fig:discCMCone}; we show all other properties of this model considered here in Fig.~\ref{fig:discwind}. The figure shows a nearly full coupling where $\fdrag\simeq\fradd-\fgravd$ and $\fin<\fdrag$ at all radii, except at dust fronts (e.g. $r\approx10\,\Rs$) where the difference is smaller. There is no clear decoupling as \citet{McGSt:92} find (cf. fig.~4), plausibly because grains are larger. The influence of decoupled phases ought to be small, as forces are weak at the large radii where decoupling occurs. Effects might be stronger at much larger radii, as \citealt{BeFr:83} find.

\begin{figure}
\includegraphics{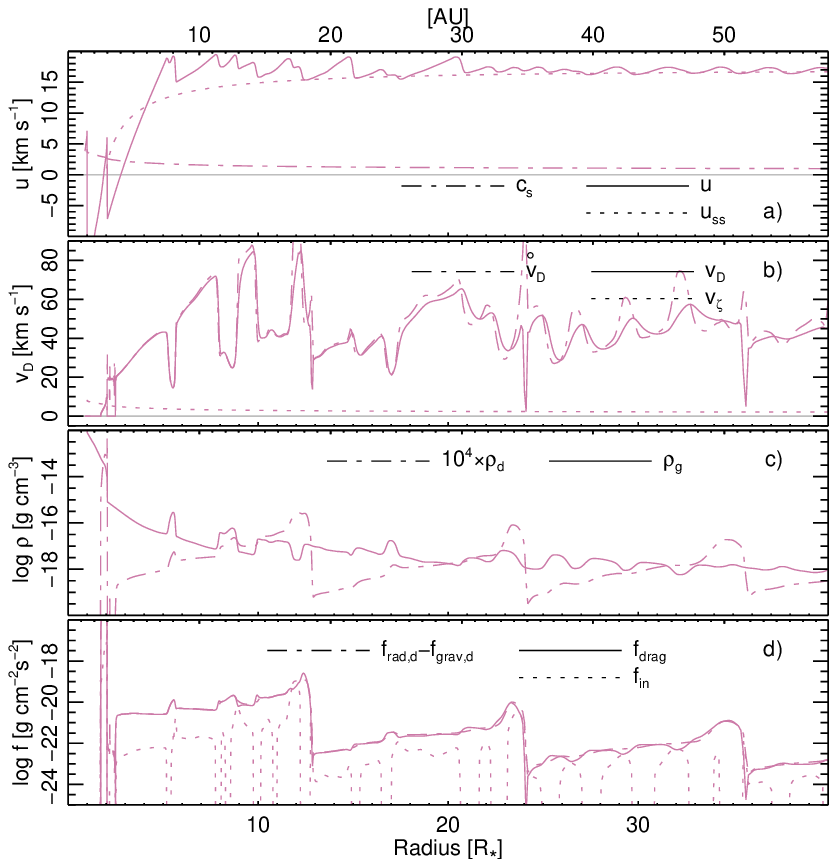}
\caption{Radial structure of a snapshot of the drift model L3.85T30E88 for the full modelled region. Panels a, b, and c correspond to panels a, c, and b in Fig.~\ref{fig:discwind}. Panel d corresponds to Fig.~\ref{fig:discCMCone}.\label{fig:discCMCtwo}}
\end{figure}

We also show the radial structure of model L3.85T30E88 where the decoupling is somewhat stronger, see Fig.~\ref{fig:discCMCtwo}. Here, $\fdrag\simeq\fradd-\fgravd$ and $\fin$ is also similar to the other terms for $r\ga24\Rs$, indicating more decoupled phases. As expected $\vDeq\simeq\vD$ for all radii, whilst there are some differences in the same outer parts.

Complete momentum coupling appears to be a suitable approximation in our models. It seems justified, however, to check this condition anew in future models where grains of different size move at separate drift velocities.

\bsp	
\label{lastpage}
\end{document}